\documentclass[aps,prx,twocolumn,superscriptaddress,nobalancelastpage]{revtex4-2}
\usepackage{preamble}

\begin{document}
\title{Diagrammatic expressions for steady-state distribution and static responses in population dynamics}
\author{Koya Katayama}
\email{koya.katayama@ubi.s.u-tokyo.ac.jp}
\affiliation{Department of Physics, The University of Tokyo, 7-3-1 Hongo, Bunkyo-ku, Tokyo 113-0033, Japan}
\author{Ryuna Nagayama}
\affiliation{Department of Physics, The University of Tokyo, 7-3-1 Hongo, Bunkyo-ku, Tokyo 113-0033, Japan}
\author{Sosuke Ito}
\affiliation{Department of Physics, The University of Tokyo, 7-3-1 Hongo, Bunkyo-ku, Tokyo 113-0033, Japan}
\affiliation{Universal Biology Institute, The University of Tokyo, 7-3-1 Hongo, Bunkyo-ku, Tokyo 113-0033, Japan}
\date{\today}

\begin{abstract}
One of the fundamental questions in population dynamics is how biological populations respond to environmental perturbations. In population dynamics, the mean fitness and the fraction of a trait in the steady state are important because they indicate how well the trait and the population adapt to the environment. In this study, we examine the parallel mutation-reproduction model, which is one of the simplest models of an evolvable population. As an extension of the Markov chain tree theorem, we derive diagrammatic expressions for the static responses of the mean fitness and the steady-state distribution of the population. For the parallel mutation-reproduction model, we consider self-loops, which represent trait reproduction and are excluded from the Markov chain tree theorem for the linear master equation.
To generalize the theorem, we introduce the concept of rooted $0$/$1$ loop forests, which generalize spanning trees with loops. We demonstrate that the weights of rooted $0$/$1$ loop forests yield the static responses of the mean fitness and the steady-state distribution.
Our results provide exact expressions for the static responses and the steady-state distribution. Additionally, we discuss approximations of these expressions in cases where reproduction or mutation is dominant. We provide numerical examples to illustrate these approximations and exact expressions.
\end{abstract}

\maketitle

\section{Introduction}
Understanding how biological populations respond to environmental changes is fundamental to biology because evolution itself can be viewed as a long-term response to these changes~\cite{fisher,crow2017introduction}. The stability of biological populations in the face of perturbations, such as climate change, is directly linked to their ability to respond to and adapt to shifting environments~\cite{sala2000global, gardner1970connectance, may1972will, allesina2012stability}.
In practical terms, understanding how viruses and cancers resist treatment and environmental stress remains an urgent challenge in the field of human health~\cite{balaban2004bacterial, aldridge2012asymmetry, wakamoto2013dynamic, maltas2025dynamic, markov2023evolution, furusawa2015global}.
Moreover, it has been a key challenge in agriculture to understand how pests acquire tolerance to pesticides~\cite{van1962integration, georghiou1972evolution}.
Substantial efforts have accordingly been made to quantify how various properties of biological populations respond to external changes.

Two of the most fundamental quantities in population dynamics are the mean fitness and the fraction of a trait.  
The mean fitness, defined as the growth rate of the total population, reflects how well a population is adapted to its environment. Because the mean fitness is experimentally accessible, the mean fitness is widely discussed not only theoretically~\cite{crow2017introduction, baake1997ising, hermisson2002mutation, baake2007mutation, saakian2007new, saakian2008dynamics, munoz2009solution, bratus2014linear, SEMENOV20151, Euler, lotka1907relation, powell1956growth, pigolotti2021generalized, fisher, EWENS1989167, frank1997price, eigen1989molecular, sughiyama2015pathwise, miyahara2022steady, leibler2010individual, kobayashi2015fluctuation, nozoe2017inferring, genthon2021universal, garcia2019linking, genthon2020fluctuation} but also experimentally~\cite{powell1956growth, scott2010interdependence, wang2010robust, lambert2015quantifying, hashimoto2016noise, nozoe2017inferring}.
The fraction of a trait indicates how well it adapts to the environment~\cite{crow2017introduction, eigen1989molecular,
furusawa2015global, markov2023evolution}, and this fraction is also experimentally accessible~\cite{lenski2017experimental,nozoe2017inferring}. These quantities can be discussed theoretically using the parallel mutation–reproduction model~\cite{crow2017introduction}, which is one of the simplest population genetics models described by a nonlinear master equation. Since then, it has been applied to ecosystems and population dynamics~\cite{baake1997ising, saakian2007new, saakian2008dynamics, munoz2009solution, bratus2014linear, SEMENOV20151, hermisson2002mutation, baake2007mutation}.

Although the parallel mutation-reproduction model is important, its mathematical treatment is less developed than that of the linear master equation because it is nonlinear. 
For example, it requires computing the eigenvalues and eigenvectors of a matrix to express the steady-state distribution and the static response of mean fitness, as discussed in Ref.~\cite{hermisson2002mutation}.
Here, the steady-state distribution is the fraction of a trait in the steady state, and the static response of mean fitness is the change in mean fitness between two different steady states.
Therefore, an analytical discussion may be difficult without considering the eigenvalue problem for a specific matrix. Indeed, many previous studies~\cite{baake1997ising, saakian2007new, saakian2008dynamics, munoz2009solution, bratus2014linear, SEMENOV20151} have assumed specific functional forms for the fitness landscape or the mutation rate. 

In contrast, the steady-state distribution and the static responses in the steady state are well understood for the linear master equation. For example, the steady-state distribution can be described by a canonical distribution if the detailed balance condition is satisfied~\cite{van1992stochastic}. Even when the detailed balance condition is not satisfied, the steady-state distribution can be graphically described using the weights of the spanning tree on Markov networks, thanks to the Markov chain tree theorem~\cite{moon1970counting, chaiken1982combinatorial, Cayley1856, Sylvester_James_Joseph_(1814–1897)_On_1908, borchardt1860ueber, Kirchhoff, maxwell1892treatise, hill1966studies, schnakenberg1976network} (also known as the matrix tree theorem) for the linear master equation. Recently, the static responses in the non-equilibrium steady state have also been discussed in stochastic thermodynamics based on the Markov chain tree theorem~\cite{owen2020universal, fernandes2023topologically}. Thus, due to the absence of a mathematical treatment similar to that established for the linear master equation in the parallel mutation-reproduction model, it has been difficult to handle analytically the steady-state distribution and the static responses of the mean fitness in the steady state for general cases, despite the importance of these quantities in population dynamics.

To address this issue, we extend the Markov chain tree theorem to the parallel mutation–reproduction model, and derive exact expressions for the steady-state distribution and {the} responses of the mean fitness in the steady state. To achieve this, we propose a graph called {a} \textit{rooted $0$/$1$ loop forest}, which is a generalization of a spanning tree. In terms of a loop in a rooted $0$/$1$ loop forest, the effect of self-reproduction can be naturally described. We thus obtain diagrammatic expressions for the steady-state distribution and the static responses of the mean fitness using the weights of rooted $0$/$1$ loop forests, as a generalization of the Markov chain tree theorem. We also discuss approximate expressions that do not use all weights, in cases where {mutation or natural selection} is dominant. We verify the validity of these approximate expressions and confirm the exact diagrammatic expressions numerically.

This paper is organized as follows. In Sec.~\ref{sec:Setup and Preliminariesy}, we introduce the parallel mutation-reproduction model and present some results from Ref.~\cite{hermisson2002mutation}.
In Sec.~\ref{sec:graph_theory}, we introduce some concepts from graph theory. In particular, we propose a new type of graph called a rooted $0$/$1$ loop forest.
In Sec.~\ref{sec:main_results} we present the main results and provide some examples.
In Sec.~\ref{sec:approximation}, we  approximate the static responses of the mean fitness and the steady-state distribution in cases where {mutation or natural selection} is dominant.
In Sec.~\ref{sec:comb_therapy}, we demonstrate how our results can be used to control the mean fitness of a population.

\section{Setup}
\label{sec:Setup and Preliminariesy}
Here, we introduce the parallel mutation-reproduction model, as well as quantities such as the steady-state distribution and the static responses of the mean fitness.

\subsection{The parallel mutation–reproduction model}
\begin{figure}[htbp]
    \centering
    \includegraphics[width=\linewidth]{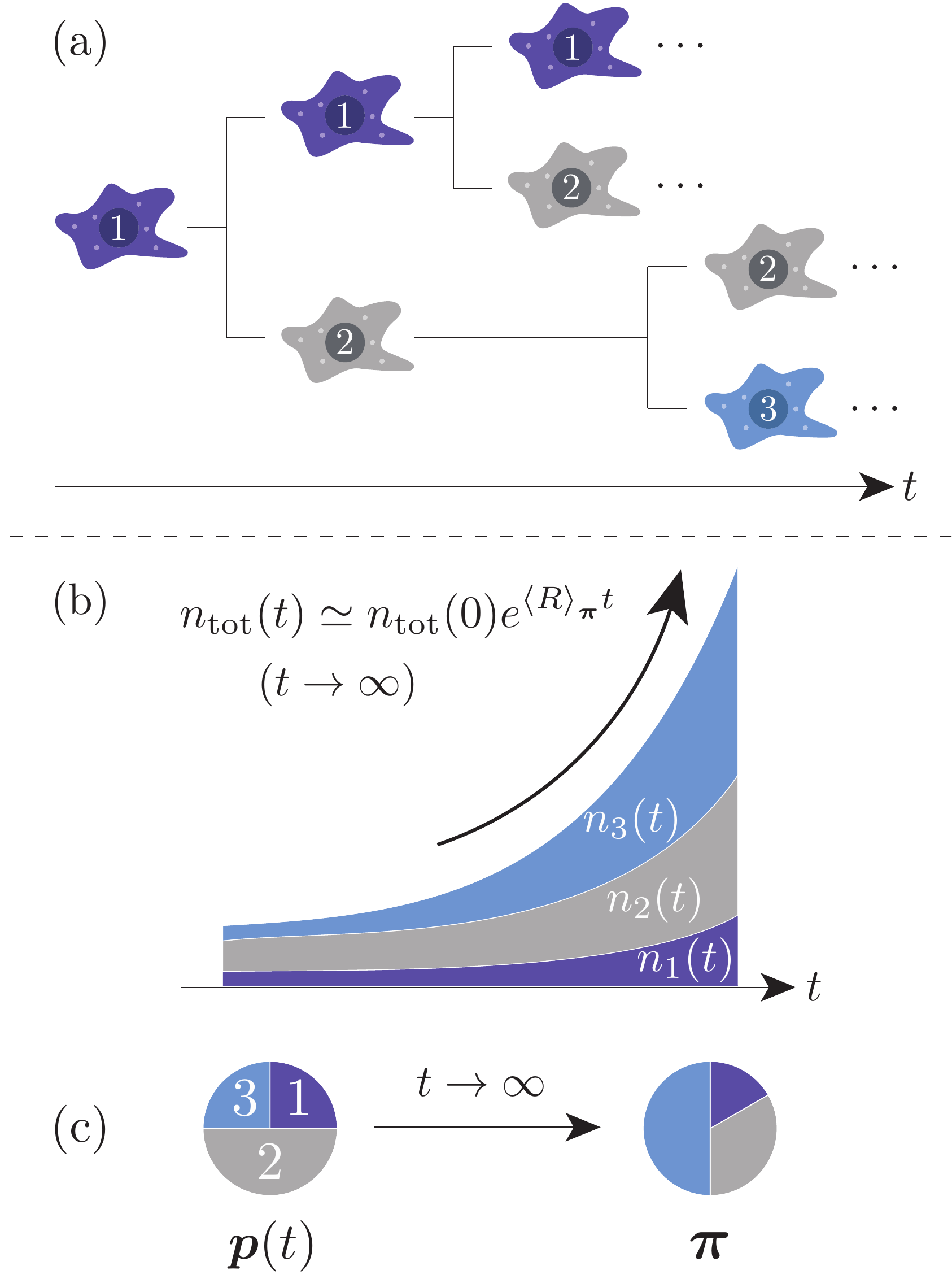}
    \caption{Schematic illustration of the model. (a) Each individual reproduces, and its offspring may mutate into a different trait.
    (b) In the long-time limit, the total number of individuals in the population grows exponentially with time, and $\langle R \rangle_{\bm \pi}$ is regarded as the reproductive rate of the total number of the individuals in the population.
    (c) The distribution of the population $\bm{p}(t)$ converges to the unique steady state $\ppi$ in the limit of $t\to\infty$.}
    \label{fig:set_up}
\end{figure}

We explain the parallel mutation-reproduction model~\cite{crow2017introduction} that we are dealing with in this paper.
Suppose that a population consists of $N$ subpopulations, and each subpopulation has a different trait.
Each individual in the population reproduces itself and mutates into a different trait (Fig.~\ref{fig:set_up}(a)).
Although other factors such as migration or predation may affect the dynamics of the population, we neglect them for simplicity.
Let $n_i(t)\ge 0$ denote the number of individuals in the subpopulation with {trait} $i\in\{1,\cdots, N\}$ at time $t$.
In the following, we assume that $n_i(t)$ is large enough to be treated as continuous number.
Let $R_i\in\mathbb{R}$ denote the reproductive rate (or the fitness) of trait $i$, and $M_{ij}\ge 0\, (i\ne j)$ denote the mutation rate from trait $j$ to trait $i$.
As explained in Appendix~\ref{app:derive_equation}, $R_i$ incorporates the net effect of reproduction and death of trait $i$. 
In addition, $M_{ij}$ incorporates the combined effect of the reproduction of trait $j$ and the mutation of its descendants.
Then, we can write down the equation that describes the dynamics of $n_i(t)$ as follows,
\begin{align}\label{eq:dn_dt}
    \frac{d}{dt}n_i(t)=R_in_i(t)+\sum_{j(\ne i)}(M_{ij}n_j(t)-M_{ji}n_i(t)).
\end{align}
If there is competition for food or predator-prey interaction between individuals, $\{R_i\}_i$ may depend on $\bm{n}(t):=(n_1(t),\ \cdots,\ n_N(t))^\top$.
However, in this paper we assume that $\{R_i\}_i$ are independent of $\bm{n}(t)$.
We also assume that all $R_i$ and $M_{ij}$ do not depend directly on time $t$, which means that the environment is stationary.
For simplicity, we rewrite Eq.~\eqref{eq:dn_dt} in the vector form.
To do this, we define $M_{ii}:=-\sum_{j(\ne i)}M_{ji}$, $\MM:=(M_{ij})\in\mathbb{R}^{N\times N}$, and $\RR:=\mathrm{diag}(R_1,\ \cdots,\ R_N)\in\mathbb{R}^{N\times N}$, where $\mathbb{R}^{N\times N}$ denotes the set of all 
$N\times N$ real matrices.
Then, Eq.~\eqref{eq:dn_dt} can be rewritten as
\begin{align}\label{eq:dn_dt_matrix}
    \frac{d}{dt}\bm{n}(t)=(\RR+\MM)\bm{n}(t).
\end{align}
We also assume that the matrix $\MM$ is irreducible, which means that for any pair of $i$ and $j$, trait $i$ can mutate into trait $j$ in finite steps.

In this paper, we focus on the \textit{distribution} of the traits rather than the number of them.
Defining the total number of the population as $n_{\mathrm{tot}}(t):=\sum_in_i(t)$, the distribution of the traits is defined as $\bm{p}(t):=(p_1(t),\ \cdots,\ p_N(t))^\top$ with $p_i(t):=n_i(t)/n_{\mathrm{tot}}(t)$.
From Eq.~\eqref{eq:dn_dt}, we obtain the equation that describes the dynamics of $\bm{p}(t)$ as
\begin{align}\label{eq:dp_dt}
    \frac{d}{dt}\bm{p}(t)=\left(\RR-\ev{R}_{\bm{p}(t)}\mathsf{I}+\MM\right)\bm{p}(t),
\end{align}
where $\ev{R}_{\bm{p}(t)}:=\sum_iR_ip_i(t)$ is the \textit{mean fitness} of the population at time $t$, and $\mathsf{I}\in\mathbb{R}^{N\times N}$ is the identity matrix (see also Appendix~\ref{app:from_n_to_p} for the derivation). 
The term $-\ev{R}_{\bm{p}(t)}\bm{p}(t)$ in Eq.~\eqref{eq:dp_dt} ensures that $\bm{p}(t)$ remains normalized over time, i.e., $\sum_i p_i(t)=1$.
Here, Eq.~\eqref{eq:dp_dt} is considered a nonlinear master equation because the term $-\ev{R}_{\bm{p}(t)}\bm{p}(t)$ is a nonlinear function of $\bm{p}(t)$.

Note that the mean fitness $\ev{R}_{\bm{p}(t)}$ can be interpreted as the reproductive rate of the total number of individuals in the population because the equation 
\begin{align}\label{eq:n_tot}
    \frac{d}{dt}n_{\mathrm{tot}}(t)=\ev{R}_{\bm{p}(t)}n_{\mathrm{tot}}(t)
\end{align}
is obtained by summing Eq.~\eqref{eq:dn_dt} over $i$. 

We note that Eq.~\eqref{eq:dp_dt} reduces to the linear master equation 
\begin{align}
    \frac{d}{dt}\bm{p}(t)=\MM\bm{p}(t),
    \label{mastereq}
\end{align}
when all reproductive rates are equal because $\ev{R}_{\bm{p}(t)}=R_i$ holds for any $i$ and {$t$}, and $\RR=\ev{R}_{\bm{p}(t)}\II$ is obtained.

\subsection{Steady-state distribution}
\label{subsec:pi}
We focus on the steady state of the population, which is one of the main topics of this paper. An expression for the steady-state distribution and its properties are obtained by applying the Perron--Frobenius theorem~\cite{pillai2005perron} (see also Appendix~\ref{app:pi}).

We clarify what is the steady state in the parallel mutation-reproduction model.
Equation~\eqref{eq:n_tot} implies that the total number of the population could diverge in the limit of ${t\to\infty}$ due to the reproduction (Fig.~\ref{fig:set_up} (b)).
In contrast, the irreducibility of $\MM$ and the Perron--Frobenius theorem ensure that the distribution of the population $\bm{p}(t)$ converges to the unique distribution $\ppi\in\mathbb{R}^N$, which is the \textit{steady-state distribution} (Fig.~\ref{fig:set_up}(c)),
\begin{align}\label{eq:pi}
    \ppi:=\lim_{t\to\infty}\bm{p}(t).
\end{align}
For the details on the existence and uniqueness of $\ppi$, see Appendix~\ref{app:pi}. Equation~\eqref{eq:pi} implies that the mean fitness also converges to the unique value $\meanfit$: $
    \lim_{t\to\infty}\ev{R}_{\bm{p}(t)}=\meanfit$.
This $\meanfit$ can be interpreted as follows.
The solution of Eq.~\eqref{eq:n_tot} is obtained as $n_{\mathrm{tot}}(t)=n_{\mathrm{tot}}(0)\exp(\int_0^{t}\ev{R}_{\bm{p}(t')} dt')$. By considering the long-time limit, we obtain 
\begin{align}
    \lim_{t\to\infty}\frac{1}{t}\ln\frac{n_{\mathrm{tot}}(t)}{n_{\mathrm{tot}}(0)}=\meanfit.
\end{align}
Roughly speaking, this equality means that the total number of the population $n_{\mathrm{tot}}(t)$ increases exponentially as $n_{\mathrm{tot}}(t)\simeq n_{\mathrm{tot}}(0)e^{\meanfit t}$ in the long-time limit (Fig.~\ref{fig:set_up}(b)).
Therefore, we can interpret the mean fitness $\meanfit$ as the reproductive rate of the total number of the individuals in the population in the long-time limit.

Perron--Frobenius theorem also tells us that every component of $\ppi$ is strictly positive, i.e., $\pi_i>0$
for any $i$ (see Appendix~\ref{app:pi} for the details). This fact means that no trait will go extinct in the long-time limit if $\MM$ is irreducible.  

The steady-state distribution $\ppi$ can be regarded as the eigenvector of $\RR+\MM$.
In fact, combining Eq.~\eqref{eq:dp_dt} and the fact that $\ppi$ is time-independent, we obtain
\begin{align}\label{eq:right_eigvec}
    (\RR+\MM)\ppi=\meanfit\ppi,
\end{align}
which means that $\ppi$ is the eigenvector of $\RR+\MM$ with eigenvalue $\meanfit$.

The left eigenvector of $\RR+\MM$ corresponding to the eigenvalue $\meanfit$ also plays an important role in population dynamics.
The Perron--Frobenius theorem implies that there exists the unique vector $\zzeta\in\mathbb{R}^N$ which satisfies the following equations (see also Appendix~\ref{app:pi} for the details),
\begin{align}   \zzeta^\top(\RR+\MM)&=\meanfit\zzeta^\top,\label{eq:left_eigvec}\\
    \zzeta^\top\ppi&=1,\label{eq:normal}
\end{align}
and $\zeta_i>0$ for any $i$.
In other words, $\zzeta$ is the left eigenvector of $\RR+\MM$ with eigenvalue $\meanfit$ normalized by $\ppi$ and whose components are all positive.
Note that, in the case of the linear master equation [Eq.~(\ref{mastereq})], $\zzeta$ is the vector whose every component is $1$, i.e., $\zzeta=\bm{1}:=(1, \cdots, 1)^\top$.
In fact, $\bm{1}^\top\MM=\bm{0}^\top=\bm{1}^\top(\meanfit\II-\RR)$ is satisfied because $\sum_i M_{ik}=0$ for any $k$. $\bm{1}^\top\ppi=\sum_i\pi_i=1$ is also satisfied. The positivity of the elements of $\bm 1$ is also obviously satisfied.

\subsection{Static responses of the mean fitness}
\label{subsec:resp}
Here we discuss the static responses of the mean fitness in the steady state, which are also the main topics of this paper.
Since Eq.~\eqref{eq:dn_dt} contains two kinds of parameters, $R_i$ and $M_{ij}$, we can consider two kinds of the static responses $\partial _{R_i}\meanfit := \partial \meanfit/ \partial R_i$ and $\partial_{M_{ij}}\meanfit$, which are important quantities that explain the variation in the mean fitness due to changes in the parameters.
{
We note that the general static response of the mean fitness can be expressed as a linear combination of these quantities.
Thus, it suffices to consider only $\partial _{R_i}\meanfit$ and $\partial_{M_{ij}}\meanfit$.
Indeed, if the reproductive rates and mutation rates depend on a parameter $\theta\in\mathbb{R}$ as $\{R_i(\theta)\}$ and $\{M_{ij}(\theta)\}$, the static response of the mean fitness with respect to $\theta$ is given by $\partial_{\theta} \meanfit=\sum_i\partial_{\theta}R_i(\theta) \partial_{R_i}\meanfit+\sum_{i, j(\ne i)}\partial_{\theta} M_{ij}(\theta)\partial_{M_{ij}}\meanfit$.}

\begin{figure*}[hbtp]
    \centering
    \includegraphics[width=\linewidth]{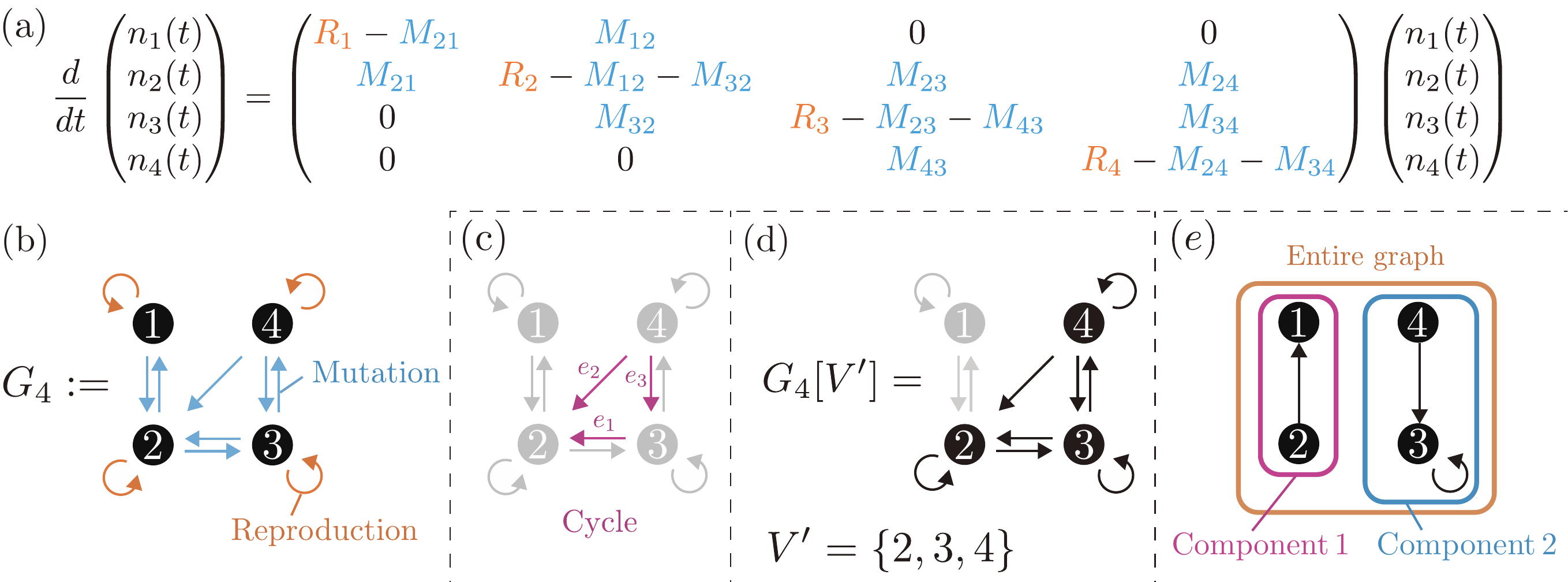}
    \caption{Example of the basic graph, a cycle, an induced subgraph, and a component. (a) An example of Eq.~\eqref{eq:dn_dt_matrix} for a population with four traits. The mutation rates, $M_{13}$, $M_{14}$, $M_{31}$, $M_{41}$, and $M_{42}$, are set to zero. 
    (b) The basic graph, denoted by $G_4$, corresponding to the population dynamics shown in (a). At each vertex, there is a loop representing reproduction. Also, if the mutation rate $M_{ij}$ is nonzero, there is a directed edge from vertex $j$ to $i$. 
    (c) An example of a cycle in the graph shown in (b). 
    This cycle $(e_1, e_2, e_3)$ consists of three edges: $e_1=(2\leftarrow 3)$, $e_2=(2\leftarrow 4)$, and $e_3=(3\leftarrow 4)$. Here, a trail is {$3-2-4-3$}.
    As this example illustrates, the edges in a cycle do not need to be oriented in the same direction along the cycle. 
    (d) An example of an induced subgraph $G_4[V'] \subseteq G_4$ for $V'=\{2, 3, 4\}$. $G_4[V']$ is obtained from $G_4$ by removing vertex $1$ and the edges with at least one endpoint at vertex $1$: $(1\leftarrow 1)$, $(1\leftarrow 2)$, and $(2\leftarrow 1)$. 
    (e) An example of components. The entire graph consists of two components: component 1 and component 2.}    
    \label{fig:Dynamics_Graph}
\end{figure*} 

The static responses $\partial _{R_i}\meanfit$ and $\partial_{M_{ij}}\meanfit$ are related to the left and right eigenvectors with the eigenvalue $\meanfit$.
Interestingly, Ref.~\cite{hermisson2002mutation} shows that the following equations
\begin{align}
  \frac{\partial\meanfit}{\partial R_i}&=\zeta_i\pi_i,\label{eq:resp_R}\\
  \frac{\partial\meanfit}{\partial M_{ij}}&=(\zeta_i-\zeta_j)\pi_j,\label{eq:resp_M}
\end{align}
hold for $i \neq j$ (see also  Appendix~\ref{app:resp}).
While these expressions give the rigorous forms of the static responses of the mean fitness, it is difficult to express these static responses explicitly by using only the given parameters, $\{R_i\}$ and $\{M_{ij}\}$.
This is because we need to solve the eigenvalue problem of the matrix $\RR+\MM$ to obtain the explicit forms of $\zeta_i\pi_i$ and $\zeta_i\pi_j$.
However, {in this paper, we will give diagrammatic expressions for the static responses of the mean fitness}. The diagrammatic {expressions include} the terms not only $\{R_i\}$ and $\{M_{ij}\}$ but also the mean fitness $\meanfit$. 

Note that $\{ \partial_{R_i}\meanfit \}$, or equivalently $\{\zeta_i\pi_i \}$, can be regarded as a probability distribution.
From Eq.~\eqref{eq:normal} and the positivity of $\pi_i$ and $\zeta_i$,  we obtain $\zeta_i\pi_i  >0$ and $\sum_i \zeta_i\pi_i=1$.
Therefore, $\{\zeta_i\pi_i \}$ is referred to as the \textit{ancestral distribution}~\cite{hermisson2002mutation}.
Combining Eq.~\eqref{eq:resp_R} and the positivity of $\zeta_i\pi_i$, we find $0<\partial_{R_i}\meanfit<1$ for any $i$.
Therefore, the mean fitness in the long-term limit is monotonically increasing with respect to any increase in the reproductive rate.

\section{Graph theory}
\label{sec:graph_theory}
In this section, we explain the graph theory for the parallel mutation-reproduction model. In particular, we newly propose a \textit{rooted} $0$/$1$ \textit{loop forest}, which plays a key role in the main result.

\subsection{Basic graph}
We introduce a \textit{basic graph}, which is a directed graph representing the population structure described in Eq.~\eqref{eq:dn_dt}. The term ``basic graph'' is introduced in Ref.~\cite{schnakenberg1976network} for the linear master equation, and we use the same term for the parallel mutation-reproduction model as a generalization of the basic graph in Ref.~\cite{schnakenberg1976network}.
The construction of a basic graph, denoted by $G=(V(G), E(G))$, is outlined as follows (see Fig.~\ref{fig:Dynamics_Graph}(a) and (b) for an example of the dynamics and the corresponding basic graph).
The vertex set of $G$, denoted by $V(G)$, is defined as the set of the traits in the population, i.e., $V(G):=\{1, \cdots, N\}$.
In other words, each vertex of $G$ represents a trait in the population.
Next, we define the edge set of $G$, denoted by $E(G)$, as follows.
Let $(i\leftarrow j)$ be a directed edge from vertex $j$ to vertex $i$.
In particular, $(i\leftarrow i)$ represents a loop at vertex $i$.
Using this notation, $E(G)$ is defined as
\begin{align}\label{eq:def_E}
    E(G):=&\{(i\leftarrow i)\mid i\in V(G)\}\notag\\
    &\cup\{(i\leftarrow j)\mid M_{ij}\ne 0,\, i \in V(G),\, j(\ne i)\in V(G)\}.
\end{align}
A loop at vertex $i$ represents the reproduction of trait $i$, and a directed edge from vertex $j$ to $i$ represents the mutation from trait $j$ to $i$.
Note that if the mutation from trait $j$ to $i$ never occurs, i.e., $M_{ij}=0$ holds, then $(i\leftarrow j)$ does not belong to $E(G)$.

\subsection{Cycle}
We introduce the concept of a \textit{cycle}~\cite{godsil2013algebraic}. A cycle is a non-empty trail (and is not a loop) in which only the first and last vertices are the same. A sequence of edges $(e_1,\ \cdots, e_L)\, (L\ge 3)$ is called a cycle if starting from a vertex, it is possible to return to the same vertex by following all the edges $(e_1,\ \cdots, e_L)$ sequentially without visiting the same vertex twice. 
Here, ``following'' an edge $e$ means moving along it either in the given direction or in the opposite direction. 
Thus, edge directions are not taken into account when defining a cycle, and a loop is not included in a cycle. 
See Fig.~\ref{fig:Dynamics_Graph}(c) for an example.

\begin{figure*}[bhtp]
    \centering
    \includegraphics[width=\linewidth]{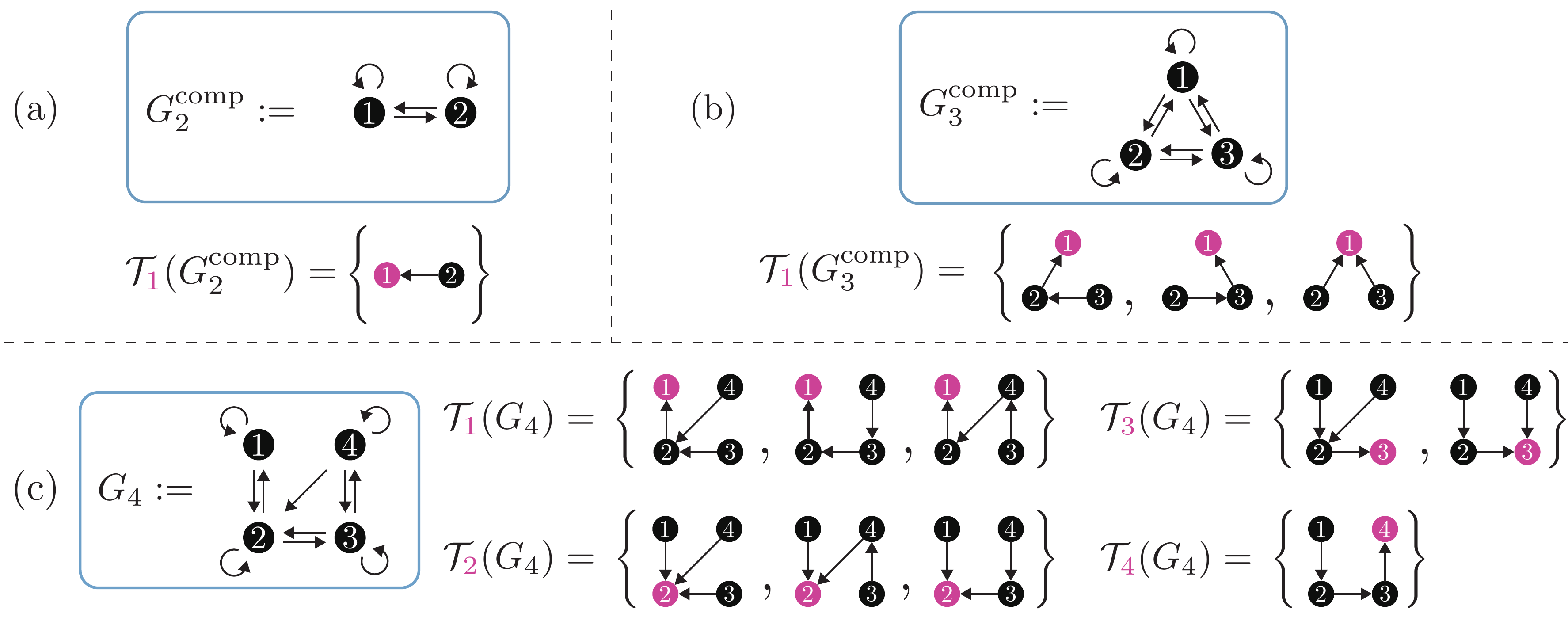}
    \caption{Examples of basic graphs and their rooted spanning trees. 
    (a) Complete graph on two vertices with loops $G^{\mathrm{comp}}_2$ and $\mathcal{T}_1(G^{\mathrm{comp}}_2)$. 
    (b) Complete graph on three vertices with loops $G^{\mathrm{comp}}_3$ and $\mathcal{T}_1(G^{\mathrm{comp}}_3)$. 
    (c) Graph on four vertices with loops $G_4$, which is not complete. 
    We show all sets of rooted spanning trees.}
    \label{fig:Graph_tree}
\end{figure*}

\subsection{Subgraph and component}
We introduce a \textit{subgraph}.
A subgraph of a graph $G$ is defined as a graph $H$ such that $V(H)\subseteq V(G)$ and $E(H)\subseteq E(G)$~\cite{godsil2013algebraic}, and we write $H\subseteq G$.

An \textit{induced subgraph} is a special kind of subgraph.
Given a graph $G$ and a subset of its vertex set $V'\subseteq V(G)$, the subgraph of $G$ induced by $V'$, denoted by $G[V']$, is defined as the graph whose vertex set is $V'$ and whose edge set consists of all edges in $E(G)$ whose endpoints are both in $V'$.
In other words, $G[V']$ is the subgraph obtained from $G$ by removing all vertices in $V(G)\setminus V'$ and all edges with at least one endpoint in $V(G)\setminus V'$.
See Fig.~\ref{fig:Dynamics_Graph}(d) for an example of an induced subgraph.

A \textit{component} is also a special kind of subgraph. A component of a graph $G$ is a connected subgraph $H\subseteq G$ that is not part of any larger connected subgraph~\cite{west2001introduction} (see Fig.~\ref{fig:Dynamics_Graph}(e) for an example).
Roughly speaking, a component of $G$ is an \textit{island} in $G$.

\subsection{Union of graphs}
We explain the union of graphs.
Suppose that there are $M$ graphs $H^{(\alpha)},\, \alpha\in\{1, \cdots, M\}$.
The \textit{union} of $\{H^{(\alpha)}\}_\alpha$ is defined as the graph with the union of the vertices and the union of the {edges},
\begin{align}
    \bigcup_{\alpha=1}^{M} H^{(\alpha)}:=\left(\bigcup_{\alpha=1}^{M}V(H^{(\alpha)}), \bigcup_{\alpha=1}^{M}E(H^{(\alpha)})\right).
\end{align}
We also use the notation $H^{(1)} \cup H^{(2)}$ for the union of $H^{(1)}$ and $H^{(2)}$, which is defined as $H^{(1)} \cup H^{(2)}:= (V(H^{(1)}) \cup V(H^{(2)}) , E(H^{(1)}) \cup E(H^{(2)}) )$. Moreover, when a graph $H$ consists of only one edge $e$, we sometimes identify $H$ with $e$. Thus, we also use the notation $(i \leftarrow j) \cup (k\leftarrow l)$ for the union of graphs expressed by the edges $(i \leftarrow j)$ and $(k\leftarrow l)$. For example, $\bigcup_{j(\neq i)} (j \leftarrow j)$ represents the graph that has a loop at every vertex except vertex $i$.

\subsection{Rooted spanning tree}
\label{subsec:spanning_tree}
We explain {\it rooted spanning trees}, which play a key role in the analysis of the linear master equation.
Given a basic graph $G$ and its subgraph $H$, we call $H$ a \textit{spanning tree of $G$ rooted at vertex $i$} if $H$ satisfies all of the following conditions~\cite{schnakenberg1976network}:
\begin{enumerate}
    \item $H$ contains no cycles.
    \item $H$ contains no loops.
    \item $V(H)=V(G)$.
    \item $H$ is connected.
    \item All edges in $H$ are directed {to} vertex $i$.
\end{enumerate}
If $H$ satisfies all of the above conditions except the fifth condition, then we call $H$ a {\it spanning tree}.
A graph $H$ is simply called a {\it rooted spanning tree} if it is a spanning tree rooted at a certain vertex.
Note that the first and fourth conditions guarantee the existence and uniqueness of the path between any two vertices in $H$, which allows us to determine the direction of each edge in the fifth condition. In addition to the first, second, and third, and {fifth} conditions above, it is known that a {rooted} spanning tree can be characterized by the following condition~\cite{west2001introduction},
\begin{enumerate}
\setcounter{enumi}{5}
    \item $H$ has $|V(G)|-1$ edges,
\end{enumerate}
where $|V(G)|$ denotes the number of elements in $V(G)$.

\begin{figure*}[htbp]
    \centering
    \includegraphics[width=\linewidth]{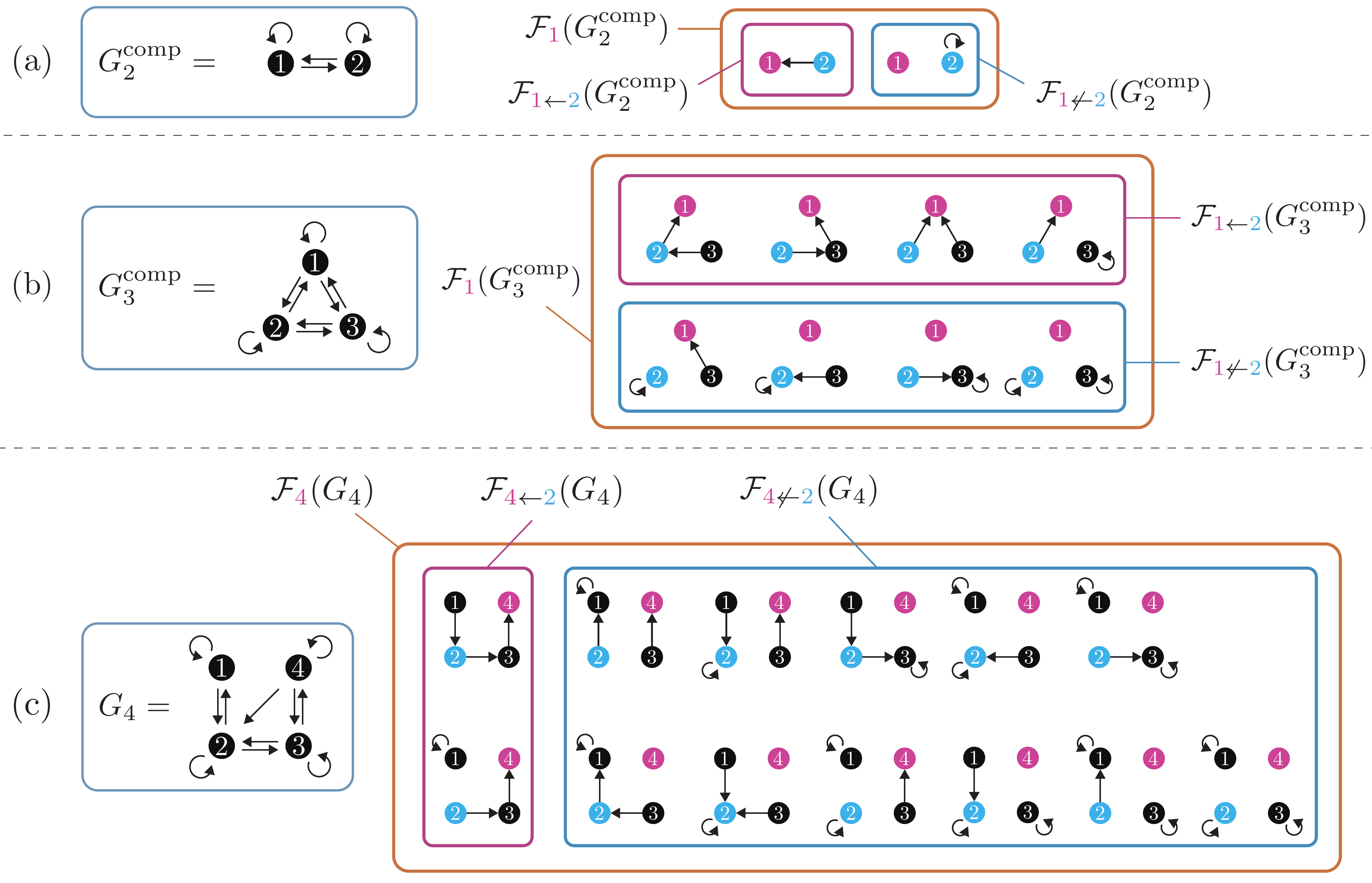}
    \caption{Examples of rooted $0$/$1$ loop forests. 
    (a) $\mathcal{F}_1(G^{\mathrm{comp}}_2)$, $\mathcal{F}_{1\leftarrow 2}(G^{\mathrm{comp}}_2)$, and $\mathcal{F}_{1 \not\leftarrow 2}(G^{\mathrm{comp}}_2)$.
    (b) $\mathcal{F}_1(G^{\mathrm{comp}}_3)$, $\mathcal{F}_{1\leftarrow 2}(G^{\mathrm{comp}}_3)$, and $\mathcal{F}_{1 \not\leftarrow 2}(G^{\mathrm{comp}}_3)$.
    (c) $\mathcal{F}_4(G_4)$, $\mathcal{F}_{4\leftarrow 2}(G_4)$, and $\mathcal{F}_{4 \not\leftarrow 2}(G_4)$. We can check that any rooted $0$/$1$ loop forest of graph $G$ has $|V(G)| -1$ edges, and $\mathcal{F}_i(G) = \mathcal{F}_{i\leftarrow j} (G) \cup \mathcal{F}_{i \not \leftarrow j} (G)$ and $\mathcal{F}_{i\leftarrow j} (G) \cap \mathcal{F}_{i \not \leftarrow j} (G)=\emptyset$ are satisfied.
    }
    \label{fig:Graph_forest}
\end{figure*}

Let $\mathcal{T}_i(G)$ denote the set of all spanning trees of $G$ rooted at vertex $i$.
For any $H\in\mathcal{T}_i(G)$, we call vertex $i$ the root of $H$. See Fig.~\ref{fig:Graph_tree} for examples of rooted spanning trees. Note that the set $\mathcal{T}_i(G)$ is used for a diagrammatic expression for the steady-state distribution in the linear master equation~\cite{schnakenberg1976network}. 

\subsection{Rooted $0$/$1$ loop forest}
\label{subsec:rooted_loop_forest}
We propose a new class of graphs called {\it rooted $0$/$1$ loop forests}, which play a central role in this paper. 
Given a basic graph $G$ and its subgraph $H$, $H$ is called a {\textit{$0$/$1$ loop forest of $G$ rooted at vertex $i$}} if $H$ satisfies all of the following conditions:
\begin{enumerate}
    \item $H$ contains no cycles.
    \item $V(H)=V(G)$.
    \item In each component of $H$ that does not contain vertex $i$, there is exactly one loop.
    \item In the component of $H$ that contains vertex $i$, there are no loops. 
    \item In the component of $H$ that contains vertex $i$, all edges are directed to vertex $i$.
    In each component of $H$ that does not contain vertex $i$, all edges are directed to the vertex with the loop.
\end{enumerate}
If $H$ satisfies all of {the} above conditions except the fifth condition for a certain vertex $i$, we call $H$ a {\it $0$/$1$ loop forest}. Thus, a $0$/$1$ loop forest consists of one component with no loops and {$N_{\mathrm{c}}-1$} components with one loop each where {$N_{\mathrm{c}}$} is the total number of components. In graph theory, a {\it forest} refers to a disjoint union of trees with no loops.  A $0$/$1$ loop forest is obtained by adding a loop to each component of a forest, except for one component.
For a $0$/$1$ loop forest, the root must be chosen from a vertex of a component that has no loops. A graph $H$ is also simply called a {\it rooted $0$/$1$ loop forest} if it is a $0$/$1$ loop forest rooted at a certain vertex. Note that the {first and third} conditions allow us to determine the directions of the edges in the fifth condition.

Let $\mathcal{F}_i(G)$ denote the set of all $0$/$1$ loop forests of $G$ rooted at vertex $i$. For any $H\in\mathcal{F}_i(G)$, we call vertex $i$ the root of $H$.
We call $H\in\mathcal{F}_i(G)$ an {\textit{$(i,j)$-connected $0$/$1$ loop forest of $G$ rooted at vertex $i$}} if $H$ satisfies the following condition:
\begin{enumerate}
\setcounter{enumi}{5}
    \item Vertex $i$ and vertex $j$ are connected in $H$. 
\end{enumerate}
We call $H\in\mathcal{F}_i(G)$ an {\textit{$(i,j)$-disconnected $0$/$1$ loop forest of $G$ rooted at vertex $i$}} if $H$ satisfies the following condition:
\begin{enumerate}
\setcounter{enumi}{6}
    \item Vertex $i$ and vertex $j$ are not connected in $H$.
\end{enumerate}

We define two subsets of $\mathcal{F}_i(G)$, $\mathcal{F}_{i\leftarrow j}(G)$ and $\mathcal{F}_{i\not\leftarrow j}(G)$, as follows. Let $\mathcal{F}_{i\leftarrow j}(G)$ denote the set of all $(i,j)$-connected $0$/$1$ loop forests of $G$ rooted at vertex $i$. Let $\mathcal{F}_{i\not\leftarrow j}(G)$ denote the set of all $(i,j)$-disconnected $0$/$1$ loop forests of $G$ rooted at vertex $i$. Because all graphs in $\mathcal{F}_i(G)$ can  be divided into two subsets based on whether vertex $i$ and vertex $j$ are connected or not, $\mathcal{F}_{i}(G)=\mathcal{F}_{i\leftarrow j}(G)\cup\mathcal{F}_{i\not\leftarrow j}(G)$ and $\mathcal{F}_{i\leftarrow j}(G)\cap\mathcal{F}_{i\not\leftarrow j}(G)=\emptyset$ are satisfied. We regard $\mathcal{F}_{i\leftarrow i}(G)$ and $\mathcal{F}_{i \not\leftarrow i}(G)$ as $\mathcal{F}_{i\leftarrow i}(G)=\mathcal{F}_{i}(G)$ and $\mathcal{F}_{i \not\leftarrow i}(G)=\emptyset$, respectively. This is because we consider vertex $i$ and vertex $i$ to be trivially connected. 

For any {basic graph}, we can construct all rooted $0$/$1$ loop forests as discussed in Appendix~\ref{app:enumeration}. 
See also Fig.~\ref{fig:Graph_forest} for examples of rooted $0$/$1$ loop forests for specific graphs.

We consider the definition of rooted $0$/$1$ loop forests in a constructive way. Suppose that a rooted $0$/$1$ loop forest $H\in\mathcal{F}_i(G)$ is given.
Let $N_{\mathrm{c}}$ be the number of components of $H$, $V(H^{(\alpha)})$ be the vertex set of the $\alpha$-th component, and $H^{(1)}$ be the component of $H$ that contains {vertex} $i$. $H^{(1)}$ is a spanning tree of the induced subgraph $G[V(H^{(1)})]$ rooted at {vertex} $i$ because $H^{(1)}$ {contains neither cycles nor loops, is connected, and all its edges} are directed to {vertex} $i$. Thus, $H= H^{(1)}$ when $N_{\mathrm{c}}=1$.
For any $\alpha\in\{2,\cdots, N_{\mathrm{c}}\}$ ($N_{\mathrm{c}} \geq 2$), the $\alpha$-th component contains a rooted spanning tree of the induced subgraph $G[V(H^{(\alpha)})]$, denoted by $H^{(\alpha)}$.
Each $\alpha$-th component has the unique loop at its root, denoted by $r^{(\alpha)}$.
Therefore, when $N_{\mathrm{c}} \geq 2$, we can rewrite $H$ as
\begin{align}\label{eq:H_union}    
H=H^{(1)}\cup\bigcup_{\alpha=2}^{N_{\mathrm{c}}}\left(H^{(\alpha)}\cup (r^{(\alpha)}\leftarrow r^{(\alpha)})\right).
\end{align}
Based on this expression, we can show that any rooted $0$/$1$ loop forest of $G$ has $|V(G)|-1$ edges because $\bigcup_{\alpha=1}^{N_{\mathrm{c}}}H^{(\alpha)}$ has $|V(G)|-N_{\mathrm{c}}$ edges and $\bigcup_{\alpha=2}^{N_{\mathrm{c}}}(r^{(\alpha)}\leftarrow r^{(\alpha)})$ has $N_{\mathrm{c}}-1$ edges.
 
We state a relationship between rooted spanning trees and rooted $0$/$1$ loop forests.
$\mathcal{F}_{i\leftarrow j}(G)$ contains all spanning trees of $G$ rooted at vertex $i$, i.e., $\mathcal{T}_i(G)\subseteq\mathcal{F}_{i\leftarrow j}(G)$ for any vertex $i$ {and vertex $j$}. 
This is because $H\in\mathcal{T}_i(G)$ can be regarded as the rooted $0$/$1$ loop forest for $N_{\mathrm{c}}=1$, and any vertex $i$ and vertex $j$ are connected for any rooted spanning tree $H\in\mathcal{T}_i(G)$.
For example, we can check $\mathcal{T}_i(G)\subseteq\mathcal{F}_{i\leftarrow j}(G)$ by comparing Figs.~\ref{fig:Graph_tree} and ~\ref{fig:Graph_forest}.

Rooted $0$/$1$ loop forests can be considered as a natural generalization of rooted spanning trees. 
Each rooted spanning tree represents the influence of the other vertices on a given root.
The parallel mutation-reproduction model involves a contribution of reproduction, which can influence the distribution at any vertex. 
Therefore, each loop can be regarded as representing the influence of the vertex with the loop on any vertex.
Consequently, each rooted $0$/$1$ loop forest also represents the influence of the other states on a given root within the parallel mutation-reproduction model. 
 
\subsection{Weight of a graph}
We introduce the \textit{weight} of a graph which is applicable to a rooted $0$/$1$ loop forest. This weight of a graph is a generalization of the weight of a rooted spanning tree, which provides a diagrammatic expression for the steady-state distribution for the linear master equation~\cite{hill1966studies, schnakenberg1976network}. In the main results, we use this weight to obtain diagrammatic expressions for the steady-state distribution and the static responses in the parallel mutation-reproduction model.

\begin{figure}[htbp]
    \centering
    \includegraphics[width=\linewidth]{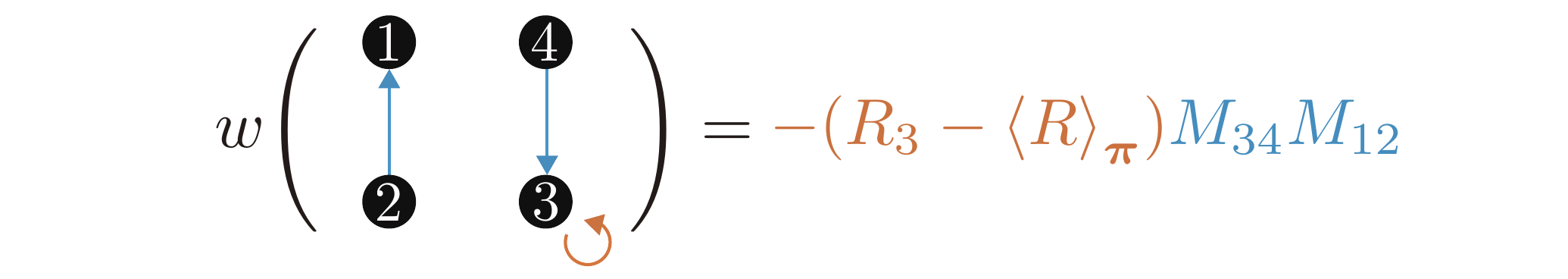}
    \caption{An example of the weight of a graph. 
    We consider a graph with two directed edges and one loop: $(1\leftarrow 2)$, $(3\leftarrow 4)$, and $(3\leftarrow 3)$. The weights of the edges are given by $\tilde{w}((1\leftarrow 2))=M_{12}$, $\tilde{w}((3\leftarrow 4))=M_{34}$, and $\tilde{w}((3\leftarrow 3))=-(R_3-\meanfit)$.
    Therefore, the weight of this graph is the product of these values: $-(R_3-\meanfit)M_{34}M_{12}$.}
    \label{fig:weight}
\end{figure}

We define the weight of a graph $H$ as
\begin{align}
    w(H):=\prod_{e\in E(H)}\tilde{w}(e),
\end{align}
where $\tilde{w}(e)$ is the weight of an edge $e$ defined as
\begin{equation}
\tilde{w}(e) := 
\left\{ 
\begin{aligned}
  &-(R_i - \meanfit) && \text{if} \: e = (i \leftarrow i), \\
  &M_{ij} && \text{if} \: e = (i \leftarrow j), \, i \neq j.
\end{aligned}
\right.
\end{equation}
Thus, the weight of a graph $H$ is given by the product of the weights of all its edges $E(H)$. See Fig.~\ref{fig:weight} for an example.

\section{Main results}
\label{sec:main_results}
We present the main results, which are diagrammatic expressions for the steady-state distribution and the static responses, and illustrate them with some examples. 
\subsection{Generalization of the Markov chain tree theorem}
We first state a lemma that plays an important role in the derivation of the main results. The lemma is a generalization of the Markov chain tree theorem~\cite{hill1966studies,schnakenberg1976network}, and provides a diagrammatic expression for $\zeta_j\pi_i$.
As discussed in Eqs.~\eqref{eq:resp_R} and \eqref{eq:resp_M}, $\zeta_j\pi_i$ characterizes the static {responses} of the mean fitness.
\begin{lem}\label{lem:Lem1} 
Suppose that {a} basic graph $G$ is given.
Then, for any $i \in V(G)$ and any $j\in V(G)$, $\zeta_j\pi_i$ is given by
\begin{align}\label{eq:Lem1}
    \zeta_j \pi_i =\frac{1}{Z}\sum_{H\in\mathcal{F}_{i\leftarrow j}(G)}w(H),
\end{align}
where $Z:=\sum_{i \in V(G)}\sum_{H\in\mathcal{F}_i(G)}w(H)$.
\end{lem}
See Appendix~\ref{app:proof_main} for the proof of this lemma. 

\begin{figure*}[hbtp]
    \centering
    \includegraphics[width=\linewidth]{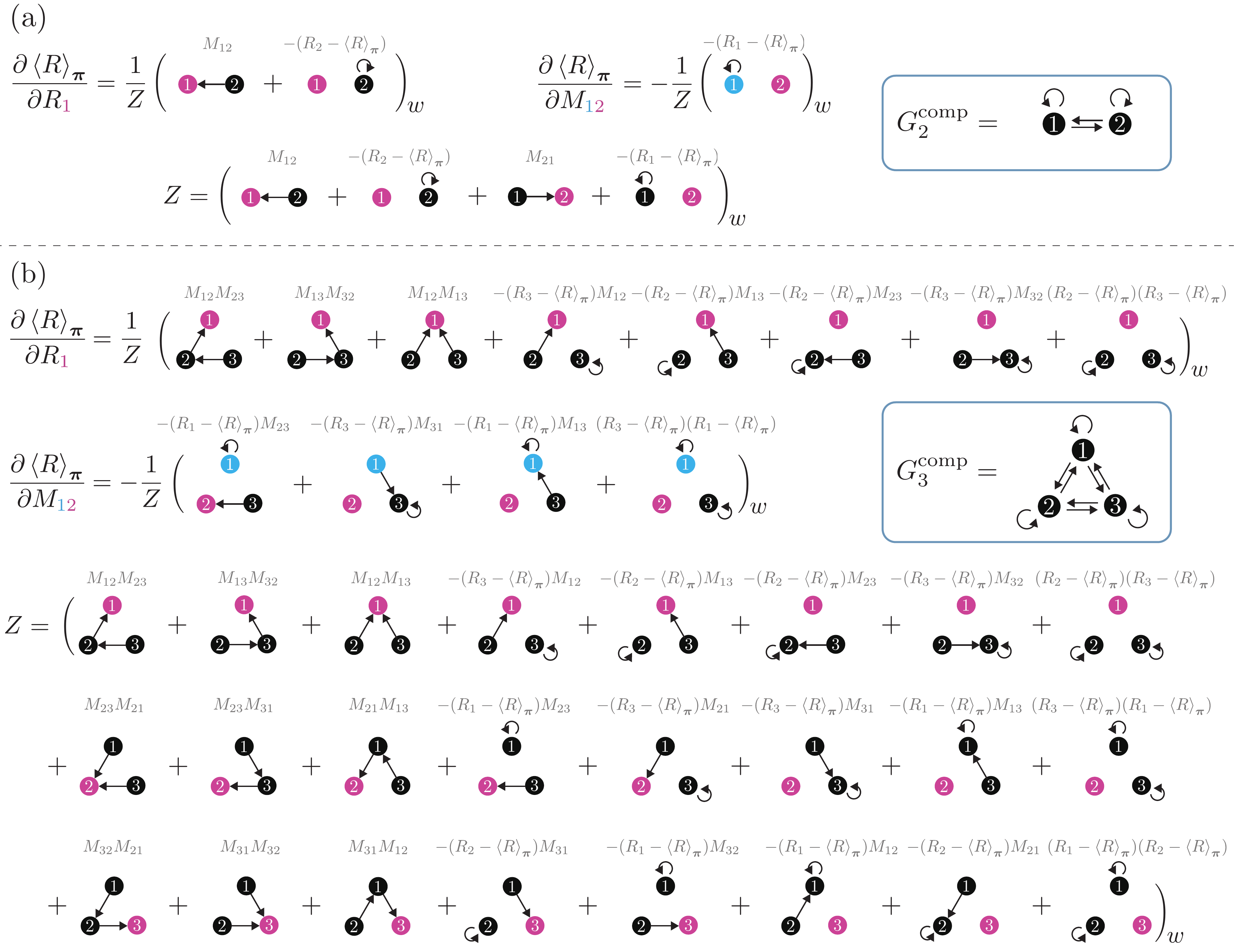}
    \caption{Examples of Theorem~\ref{thm:Th1} and Theorem~\ref{thm:Th2}.
    In this figure, we use the notation $(H+H'+\cdots)_w:=w(H)+ w(H')+ \cdots$.
    The weight of each graph is shown in gray.
    (a) Diagrammatic expressions for {$\partial_{R_1}\meanfit =\sum_{H\in\mathcal{F}_1(G^{\mathrm{comp}}_2)}w(H)/Z$} and {$\partial_{M_{12}}\meanfit =-\sum_{H\in\mathcal{F}_{2\not\leftarrow 1}(G^{\mathrm{comp}}_2)}w(H)/Z$} with $Z=\sum_{i=1}^2\sum_{H \in\mathcal{F}_i(G^{\mathrm{comp}}_2)}w(H)$ . 
    (b) Diagrammatic expressions for {$\partial_{R_1}\meanfit =\sum_{H\in\mathcal{F}_1(G^{\mathrm{comp}}_3)}w(H)/Z$} and {$\partial_{M_{12}}\meanfit=-\sum_{H\in\mathcal{F}_{2\not\leftarrow 1}(G^{\mathrm{comp}}_3)}w(H)/Z$} with $Z=\sum_{i=1}^3\sum_{H \in\mathcal{F}_i(G^{\mathrm{comp}}_3)}w(H)$. 
    }
    \label{fig:Thm_1_2_ex}
\end{figure*}

As shown in Appendix~\ref{app:max_number}, the number of elements in $\mathcal{F}_i(G)$, $|\mathcal{F}_i(G)|$, is at most $2(N+1)^{N-2}$ with $N=|V(G)|$. Thus, the number of terms appearing in $\sum_{H\in\mathcal{F}_{i\leftarrow j}(G)}w(H)$ is at most $2(N+1)^{N-2}$ because $|\mathcal{F}_i(G)| \geq |\mathcal{F}_{i\leftarrow j}(G)|$.
The number of terms appearing in $Z$ is at most $2N(N+1)^{N-2}$.

When all reproductive rates are equal ($\mathsf{R} = \ev{R}_{\bm{p}(t)} \mathsf{I}$), Lemma~\ref{lem:Lem1} reduces to the Markov chain tree theorem as follows.
\begin{onlyname}[\textbf{Markov chain tree theorem.}] 
Suppose that {a} basic graph $G$ is given and $R_i=R_j$ holds for any $i \in V(G)$ and any $j\in V(G)$.
Then, for any $i\in V(G)$, the steady-state distribution $\pi_i$ is given by
\begin{align} \label{eq:Markovchaintreetheorem}
    \pi_i=\frac{1}{Z'}\sum_{H\in\mathcal{T}_{i}(G)}w(H),
\end{align}
where $Z':=\sum_{i\in V(G)}\sum_{H\in\mathcal{T}_i(G)}w(H)$.
\end{onlyname}

This theorem is proved from Lemma~\ref{lem:Lem1} as follows.
When $R_i=R_j$ holds for any pairs $(i,j)$, $\zzeta=\bm{1}^\top$ holds as mentioned in Sec.~\ref{subsec:pi}. 
Thus, the left-hand side of Eq.~\eqref{eq:Lem1} reduces to $\pi_i$.
In this case, $R_i-\meanfit=0$ also holds for any $i$. 
Therefore, if $H\in\mathcal{F}_{i\leftarrow j}(G)$ has at least one loop, the weight of $H$ vanishes, i.e., $w(H)=0$. Only the graphs without loops contribute to the right-hand side of Eq.~\eqref{eq:Lem1}, and such graph $H$ in $\mathcal{F}_{i}(G)$ or $\mathcal{F}_{i\leftarrow j}(G) (\subseteq \mathcal{F}_{i}(G))$ is regarded as a rooted spanning tree $H \in \mathcal{T}_i(G)$ because Eq.~(\ref{eq:H_union}) implies that a graph $H \in \mathcal{F}_{i}(G)$ should be a rooted spanning tree if $N_{\mathrm{c}} = 1$, otherwise the graph should contain at least one loop.
Thus, we obtain $\sum_{H\in\mathcal{F}_{i\leftarrow j}(G)}w(H)=\sum_{H\in\mathcal{T}_i(G)}w(H)$ and 
$Z =Z'$. Thus, Lemma~\ref{lem:Lem1} [Eq.~(\ref{eq:Lem1})] reduces to the Markov chain tree theorem [Eq.~(\ref{eq:Markovchaintreetheorem})].

\subsection{Diagrammatic expressions for 
the static responses}
We derive diagrammatic expressions for $\partial_{R_i}\meanfit$ and $\partial_{M_{ij}}\meanfit$ from Lemma~\ref{lem:Lem1}. First, we discuss a diagrammatic expression for $\partial_{R_i}\meanfit$:
\begin{thm}\label{thm:Th1} 
Suppose that {a} basic graph $G$ is given.
Then, for any $i\in V(G)$, the static response of the mean fitness to a perturbation in the reproductive rate is given by
\begin{align}\label{eq:Th1}
    \frac{\partial\meanfit}{\partial R_i}=\frac{1}{Z}\sum_{H\in\mathcal{F}_i(G)}w(H),
\end{align}
where $Z=\sum_{i \in V(G)}\sum_{H \in\mathcal{F}_i(G)}w(H)$.
\end{thm}
To prove this theorem, we consider $j=i$ in Eq.~\eqref{eq:Lem1}. The left-hand side of Eq.~\eqref{eq:Lem1} becomes $\zeta_i \pi_i= \partial_{R_i}\meanfit$ as discussed in Eq.~\eqref{eq:resp_R}.
The right-hand side coincides with $Z^{-1}\sum_{H\in\mathcal{F}_i(G)}w(H)$ because $\mathcal{F}_{i \leftarrow i}(G)=\mathcal{F}_i(G)$.

This theorem provides an expression for $\partial_{R_i}\meanfit$ by $\{R_i\}$, $\{M_{ij}\}$, and $\meanfit$.
In other words, it allows us to predict the static response of the mean fitness to a perturbation in the reproductive rate using only the physically interpretable and measurable quantities.
It can be more useful than the expression $\zeta_i\pi_i$, because it is difficult to interpret the physical meaning only from $\zeta_i$ and $\pi_i$, which are given by eigenvectors of $\RR + \MM$. We show two simple examples of Theorem~\ref{thm:Th1} in Fig.~\ref{fig:Thm_1_2_ex}.

Similarly, we can also obtain a diagrammatic expression for $\partial_{M_{ij}}\meanfit$:
\begin{thm}\label{thm:Th2} 
Suppose that {a} basic graph $G$ is given.
Then, for any $i\in V(G)$ and $j\in V(G)$, the static response of the mean fitness to a perturbation in the mutation rate is given by
\begin{align}\label{eq:Th2}
    \frac{\partial\meanfit}{\partial M_{ij}}=-\frac{1}{Z}\sum_{H\in\mathcal{F}_{j\not\leftarrow i}(G)}w(H),
\end{align}
where $Z=\sum_{i \in V(G)} \sum_{H \in\mathcal{F}_i(G)}w(H)$.
\end{thm}
To prove this theorem, we consider Eqs.~\eqref{eq:resp_M} and \eqref{eq:Lem1}.
From Eqs.~\eqref{eq:resp_M} and \eqref{eq:Lem1}, we obtain
\begin{align}
    \frac{\partial\meanfit}{\partial M_{ij}}&=\frac{1}{Z}\left(\sum_{H\in\mathcal{F}_{j\leftarrow i}(G)}w(H)-\sum_{H\in\mathcal{F}_j(G)}w(H)\right)\notag\\
    &=-\frac{1}{Z}\sum_{H \in\mathcal{F}_{j\not\leftarrow i}(G)}w(H),
\end{align}
where we used $\mathcal{F}_{j}(G)=\mathcal{F}_{j\leftarrow i}(G)\cup\mathcal{F}_{j\not\leftarrow i}(G)$ and $\mathcal{F}_{j\leftarrow i}(G)\cap\mathcal{F}_{j\not\leftarrow i}(G)=\emptyset$.

This theorem also provides an expression for {$\partial_{M_{ij}}\meanfit$} by physically interpretable and measurable quantities.  We also show two simple examples of Theorem~\ref{thm:Th2} in Fig.~\ref{fig:Thm_1_2_ex}.

Theorems~\ref{thm:Th1} and \ref{thm:Th2} provide expressions not only for $\partial _{R_i}\meanfit$ and $\partial_{M_{ij}}\meanfit$, but also for general static responses of the mean fitness.
This is because, as mentioned in Sec.~\ref{subsec:resp}, any static response of the mean fitness can be expressed as a linear combination of $\partial _{R_i}\meanfit$ and $\partial_{M_{ij}}\meanfit$.

\subsection{Diagrammatic expression for the steady-state distribution}
Lemma~\ref{lem:Lem1} also provides a diagrammatic expression for {the ratio of the steady-state distribution}:
\begin{thm}\label{thm:Th3} 
Suppose that a basic graph $G$ is given.
Then, for any $i\in V(G)$ and $j\in V(G)$, the ratio of the steady-state distribution is given by
\begin{align}\label{eq:Th3}
    \frac{\pi_j}{\pi_i}=\frac{\sum_{H'\in\mathcal{F}_{j\leftarrow i}(G)}w(H')}{\sum_{H\in\mathcal{F}_{i}(G)}w(H)}.
\end{align}
\end{thm}
To prove this theorem, we give a diagrammatic expression for the ratio of the steady-state distributions $\pi_j/\pi_i$.
Substituting Eq.~\eqref{eq:Lem1} into the right-hand side of $\pi_j/\pi_i=(\zeta_i\pi_j)/(\zeta_i\pi_i)$, we obtain Eq.~\eqref{eq:Th3} because $\mathcal{F}_{i\leftarrow i}(G) = \mathcal{F}_{i}(G)$.

Moreover, we can obtain a diagrammatic expression for $\ppi$ itself:
\begin{cor}\label{cor:pi}
Suppose that {a} basic graph $G$ is given.
Then, for any $i\in V(G)$, the steady-state distribution is given by
\begin{align}\label{eq:cor_pi}
\pi_i=\frac{\sum_{H'\in\mathcal{F}_{i}(G)}w(H')}{\sum_{j \in V(G)}\sum_{H\in\mathcal{F}_{j\leftarrow i}(G)}w(H)}.
\end{align}
\end{cor}
To prove this theorem, we substitute Eq.~\eqref{eq:Th3} into the right-hand side of $\pi_i=\{\sum_{j\in V(G)} (\pi_j/\pi_i)\}^{-1}$.

\begin{figure*}[htbp]
    \centering
    \includegraphics[width=\linewidth]{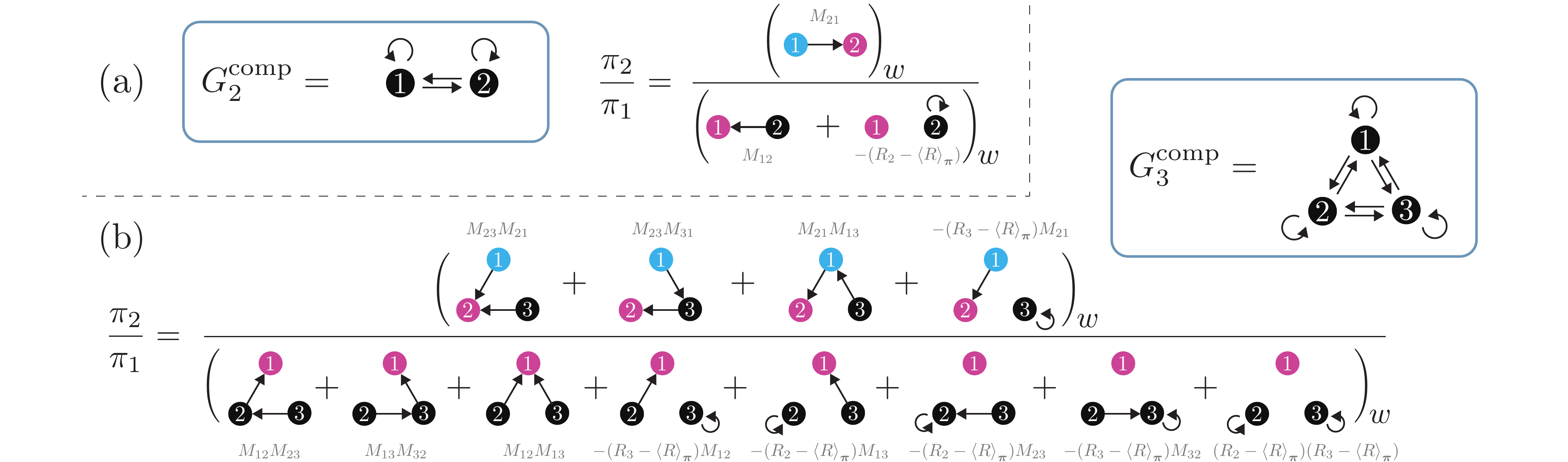}
    \caption{Examples of Theorem 3.
    In this figure, we use the notation $(H+H'+\cdots)_w:=w(H)+ w(H')+ \cdots$.
    The weight of each graph is shown in gray.
    (a) A diagrammatic expression for $\pi_2/\pi_1 =[\sum_{H'\in\mathcal{F}_{2\leftarrow 1}(G^{\mathrm{comp}}_2)}w(H')]/ [\sum_{H\in\mathcal{F}_{1}(G^{\mathrm{comp}}_2)}w(H)]$. 
    (b) A diagrammatic expression for $\pi_2/\pi_1=[\sum_{H'\in\mathcal{F}_{2\leftarrow 1}(G^{\mathrm{comp}}_3)}w(H')]/ [\sum_{H\in\mathcal{F}_{1}(G^{\mathrm{comp}}_3)}w(H)]$.
    }
    \label{fig:Thm_3_ex}
\end{figure*}

This theorem also provides an expression for the steady-state distribution by physically interpretable and measurable quantities. See Fig.~\ref{fig:Thm_3_ex} for some examples of Theorem~\ref{thm:Th3}.

\section{Applications for two limiting cases}
\label{sec:approximation}
While our results provide exact expressions for the static responses and steady-state distributions, the computation with counting graphs can be complicated. Here, we consider approximations for special situations where either {mutation or natural selection} is dominant.

\subsection{Mutation-dominant case}
{
We consider the case where the effect of mutation is dominant compared to that of natural selection, and we derive approximate expressions for the steady-state distribution and the static responses of the mean fitness.
Let $\epsilon$ be a small parameter satisfying $0<\epsilon\ll\min_{i, j(\ne i)| M_{ij}\ne 0}M_{ij}$.
Then, we assume $\max_{i, j}|R_i-R_j|=\mathcal{O}(\epsilon)$, where $\mathcal{O}(\cdot)$ denotes big O notation.
}
In other words, we assume that mutation rates are much higher than the differences in reproductive rates, or that all reproductive rates are nearly equal; that is, we assume neutral evolution.
We call this the mutation-dominant case.

We rewrite this condition so that it is suitable for our diagrammatic expressions.
Since $|R_i-\ev{R}_{\bm{\pi}}|\le\max_{j, k}|R_j-R_k|$ holds, we obtain
\begin{align}\label{eq:mutation_dominant}
    \frac{|R_k-\meanfit|}{M_{ij}}=\mathcal{O}(\epsilon)
\end{align}
for any $i$, $j(\ne i)$, and $k$. 
This means that the weight of a directed edge between vertices dominates the weight of a loop.

By applying Eq.~\eqref{eq:mutation_dominant}, we approximate $\ppi$ in Corollary~\ref{cor:pi}. 
Let $\mathcal{F}^{[n_{\rm l}]}_{i}(G) (\subseteq \mathcal{F}_{i}(G) )$ and $\mathcal{F}^{[n_{\rm l}]}_{i \leftarrow j}(G) (\subseteq \mathcal{F}_{i \leftarrow j}(G))$ be the sets of graphs in $\mathcal{F}_{i}(G)$ and $\mathcal{F}_{i \leftarrow j}(G)$ that contain exactly $n_{\rm l}$ loops, respectively. 
Equation~\eqref{eq:mutation_dominant} implies that the fewer loops in $H$, the heavier the weight $w(H)$. Thus, the leading and second-leading terms in the numerator of $\pi_i$ are 
$\sum_{H\in\mathcal{F}^{[0]}_i(G)}w(H)=\sum_{H\in\mathcal{T}_i(G)}w(H)$ and $\sum_{H\in\mathcal{F}^{[1]}_{i}(G)}w(H)$, respectively.
The leading and second-leading terms in the denominator of $\pi_i$ are $\sum_{j \in V(G)}\sum_{H\in\mathcal{T}_j(G)}w(H)=Z'$ and $\sum_{j \in V(G)}\sum_{H\in\mathcal{F}^{[1]}_{j\leftarrow i}(G)}w(H)$, respectively. Thus, we obtain
\begin{align}    
\pi_i=&\frac{\sum_{H\in\mathcal{T}_i(G)}w(H)} {Z'} + \frac{\sum_{H\in\mathcal{F}^{[1]}_i(G)}w(H)} {Z'}\notag\\
&-\frac{\sum_{j\in V(G)}\sum_{H\in\mathcal{F}^{[1]}_{j\leftarrow i}(G)} \sum_{H'\in\mathcal{T}_i(G)} w(H)w(H') }{(Z')^2} \notag\\
&+\mathcal{O}(\epsilon^2).
\end{align}
The leading term of $\pi_i$ is $\sum_{H\in\mathcal{T}_i(G)}w(H)/Z' =\mathcal{O}(1)$, which is consistent with the Markov chain tree theorem [Eq.~\eqref{eq:Markovchaintreetheorem}]. 

\begin{figure*}[hbtp]
    \centering
    \includegraphics[width=\linewidth]{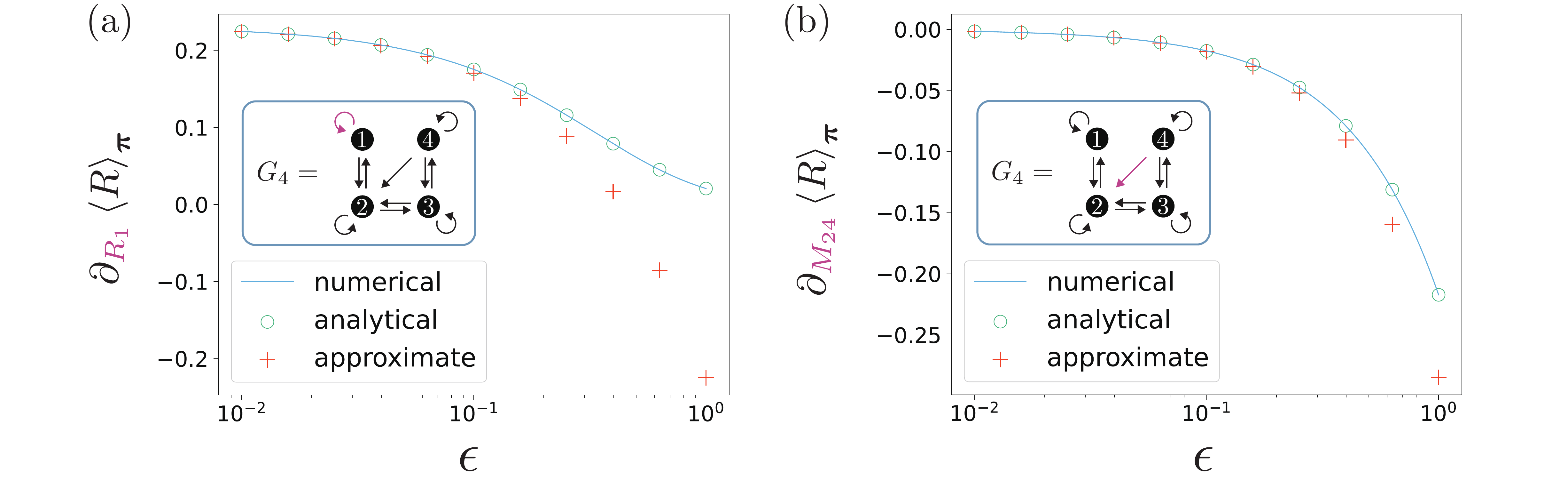}
    \caption{Numerical, analytical, and approximate evaluations of (a) $\partial_{R_1}\meanfit$ and (b) $\partial_{M_{24}}\meanfit$.
    The basic graph of the population is given by $G_4$, shown in the inset.
    The matrices $\RR$ and $\MM$ are given in Eq.~\eqref{eq:R_M_M_dominant}.
    }
    \label{fig:numerical_M_dominant}
\end{figure*}

We also approximate $\partial_{R_i}\meanfit=\sum_{H\in\mathcal{F}_i(G)}w(H)/Z$ [Theorem~\ref{thm:Th1}] in the mutation-dominant case. The leading and second-leading terms in the numerator of $\partial_{R_i}\meanfit$ are $\sum_{H\in\mathcal{T}_i(G)}w(H)$ and $\sum_{H\in\mathcal{F}^{[1]}_i(G)}w(H)$, respectively.
The leading and second-leading terms in $Z$ are $\sum_{i\in V(G)}\sum_{H\in\mathcal{T}_i(G)}w(H)=Z'$ and $\sum_{i\in V(G)}\sum_{H\in\mathcal{F}^{[1]}_i(G)}w(H)$, respectively.
Therefore, we have
\begin{align}\label{eq:del_meanfit_del_R_M_dominant}
    \frac{\partial\meanfit}{\partial R_i}=&\frac{\sum_{H\in\mathcal{T}_i(G)}w(H)}{Z'}+\frac{\sum_{H\in\mathcal{F}^{[1]}_i(G)}w(H)}{Z'}\notag\\
    &-\frac{\sum_{j\in V(G)}\sum_{H\in\mathcal{F}^{[1]}_j(G)}\sum_{H'\in\mathcal{T}_i(G)} w(H)w(H')}{(Z')^2}\notag\\
    &+\mathcal{O}(\epsilon^2).
\end{align}
Note that the leading term of $\partial_{R_i}\meanfit$ is $\sum_{H\in\mathcal{T}_i(G)}w(H)/Z' =\mathcal{O}(1)$, which is identical to that of $\pi_i$.

If $R_i = \meanfit$ is satisfied for any $i$ and Eq.~\eqref{eq:dp_dt} is considered to be the linear master equation, $\partial_{R_i}\meanfit=\pi_i$ holds exactly because the weight of any graph containing at least one loop vanishes, as mentioned below Eq.~\eqref{eq:Markovchaintreetheorem}. This fact can also be understood as the situation in which $\zeta_i =1$ for any $i$ in Eq.~\eqref{eq:resp_R}. Indeed, the left eigenvector for the linear master equation is $\boldsymbol{\zeta}= \boldsymbol{1}$, and $\partial_{R_i}\meanfit=\pi_i$ can be confirmed.

Similarly, we discuss an approximate expression for $\partial_{M_{ij}}\meanfit=-\sum_{H\in\mathcal{F}_{j\not\leftarrow i}(G)}w(H)/Z$ [Theorem~\ref{thm:Th2}].
Let $\mathcal{F}^{[n_{\rm l}]}_{i\not\leftarrow j}(G)(\subseteq \mathcal{F}_{i\not\leftarrow j}(G))$ be the set of graphs in $\mathcal{F}_{i\not\leftarrow j}(G)$ that contain exactly $n_{\rm l}$ loops.
Since any graph in $\mathcal{F}_{j\not\leftarrow i}(G)$ contains at least one loop, the leading term in $\sum_{H \in \mathcal{F}_{j\not\leftarrow i}(G)} w(H)$ is $\sum_{H \in \mathcal{F}^{[1]}_{j\not\leftarrow i}(G)} w(H)$.
Thus, we obtain
\begin{align}\label{eq:del_meanfit_del_M_M_dominant}
    \frac{\partial \meanfit}{\partial M_{ij}} =- \frac{1}{Z'} \sum_{H \in \mathcal{F}^{[1]}_{j\not\leftarrow i}(G)} w(H) + \mathcal{O}(\epsilon^2).
\end{align}
Note that the leading term of $\partial_{M_{ij}} \meanfit$ is $\mathcal{O}(\epsilon)$, whereas that of $\partial_{R_i} \meanfit$ is $\mathcal{O}(1)$.

If $R_i = \meanfit$ is satisfied for any $i$ and Eq.~\eqref{eq:dp_dt} is considered to be the linear master equation, $\partial_{M_{ij}}\meanfit=0$ holds exactly. We can also confirm $\partial_{M_{ij}}\meanfit=0$ because the left eigenvector for the linear master equation is $\boldsymbol{\zeta}= \boldsymbol{1}$ and $\zeta_i -\zeta_j =0$ is satisfied in Eq.~\eqref{eq:resp_M}. 

In Fig.~\ref{fig:numerical_M_dominant}, we evaluate $\partial_{R_1}\meanfit$ and $\partial_{M_{24}}\meanfit$ using three different approaches: numerical calculation, the analytical results [Theorems~\ref{thm:Th1} and \ref{thm:Th2}], and their approximations [Eqs.~\eqref{eq:del_meanfit_del_R_M_dominant} and \eqref{eq:del_meanfit_del_M_M_dominant}].
We consider a population whose basic graph $G_4$ is shown in Fig.~\ref{fig:numerical_M_dominant}.
The matrices $\RR$ and $\MM$ are given by
\begin{equation}\label{eq:R_M_M_dominant}
    \RR=\epsilon
    \begin{pmatrix}
        1&0&0&0\\
        0&3&0&0\\
        0&0&4&0\\
        0&0&0&7
    \end{pmatrix},\quad
    \MM=
    \begin{pmatrix}
        -2&2&0&0\\
        2&-9&1&5\\
        0&7&-7&1\\
        0&0&6&-6
    \end{pmatrix},
\end{equation}
with $\epsilon>0$.
Since $\RR$ is parametrized by $\epsilon$, Eq.~\eqref{eq:mutation_dominant} is satisfied.
As shown in Fig.~\ref{fig:numerical_M_dominant}, the analytical results agree with the numerical calculations in the whole range of $\epsilon$, and the approximation is valid when $0<\epsilon\ll 1$.

\subsection{Selection-dominant case}
\label{subsec:selection_dominant}
Next, we consider the case where the effect of natural selection dominates that of mutation, and derive approximate expressions for the steady-state distribution and the static responses of the mean fitness.
We define the ``fittest'' trait in the population $i_*$ as $i_*:= {\operatorname{argmax}}_i \{R_i\}$, and assume that this trait $i_*$ is unique.
Let $\epsilon$ be a small parameter satisfying $0<\epsilon\ll \min_{j(\ne i_*)}\{R_{i_*}-R_j\}$.
We then assume that $M_{ij}=\mathcal{O}(\epsilon)$ for any $i$ and $j(\ne i)$.
In other words, we assume that the effect of mutation is negligible compared to the differences in reproductive rates.
We call this the selection-dominant case.

We rewrite this condition so that it is suitable for our diagrammatic expressions.
In our diagrammatic expressions, we use the following inequality:
\begin{align}\label{eq:R_i-meanfit<-M_ii}
    R_i-\meanfit<-M_{ii},
\end{align}
which holds for any $i$.
This inequality [Eq.~\eqref{eq:R_i-meanfit<-M_ii}] always holds and
can be derived as follows.
Since $\ppi$ is the right eigenvector of $\RR+\MM$ with eigenvalue $\meanfit$,
it satisfies $(R_i+M_{ii})\pi_i+\sum_{j(\ne i)}M_{ij}\pi_j=\meanfit\pi_i$.
Dividing both sides by $\pi_i>0$ and rearranging yields $R_i-\meanfit+M_{ii}=-\sum_{j(\ne i)}M_{ij}\pi_j/\pi_i<0$, which leads to Eq.~\eqref{eq:R_i-meanfit<-M_ii}.
By setting $i=i_*$ in Eq.~\eqref{eq:R_i-meanfit<-M_ii} and subtracting $R_j$ from both sides, we obtain 
\begin{align}\label{eq:meanfit-R_j}
    \meanfit-R_j>R_{i_*}-R_j+M_{i_* i_*}
\end{align}
for any $j(\ne i_*)$.
Combining this inequality with the assumption $-M_{i_* i_*}= \sum_{j (\neq i_*)} M_{ji_*}=\mathcal{O}(\epsilon) \ll R_{i_*} -R_j$, we find $\meanfit-R_j>0$ for any $j(\ne i_*)$.
Moreover, combining Eq.~\eqref{eq:meanfit-R_j} with $M_{ij}=\mathcal{O}(\epsilon)$, we obtain
\begin{align}
    0<\frac{M_{kl}}{\meanfit-R_j}<\frac{M_{kl}}{R_{i_*}-R_j+M_{i_* i_*}}
\end{align}
for any $j(\ne i_*), k$, and $l(\ne k)$.
Since $M_{i_*i_*}=\mathcal{O}(\epsilon)$ and $M_{kl}=\mathcal{O}(\epsilon)$, we obtain
\begin{align}\label{eq:selection_dominant_condition}
    \frac{M_{kl}}{\meanfit-R_j}=\mathcal{O}(\epsilon)
\end{align}
for any $j(\ne i_*), k$, and $l(\ne k)$.
This equation means that the weight of the loop at vertex $j(\ne i_*)$ dominates the weight of a directed edge from vertex $l$ to vertex $k$.

By applying Eq.~\eqref{eq:selection_dominant_condition}, we approximate $\ppi$.
We first approximate $\pi_j/\pi_{i_*}$ for each $j(\ne i_*)$.
From Theorem~\ref{thm:Th3}, we have
\begin{align}\label{eq:pi_j/pi_i_*}
    \frac{\pi_j}{\pi_{i_*}}=\frac{\sum_{H'\in\mathcal{F}_{j\leftarrow i_*}(G)}w(H')}{\sum_{H\in\mathcal{F}_{i_*}(G)}w(H)}.
\end{align}
Note that no graph in $\mathcal{F}_{i_*}(G)$ or $\mathcal{F}_{j\leftarrow i_*}(G)$ contains the loop at vertex $i_*$.
Equation~\eqref{eq:selection_dominant_condition} implies that the more loops in $H$, the heavier the weight $w(H)$.
Thus, the leading term in the denominator of Eq.~\eqref{eq:pi_j/pi_i_*} is
$\sum_{H \in \mathcal{F}^{[N-1]}_{i_*}(G)} w(H)$ with $N = |V(G)|$, and $\mathcal{F}^{[N-1]}_{i_*}(G)$ consists of only one graph 
$\bigcup_{k \ne i_*}(k \leftarrow k)$.
The leading term in the numerator of Eq.~\eqref{eq:pi_j/pi_i_*} is $\sum_{H' \in \mathcal{F}^{[N-2]}_{j \leftarrow i_*}(G)}w(H')$, and $\mathcal{F}^{[N-2]}_{j \leftarrow i_*}(G)$ also consists of only one graph 
$(j\leftarrow i_*) \cup \bigcup_{k \ne i_*,\, j}(k \leftarrow k)$.
Combining these facts, the approximate form of $\pi_j /\pi_{i_*}$ is given by 
\begin{align}\label{eq:pi_j/pi_i_approx}
    \frac{\pi_j}{\pi_{i_*}}=\frac{M_{ji_*}}{\meanfit-R_j}+\mathcal{O}(\epsilon^2).
\end{align}
We can also derive an approximate expression for $\ppi$ as follows.
Since $\pi_{i_*}=(1+\sum_{j(\ne i_*)}\pi_j/\pi_{i_*})^{-1}$, we have
$\pi_{i_*}=(1+\mathcal{O}(\epsilon))^{-1}=1+\mathcal{O}(\epsilon)$.
By multiplying Eq.~\eqref{eq:pi_j/pi_i_approx} by $\pi_{i_*}=1+\mathcal{O}(\epsilon)$, we obtain
\begin{align}\label{eq:pi_j_approx}
    \pi_j=\frac{M_{ji_*}}{\meanfit-R_j}+\mathcal{O}(\epsilon^2)
\end{align}
for any $j(\ne i_*)$.
This equation implies that, if $M_{ji_*}\ne 0$, it follows that $\pi_j=\mathcal{O}(\epsilon)$; otherwise, $\pi_j=\mathcal{O}(\epsilon^2)$. 
Combining Eq.~\eqref{eq:pi_j_approx} with $\pi_{i_*} = 1- \sum_{j (\neq i_*)} \pi_j$, we also obtain 
\begin{align}
    \pi_{i_*}= 1- \sum_{j (\neq i_*)}\frac{M_{ji_*}}{\meanfit-R_j} +\mathcal{O}(\epsilon^2).
\end{align}

We also discuss $\partial_{R_i}\meanfit=\sum_{H\in\mathcal{F}_i(G)}w(H)/Z$ [Theorem~\ref{thm:Th1}].
First, we consider $\partial_{R_j}\meanfit$ with $j\ne i_*$.
Since some graphs in $\mathcal{F}_j(G)$ contain the loop at vertex $i_*$, we need to consider $\meanfit-R_{i_*}$.
By setting $i=i_*$ in Eq.~\eqref{eq:R_i-meanfit<-M_ii} and using $M_{i_*i_*}=\mathcal{O}(\epsilon)$ and $R_{i_*}-\meanfit>0$, we find $\meanfit-R_{i_*}=\mathcal{O}(\epsilon)$.
Therefore, the leading terms in $\sum_{H\in\mathcal{F}_j(G)}w(H)$ are $\prod_{k(\ne j)}(\meanfit-R_k)$ and $M_{ki_*}\prod_{l(\ne i_*, j)}(\meanfit-R_l)$ with $k(\ne i_*)$.
These terms are $\mathcal{O}(\epsilon)$ since they contain $\meanfit-R_{i_*}$ or $M_{ki_*}$.
However, their sum is $\mathcal{O}(\epsilon^2)$ because
\begin{align}\label{eq:cancel_out}
    \meanfit-R_{i_*}+\sum_{k(\ne i_*)}M_{ki_*}=\mathcal{O}(\epsilon^2),
\end{align}
which can be derived by combining
\begin{align}
    R_{i_*}\pi_{i_*}+\sum_{k(\ne i_*)}(M_{i_*k}\pi_k-M_{ki_*}\pi_{i_*})=\meanfit\pi_{i_*}
\end{align}
with $\pi_{i_*}=1-\mathcal{O}(\epsilon)$, $\pi_k=\mathcal{O}(\epsilon)$, and $M_{i_*k}=\mathcal{O}(\epsilon)$ for any $k(\ne i_*)$.
The leading term in $Z=\sum_{i\in V(G)}\sum_{H\in\mathcal{F}_i(G)}w(H)$ is $\prod_{k(\ne i_*)}(\meanfit-R_k)=\mathcal{O}(1)$. 
Therefore, we find
\begin{align}\label{eq:partial_R_meanfit_e^2}
    \partial_{R_j}\meanfit=\mathcal{O}(\epsilon^2)
\end{align}
for any $j(\ne i_*)$.
Combining this equation with $\partial_{R_{i_*}}\meanfit=1-\sum_{j(\ne i_*)}\partial_{R_j}\meanfit$, we also find
\begin{align}
    \partial_{R_{i_*}}\meanfit=1-\mathcal{O}(\epsilon^2).
\end{align}

We can also approximate $\partial_{M_{ij}}\meanfit=-\sum_{H\in\mathcal{F}_{j\not\leftarrow i}(G)}w(H)/Z$ [Theorem~\ref{thm:Th2}].
First, we consider $\partial_{M_{jk}}\meanfit$ with $j \ne i_*$ and $k \ne i_*, j$.
The dominant terms in $\sum_{H\in\mathcal{F}_{k\not\leftarrow j}(G)}w(H)$ are $\prod_{l(\ne k)}(\meanfit-R_l)$ and $M_{li_*}\prod_{m(\ne i_*, k)}(\meanfit-R_m)$ with $l \ne i_*$.
Although these terms are $\mathcal{O}(\epsilon)$, their sum is $\mathcal{O}(\epsilon^2)$ in the same way as in Eq.~\eqref{eq:cancel_out}.
Thus, we find
\begin{align}
    \partial_{M_{jk}}\meanfit=\mathcal{O}(\epsilon^2)
\end{align}
for $j\ne i_*$ and $k\ne j,i_*$. Next, we consider $\partial_{M_{i_*j}}\meanfit$ with $j\ne i_*$.
The leading terms in $\sum_{H\in\mathcal{F}_{j\not\leftarrow i_*}(G)}w(H)$ are $\prod_{k(\ne j)}(\meanfit-R_k)$ and $M_{ki_*}\prod_{l(\ne i_*, j)}(\meanfit-R_l)$ with $k(\ne i_*, j)$.
The sum of these terms are 
\begin{align}
    &\left(\meanfit-R_{i_*}+\sum_{k(\ne i_*, j)}M_{ki_*}\right)\prod_{l(\ne i_*, j)}(\meanfit-R_l)\notag\\
    &=-M_{ji_*}\prod_{l(\ne i_*, j)}(\meanfit-R_l)+\mathcal{O}(\epsilon^2),
\end{align}
where we used Eq.~\eqref{eq:cancel_out}.
The leading term in $Z$ is $\prod_{k(\ne i_*)}(\meanfit-R_k)$.
Thus, we find
\begin{align}\label{eq:del_R_approx}
    \frac{\partial\meanfit}{\partial M_{i_*j}}=\frac{M_{ji_*}}{\meanfit-R_j}+\mathcal{O}(\epsilon^2).
\end{align}
Finally, we consider $\partial_{M_{ji_*}}\meanfit$ with $j\ne i_*$.
The leading and second-leading terms in $\sum_{H\in\mathcal{F}_{i_*\not\leftarrow j}(G)}w(H)$ are $\sum_{H\in\mathcal{F}^{[N-1]}_{i_*\not\leftarrow j}(G)}w(H)$ and $\sum_{H\in\mathcal{F}^{[N-2]}_{i_*\not\leftarrow j}(G)}w(H)$ with $N=|V(G)|$, respectively.
The leading and second-leading terms in $Z$ are $\sum_{H\in\mathcal{F}^{[N-1]}_{i_*}(G)}w(H)$ and $\sum_{H\in\mathcal{F}^{[N-2]}_{i_*}(G)}w(H)$, respectively.
Thus, we obtain
\begin{align}
    &\frac{\partial\meanfit}{\partial M_{ji_*}}\notag\\
    &=\!- \frac{\sum_{H\in\mathcal{F}^{[N-1]}_{i_*\not\leftarrow j}(G)}w(H) \!+\!\sum_{H\in\mathcal{F}^{[N-2]}_{i_*\not\leftarrow j}(G)}w(H)}{\sum_{H\in\mathcal{F}^{[N-1]}_{i_*}(G)}w(H)\!+ \!\sum_{H\in\mathcal{F}^{[N-2]}_{i_*}(G)}w(H)} \!+ \!\mathcal{O}(\epsilon^2)
    \notag\\&=-1+\!\frac{\sum_{H\in\mathcal{F}^{[N-2]
    }_{i_*}(G) \setminus \mathcal{F}^{[N-2]}_{i_*\not\leftarrow j}(G)}w(H)}{\sum_{H\in \mathcal{F}^{[N-1]}_{i_*}(G)}w(H)}+\mathcal{O}(\epsilon^2)\notag\\
    &=-1+\frac{M_{i_*j}}{\meanfit-R_j}+\mathcal{O}(\epsilon^2),
\end{align}
where we used $\mathcal{F}^{[N-1]}_{i_*}(G)=\mathcal{F}^{[N-1]}_{i_*\not\leftarrow j}(G)$, $\mathcal{F}^{[N-2]}_{i_*\not\leftarrow j}(G)\subseteq\mathcal{F}^{[N-2]}_{i_*}(G)$, the fact that $\mathcal{F}^{[N-1]}_{i_*}(G)$ consists of one graph $\bigcup_{k(\ne i_*)}(k\leftarrow k)$, and the fact that $\mathcal{F}^{[N-2]}_{i_*}(G)\setminus\mathcal{F}^{[N-2]}_{i_*\not\leftarrow j}(G)$ consists of one graph $(i_*\leftarrow j)\cup\bigcup_{k(\ne i_*, j)}(k\leftarrow k)$.

\begin{figure}[hbtp]
    \centering
    \includegraphics[width=\linewidth]{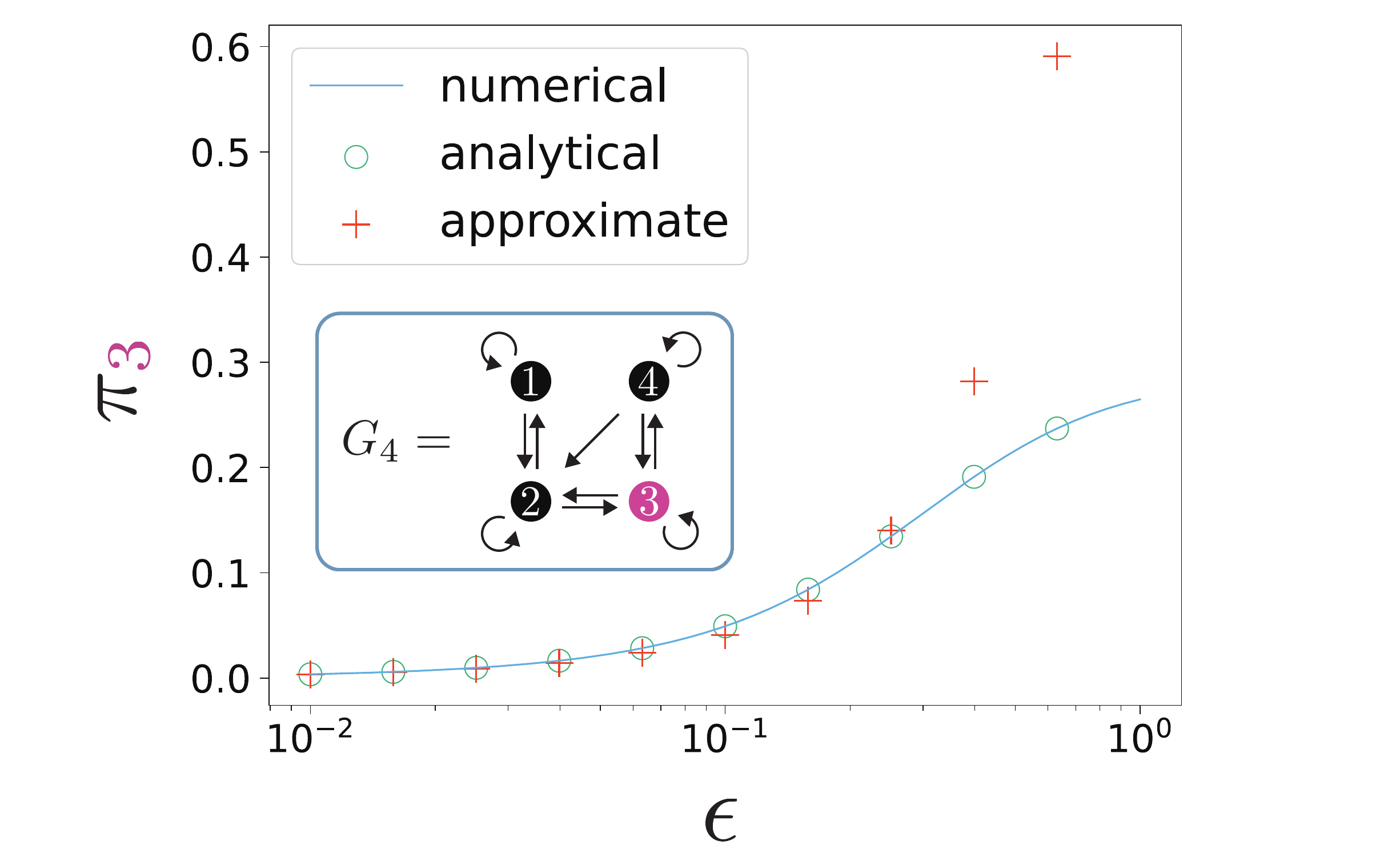}
    \caption{Numerical, analytical, and approximate evaluations of $\pi_3$.
    The basic graph of the population is given by $G_4$, shown in the inset.
    The matrices $\RR$ and $\MM$ are given in Eq.~\eqref{eq:R_M_R_dominant}.
    }
    \label{fig:numerical_R_dominant}
\end{figure}

In Fig.~\ref{fig:numerical_R_dominant}, we evaluate $\ppi$ using three different approaches: numerical {calculation}, the analytical result [Corollary~\ref{cor:pi}], and its approximation [Eq.~\eqref{eq:pi_j_approx}].
We consider a population whose basic graph $G_4$ is shown in Fig.~\ref{fig:numerical_R_dominant}.
The matrices $\RR$ and $\MM$ are given by
\begin{equation}\label{eq:R_M_R_dominant}
    \RR=
    \begin{pmatrix}
        1&0&0&0\\
        0&3&0&0\\
        0&0&4&0\\
        0&0&0&7
    \end{pmatrix},\quad
    \MM=\epsilon
    \begin{pmatrix}
        -2&2&0&0\\
        2&-9&1&5\\
        0&7&-7&1\\
        0&0&6&-6
    \end{pmatrix},
\end{equation}
with $\epsilon>0$.
In this example, $i_*=4$.
Since $\MM$ is parameterized by $\epsilon$, Eq.~\eqref{eq:selection_dominant_condition} is satisfied.
As shown in Fig.~\ref{fig:numerical_R_dominant}, the analytical result agrees with the numerical calculation in the whole range of $\epsilon$, and the approximation is valid when $0<\epsilon\ll 1$.

\section{Application for combination therapies}
\label{sec:comb_therapy}
\begin{figure*}[hbtp]
    \centering
    \includegraphics[width=\linewidth]{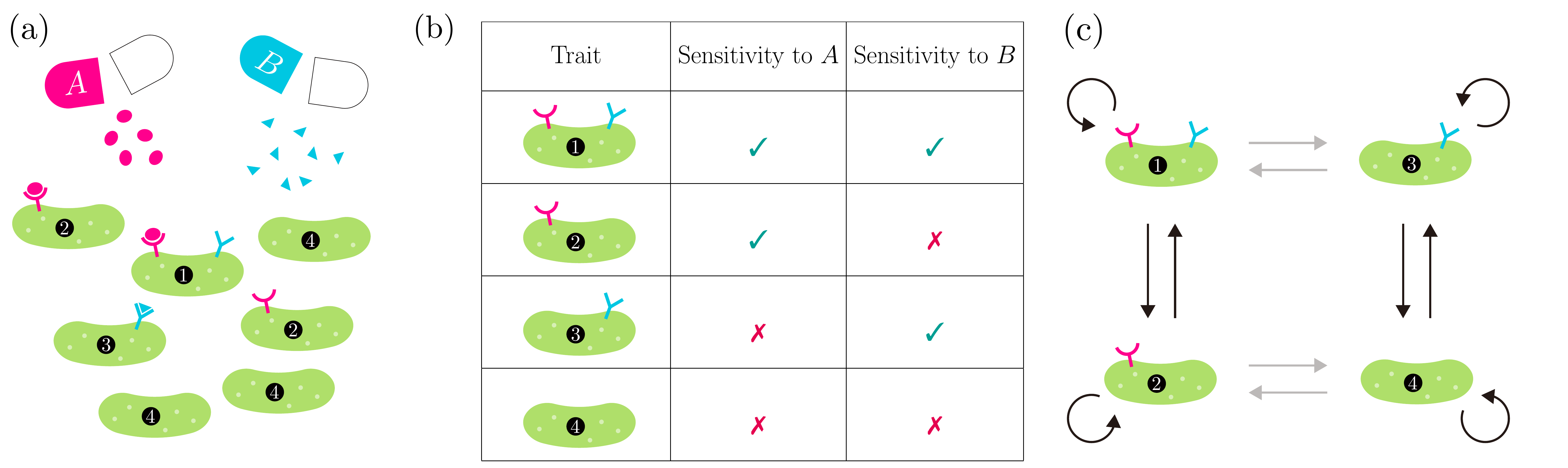}
    \caption{
    (a) We manipulate the concentrations of two inhibitors, $A$ and $B$, in order to reduce the mean fitness of a harmful population.
    When an inhibitor binds to its corresponding receptor on an individual's surface, its reproduction is inhibited.
    (b) There are four traits in the population that are classified by sensitivity to $A$ and $B$.
    A trait without a receptor for $A$ (or $B$) is not inhibited by $A$ (or $B$).
    (c) Basic graph $G$ for the population.
    The edges in $\mathcal{E}=\{(1, 3), (3, 1), (2, 4), (4, 2)\}$ are shown in gray to emphasize that their weights are of order $\mathcal{O}(\epsilon)$.
    }
    \label{fig:comb_therapy}
\end{figure*}

We demonstrate how our results can be used to control harmful populations.
Specifically, we consider how to optimally reduce the mean fitness of a harmful population through \textit{combination therapy}, in which multiple kinds of inhibitors are used simultaneously~\cite{baym2016multidrug, tyers2019drug, molina2025optimization}.
Combination therapies have been developed to prevent harmful populations, such as cancer cells~\cite{mokhtari2017combination, quinn2015pancreatic} or Gram-negative bacteria~\cite{tamma2012combination}, from acquiring resistance to inhibitors.

Let us consider a harmful population whose mean fitness we aim to reduce.
We control the mean fitness of the population by manipulating the concentrations of two kinds of inhibitors, $A$ and $B$ (see Fig.~\ref{fig:comb_therapy}(a)).
For cancer cells, $A$ and $B$ 
can be sabutoclax and minocycline~\cite{quinn2015pancreatic}, whereas for Gram-negative bacteria, they can be a $\beta$-lactam and an aminoglycoside~\cite{tamma2012combination}.
Assume that there are four traits in the population: trait $1$ that is sensitive to both $A$ and $B$, trait $2$ that is sensitive to $A$ but resistant to $B$, trait $3$ that is resistant to $A$ but sensitive to $B$, and trait $4$ that is resistant to both $A$ and $B$ (see also Fig.~\ref{fig:comb_therapy}(b)).
Here, ``sensitive" means that the reproductive rate decreases with increasing inhibitor concentration, while ``resistant" means that the reproductive rate is unaffected by inhibitor.
Each reproductive rate can be modeled using the generalized Monod equation~\cite{chou1981generalized, pena2014testing}:
\begin{gather}
    R_1(a, b)=\frac{\rho_1}{1+\kappa^A_1a+\kappa^B_1b},\quad R_2(a)=\frac{\rho_2}{1+\kappa^A_2a},\notag\\
    R_3(b)=\frac{\rho_3}{1+\kappa^B_3b},\quad R_4=\rho_4. \label{eq:Monod}
\end{gather}
Here, $a$ and $b$ denote the concentration of $A$ and $B$, respectively, the constant $\rho_i$ denotes the reproductive rate of trait $i$ in the absence of both $A$ and $B$, and the positive constants $\kappa^A_i$ and $\kappa^B_i$ quantify how strongly $A$ and $B$ inhibit the reproduction of trait $i$, respectively.
We assume that there are no mutations between traits $1$ and $4$, or between traits $2$ and $3$, i.e., $M_{14}=M_{41}=M_{23}=M_{32}=0$.
We also assume that the mutation rates associated with resistance to $A$ are much lower than those associated with $B$.
In other words, $M_{13}$, $M_{31}$, $M_{24}$, and $M_{42}$ are of order $\mathcal{O}(\epsilon)$ with $0<\epsilon\ll 1$.
Indeed, mutation rates can differ substantially between different drugs in actual experiments~\cite{david1970probability, bergval2009resistant}.
The basic graph $G$ corresponding to the population is shown in Fig.~\ref{fig:comb_therapy}(c).

Under this scenario, we find the optimal control of $a$ and $b$. 
Suppose that we change $a$ and $b$ slightly by $\delta a$ and $\delta b$, respectively.
Our goal is to identify the vector $\delta\bm{a}^*:=(\delta a^*, \delta b^*)^\top$ that most significantly reduces the mean fitness, subject to the constraint $\sqrt{(\delta a^*)^2+ (\delta b^*)^2}=1$.
The vector $\delta\bm{a}^*$ is given by
\begin{align}\label{eq:optimal}
     \begin{pmatrix}
    \delta a^* \\
    \delta b^*
    \end{pmatrix}
    =-\mathcal{N}
    \begin{pmatrix}
    \partial_a\meanfit \\
    \partial_b\meanfit
    \end{pmatrix}
\end{align}
with $\mathcal{N}:=1/\sqrt{(\partial_a\meanfit)^2+(\partial_b\meanfit)^2}$.
Indeed, $\delta\bm{a}^*$ given by Eq.~\eqref{eq:optimal} attains the following lower bound of the mean fitness change:
\begin{align}
    \begin{pmatrix}
    \delta a \\
    \delta b
    \end{pmatrix}
    \cdot
    \begin{pmatrix}
    \partial_a\meanfit \\
    \partial_b\meanfit
    \end{pmatrix}
    \ge
    -\sqrt{(\partial_a\meanfit)^2+(\partial_b\meanfit)^2},
\end{align}
where we applied the Cauchy--Schwarz inequality.
Therefore, we can obtain the expressions for $\delta\bm{a}^*$:
\begin{align}
    &\begin{pmatrix}
    \delta a^* \\
    \delta b^*
    \end{pmatrix}
    \propto
    \begin{pmatrix}    \partial_aR_1\cdot\partial_{R_1}\meanfit+\partial_aR_2\cdot\partial_{R_2}\meanfit\\\partial_bR_1\cdot\partial_{R_1}\meanfit+\partial_bR_3\cdot\partial_{R_3}\meanfit
    \end{pmatrix}\notag\\
    &\propto    
    \begin{pmatrix}    \partial_aR_1\sum_{H\in\mathcal{F}_1(G)}w(H)+\partial_aR_2\sum_{H'\in\mathcal{F}_2(G)}w(H') \\  \partial_bR_1\sum_{H\in\mathcal{F}_1(G)}w(H)+\partial_bR_3\sum_{H'\in\mathcal{F}_3(G)}w(H')
    \end{pmatrix},\label{eq:estimate_graph}
\end{align}
where we used Eq.~\eqref{eq:optimal} in the first line and Theorem~\ref{thm:Th1} in the last line.

\begin{figure}[hbtp]
    \centering
    \includegraphics[width=\linewidth]{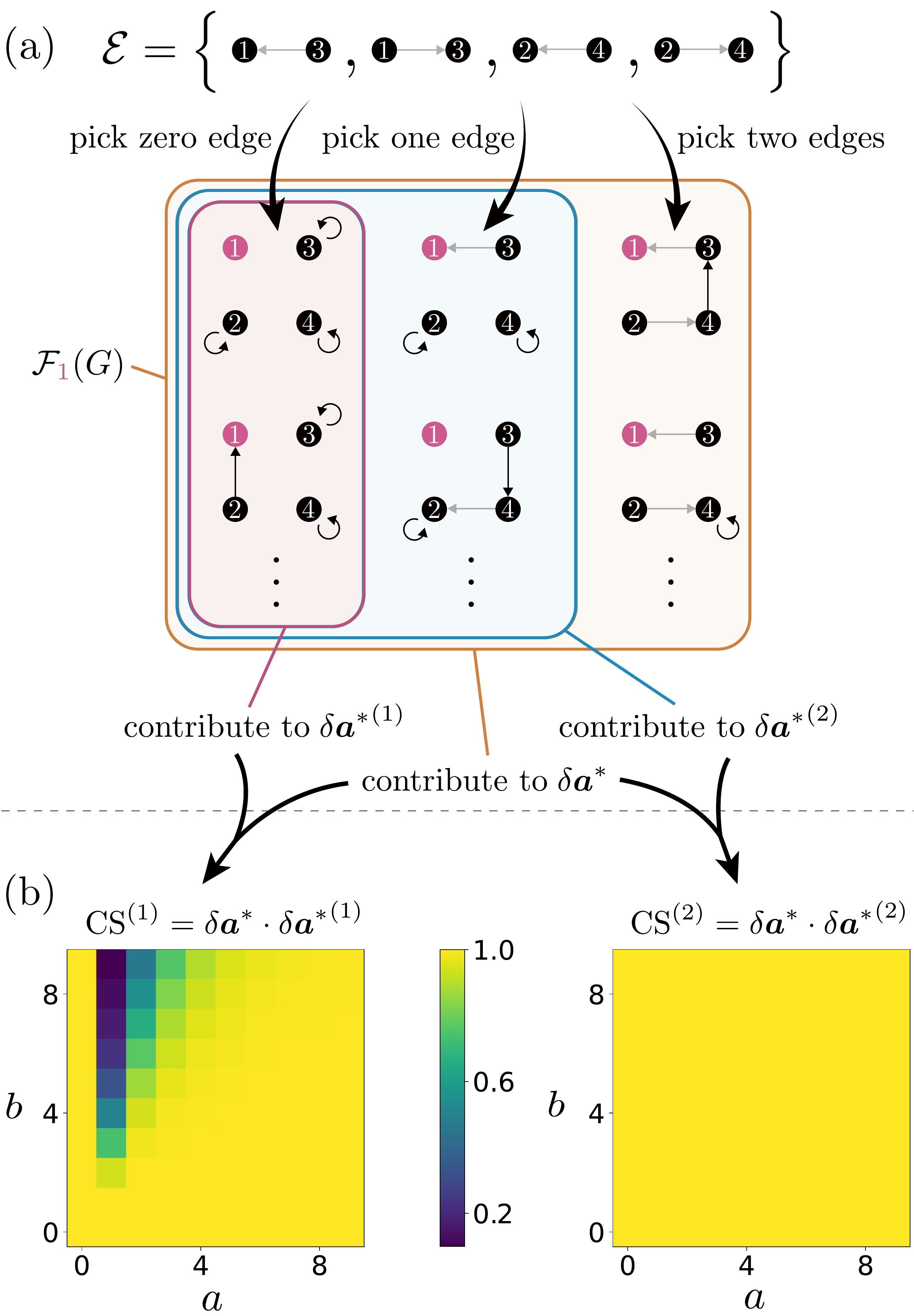}
    \caption{
    (a) Illustration of which graphs in $\mathcal{F}_1(G)$ contribute to ${\delta\bm{a}^*}^{(1)}$, ${\delta\bm{a}^*}^{(2)}$, and $\delta\bm{a}^*$.
    Graphs with no edges in $\mathcal{E}$ contribute to ${\delta\bm{a}^*}^{(1)}$, those with at most one edge in $\mathcal{E}$ contribute to ${\delta\bm{a}^*}^{(2)}$, and all graphs in $\mathcal{F}_1(G)$ contribute to $\delta\bm{a}^*$.
    (b) Cosine similarity $\mathrm{CS}^{(1)}=\delta\bm{a}^*\cdot{\delta\bm{a}^*}^{(1)}$ (left) and $\mathrm{CS}^{(2)}=\delta\bm{a}^*\cdot{\delta\bm{a}^*}^{(2)}$ (right).
    }
    \label{fig:accuracy}
\end{figure}

The diagrammatic expression in Eq.~\eqref{eq:estimate_graph} allows us to reduce the number of graphs that need to be considered when calculating $\delta\bm{a}^*$.
Since the edges in $\mathcal{E}:=\{(1, 3), (3, 1), (2, 4), (4, 2)\}$ have much lower weights, we can obtain two approximations of $\delta\bm{a}^*$ based on how many of these edges are taken into account in Eq.~\eqref{eq:estimate_graph}.
The first approximation, denoted by ${\delta\bm{a}^*}^{(1)}$, is obtained from Eq.~\eqref{eq:estimate_graph} by including only graphs that do not contain any edges in $\mathcal{E}$ (see Fig.~\ref{fig:accuracy}(a) for the case of $\mathcal{F}_1(G)$).
The second approximation, denoted by ${\delta\bm{a}^*}^{(2)}$, is obtained from Eq.~\eqref{eq:estimate_graph} by including only graphs that contain at most one edge in $\mathcal{E}$ (see Fig.~\ref{fig:accuracy}(a)).
Note that if we include graphs containing at most two edges in $\mathcal{E}$, we obtain the exact expression for $\delta\bm{a}^*$, since every graph in Eq.~\eqref{eq:estimate_graph} contains at most two edges in $\mathcal{E}$ (see also Fig.~\ref{fig:accuracy}(a)). 
We emphasize that these approximations, based on the diagrammatic approach, cannot be derived from the conventional expressions for $\partial_{R_i}\meanfit$, Eq.~\eqref{eq:resp_R}.

We demonstrate that the second approximation of $\delta\bm{a}^*$, ${\delta\bm{a}^*}^{(2)}$, is sufficiently accurate and reduces the number of graphs that need to be considered.
To this end, we evaluate the accuracy of ${\delta\bm{a}^*}^{(1)}$ and ${\delta\bm{a}^*}^{(2)}$ using the cosine similarity, defined as 
\begin{align}
    \mathrm{CS}^{(1)}:=\delta\bm{a}^*\cdot{\delta\bm{a}^*}^{(1)},\quad \mathrm{CS}^{(2)}:=\delta\bm{a}^*\cdot{\delta\bm{a}^*}^{(2)}.
\end{align}
In Fig~\ref{fig:accuracy}(b), we calculate $\mathrm{CS}^{(1)}$ and $\mathrm{CS}^{(2)}$ for $(a, b)\in [0, 9]\times [0, 9]$.
In the calculation, the parameters in Eq.~\eqref{eq:Monod} are set as
\begin{gather}
    R_1(a, b)=\frac{10}{1+a+b},\quad R_2(a)=\frac{8}{1+a},\notag\\
    R_3(b)=\frac{7}{1+b},\quad R_4=5,
\end{gather}
and mutation rates are set as
\begin{align}
    \MM &=
    \begin{pmatrix}
        -0.2&0.1&0&0\\
        0.2&-0.1&0&0\\
        0&0&-0.2&0.1\\
        0&0&0.2&-0.1
    \end{pmatrix}\notag\\
    &\quad+\epsilon
    \begin{pmatrix}
        -0.1&0&0.5&0\\
        0&-0.1&0&0.3\\
        0.1&0&-0.5&0\\
        0&0.1&0&-0.3
    \end{pmatrix}
\end{align}
with $\epsilon=0.001$.
As shown in Fig~\ref{fig:accuracy}(b), ${\delta\bm{a}^*}^{(2)}$ is sufficiently accurate, while ${\delta\bm{a}^*}^{(1)}$ is not.
Therefore, it is sufficient to consider only graphs that contain at most one edge in $\mathcal{E}$.
In general, the diagrammatic approach enables us to reduce the number of graphs that need to be considered when some mutation rates are much lower than the rest.

\section{Summary and discussion}
In this paper, we extend the Markov chain tree theorem to the parallel mutation-reproduction model [Lemma~\ref{lem:Lem1}] and derived diagrammatic expressions for the static responses of the mean fitness [Theorems~\ref{thm:Th1} and \ref{thm:Th2}] and the steady-state distribution [Theorem~\ref{thm:Th3} and Corollary~\ref{cor:pi}].
We applied our results to two limiting cases and provided a constructive approach for approximating the static responses of the mean fitness and the steady-state distribution.
Specifically,  Theorems~\ref{thm:Th1} and \ref{thm:Th2} enable the prediction of the static responses of the mean fitness based solely on information about the current state of the population. This study focuses primarily on population dynamics, but our results can be applied to various systems described by the parallel mutation-reproduction model. For example, Eq.~\eqref{eq:dn_dt} can be interpreted as the rate equation for deterministic chemical reactions involving self-replication. Therefore, our results could be useful for analyzing the static properties of these systems.

Approximate and exact expressions {for} the steady-state distribution and {the} static responses of {the} mean fitness may both be useful for controlling evolutionary dynamics. Because these expressions are given by the parameters $\MM$, $\RR$ and $\meanfit$, they provide guidelines for altering the environment to achieve the desired steady state and static responsiveness by changing the mutation rates $\MM$ and the fitness $\RR$.
Indeed, we applied our results in the context of combination therapy and reduced the number of graphs required to determine the optimal control of inhibitors.
Controlling such evolution and population dynamics will likely be useful from both an epidemiological and a biotechnological perspective.

Similar to how the conventional Markov chain tree theorem and the concept of the spanning tree are used to analyze the static properties of systems in stochastic thermodynamics~\cite{schnakenberg1976network, maes2013heat, polettini2017effective, owen2020universal, khodabandehlou2022trees, fernandes2023topologically, dal2023geometry, chun2023trade, liang2024thermodynamic, harunari2024mutual, floyd2024learning, floyd2024limits}, our result may be useful for investigating the static properties of the parallel mutation-reproduction model. In stochastic thermodynamics, various inequalities have been obtained using the Markov chain tree theorem. It may be possible to find similar inequalities in population dynamics using our generalization of the Markov chain tree theorem.

Our results may be of interest from a graph theory perspective. In order to obtain diagrammatic expressions for the steady-state distribution and the static response of the mean fitness, we must introduce a generalization of a rooted spanning tree, that is a rooted $0$/$1$ loop forest. Since a rooted $0$/$1$ loop forest generalizes a rooted spanning tree, it may be possible to generalize some concepts in graph theory related to spanning trees, such as cycles and cyclic affinities~\cite{schnakenberg1976network}, for the parallel mutation-reproduction model. Interestingly, we indeed obtain a generalization of Cayley’s
formula for the maximum number of rooted $0$/$1$ loop forests in Appendix~\ref{app:max_number}.

\begin{acknowledgments}
The authors thank Kohei Yoshimura, Naruo Ohga, and Namiko Mitarai for their suggestive comments.
R.N.\ is supported by JSPS KAKENHI Grants No.~25KJ0931.
S.I.\ is supported by JSPS KAKENHI Grants No.~22H01141, No.~23H00467, and No.~24H00834, 
JST ERATO Grant No.~JPMJER2302, 
and UTEC-UTokyo FSI Research Grant Program.
This research was supported by JSR Fellowship, the University of Tokyo.
\end{acknowledgments}

\appendix

\section{The details of Eq.~\eqref{eq:dn_dt}}
\label{app:derive_equation}
Here, we explain the details of  Eq.~\eqref{eq:dn_dt}. We assume that reproduction and death occur independently, and that mutation to another trait occurs during reproduction.
Let $r_i$ and $d_i$ be the probabilities that trait $i$ will reproduce and die in unit time, respectively.
Furthermore, let $P_{ij}$ be the probability that the immediate descendants of trait $j$ will have trait $i$. Note that $P_{ij}$ satisfies $\sum_iP_{ij}=1$.
Then, we can write down the number of trait $i$ at time $t+\Delta t$ as follows,
\begin{align}\label{eq:delta_t}
    n_i(t+\Delta t)&=(1-d_i\Delta t)n_i(t) +\Delta t\sum_{j}P_{ij}r_jn_j(t),
\end{align} 
The first term represents the individuals with trait $i$ that do not die.
The second term represents the mutation to trait $i$ from the other traits $j\neq i$ and the reproduction of trait $i$.
Taking the limit $\Delta t\to 0$ in Eq.~\eqref{eq:delta_t} and using $P_{ii}=1-\sum_{j(\ne i)}P_{ji}$, we obtain
\begin{align}
    \frac{d}{dt}n_i(t)=(r_i-d_i)n_i(t)+\sum_{j(\ne i)}(P_{ij}r_jn_j(t)-P_{ji}r_in_i(t)).
\end{align}
Therefore, defining $R_i:=r_i-d_i$ and $M_{ij}:=P_{ij}r_j\, (i\ne j)$ reproduces Eq.~\eqref{eq:dn_dt}.

\section{Derivation of Eq.~\eqref{eq:dp_dt}}
\label{app:from_n_to_p}
We derive Eq.~\eqref{eq:dp_dt} from Eq.~\eqref{eq:dn_dt}.
Using Eq.~\eqref{eq:dn_dt}, $d p_i(t)/dt$ can be computed as
\begin{align}\label{eq:derive_dp_dt}
    &\frac{d}{dt}p_i(t)=\frac{d}{dt}\frac{n_i(t)}{n_\mathrm{tot}(t)}\notag\\
    &=\frac{1}{n_\mathrm{tot}(t)}\left\{R_in_i(t)+\sum_{j(\ne i)}(M_{ij}n_j(t)-M_{ji}n_i(t))\right\}\notag\\
    &\quad-\frac{n_i(t)}{n_\mathrm{tot}(t)^2}\sum_j\Bigg\{R_jn_j(t)\Bigg.\notag\\
    &\hspace{3cm}\Bigg.+\sum_{k(\ne j)}(M_{jk}n_k(t)-M_{kj}n_j(t))\Bigg\}\notag\\
    &=R_ip_i(t)+\sum_{j(\ne i)}(M_{ij}p_j(t)-M_{ji}p_i(t))- \ev{R}_{\bm{p}(t)} p_i(t).
\end{align}
In the last equation, we used the definition of mean fitness and $\sum_j\sum_{k(\ne j)}M_{jk}n_k(t)=\sum_j\sum_{k(\ne j)}M_{kj}n_j(t)$.
Rewriting Eq.~\eqref{eq:derive_dp_dt} in the vector form yields Eq.~\eqref{eq:dp_dt}. 

\section{Relationship between the steady state and the Perron-Frobenius theorem}
\label{app:pi}
We show how the Perron--Frobenius theorem provides an expression for the steady state and its mathematical properties in Sec.~\ref{subsec:pi}. The statement of the Perron--Frobenius theorem is as follows.
\begin{onlyname}[\textbf{Perron--Frobenius theorem~\cite{pillai2005perron}.}] 
    Let $\mathsf{A}\in\mathbb{R}^{N\times N}$ be an irreducible matrix whose all components are real and non-negative,
    and $\lambda_1,\lambda_2\cdots,\lambda_N$ be the eigenvalues of $\mathsf{A}$ that are ordered as ${\rm Re}\lambda_1\ge {\rm Re}\lambda_2 \cdots\ge {\rm Re}\lambda_N$. Then, the following statements hold:
    \begin{enumerate}
        \item $\lambda_1$ is real and positive, and satisfies
        \begin{align}\label{eq:PF_max}
            |\lambda_i| \leq \lambda_1,\quad i=2,\cdots, N.
        \end{align}
        \item $\lambda_1$ is simple, i.e.,
        \begin{align}\label{eq:PF_simple}
            \lambda_1 \neq \lambda_i,\quad i=2,\cdots, N.
        \end{align}
        \item There exists an right eigenvector $\uu^{(1)}$ of $\mathsf{A}$ with eigenvalue $\lambda_1$ such that all components are positive.
    \end{enumerate}
\end{onlyname}
To apply the Perron--Frobenius theorem to our model, we define $\mathsf{A}$ as $\mathsf{A}:=\RR+\MM+a\mathsf{I}$ with $a =\max_i[-(R_i+M_{ii})]$.
For any eigenvalue of $\mathsf{A}$ denoted by $\lambda$, the corresponding eigenvector $\uu$  satisfies
\begin{align}\label{eq:eigval_shift}
(\RR+\MM)\uu=(\lambda-a)\uu,
\end{align}
which implies that the eigenvalues of $\RR+\MM$ are shifted by $-a$ from the eigenvalues of $\mathsf
{A}$ while the eigenvectors of $\RR+\MM$ are the same as the eigenvectors of $\mathsf{A}$. 
Since the matrix $\mathsf{A}$ is an irreducible matrix whose components are all real and non-negative, we can apply the Perron--Frobenius theorem to the matrix $\mathsf{A}$. We thus obtain the following corollary:
\begin{appendixcor}\label{cor:PF}
    Let $\mu_1,\mu_2\cdots,\mu_N$ be the eigenvalues of $\RR+\MM$ that are ordered as $\Re\mu_1\ge\Re\mu_2\cdots\ge\Re\mu_N$.
    Then, the following statements hold:
    \begin{enumerate}
        \item $\mu_1$ is real, and satisfies
        \begin{align}\label{eq:cor_PF_1}
            \Re\mu_i<\mu_1,\quad i=2,\cdots, N.
        \end{align}
        
        \item There exists an right eigenvector $\uu^{(1)}$ of $\RR + \MM$ with eigenvalue $\mu_1$ such that all components are positive.
    \end{enumerate}
\end{appendixcor}
The first statement of Corollary~\ref{cor:PF} is derived as follows.
From the first statement of the Perron--Frobenius theorem, we obtain
\begin{align}\label{eq:proof_cor_PF}
    \Re(\lambda_i-a)\le|\lambda_i|-a\le\lambda_1-a,\quad i=2,\cdots, N,
\end{align}
where $\lambda_1,\lambda_2\cdots,\lambda_N$ are the eigenvalues of $\mathsf{A}$ ordered as ${\rm Re}\lambda_1\ge {\rm Re}\lambda_2 \cdots\ge {\rm Re}\lambda_N$.
Combining Eq.~\eqref{eq:proof_cor_PF} with Eqs.~\eqref{eq:PF_simple} and \eqref{eq:eigval_shift},
we have $\mu_i=\lambda_i-a$ for $i=1,\cdots, N$.
In particular, we find that $\mu_1=\lambda_1-a\in\mathbb{R}$ and Eq.~\eqref{eq:cor_PF_1} hold.
The second statement is derived by combining Eq.~\eqref{eq:eigval_shift}, $\mu_1=\lambda_1-a$, and the third statement of the Perron--Frobenius theorem.

From Corollary~\ref{cor:PF}, we can prove an expression for the steady-state distribution in Sec.~\ref{subsec:pi} as follows.
For simplicity, we discuss the case where $\RR+\MM$ is diagonalizable, i.e., there exists a unitary matrix $\mathsf{U}\in\mathbb{C}^{N\times N}$ such that
\begin{align}\label{eq:spec_decomp}
    \RR+\MM&=\mathsf{U}
    \begin{pmatrix}
        \mu_1 & & \\
        & \ddots & \\
        &  & \mu_N
    \end{pmatrix}
    \mathsf{U}^\dagger \notag\\
    &=
    \begin{pmatrix}
        \uu^{(1)} &\cdots & \uu^{(N)}
    \end{pmatrix}
    \begin{pmatrix}
        \mu_1 & & \\
        & \ddots & \\
        &  & \mu_N
    \end{pmatrix}   
    \begin{pmatrix}
        \left({\vv^{(1)}}\right)^\top\\
        \vdots\\
        \left({\vv^{(N)}}\right)^\top
    \end{pmatrix}\notag\\    &=\sum_i\mu_i\uu^{(i)}\left(\vv^{(i)}\right)^\top,
\end{align}
where $\mathsf{U}^\dagger$ is the {Hermitian} conjugate of $\mathsf{U}$.
Here, $\uu^{(i)}$ and $\vv^{(i)}$ are the right and left eigenvectors of $\RR+\MM$ with eigenvalue $\mu_i$, respectively. Since $\mathsf{U}$ is a unitary matrix, 
$\uu^{(i)}$ and $\vv^{(i)}$ satisfy $\left(\vv^{(i)}\right)^\top\uu^{(j)}=(\mathsf{U}^\dagger \mathsf{U})_{ij}=\delta_{ij}$, where $\delta_{ij}$ is the Kronecker delta. 
Any initial condition $\bm{n}(0)$ is expanded as $\bm{n}(0)=\sum_ic_i\uu^{(i)}$ with $c_i\in\mathbb{C}$.
Thus, we obtain
\begin{align}\label{eq:n_decomp}
    \bm{n}(t)=e^{(\RR+\MM)t}\bm{n}(0)=\sum_ic_ie^{\mu_i t}\uu^{(i)},
\end{align}
where we used $e^{(\RR+\MM)t}=\sum_ie^{\mu_i t}\uu^{(i)}\left(\vv^{(i)}\right)^\top$ which follows from Eq.~\eqref{eq:spec_decomp} and $\left(\vv^{(i)}\right)^\top\uu^{(j)}=\delta_{ij}$.
Using Eq.~\eqref{eq:n_decomp}, we obtain the long-time limit of $\bm{p}(t)$ as
\begin{align}\label{eq:p_converge}
    \lim_{t\to\infty}\bm{p}(t)&=\lim_{t\to\infty}\frac{\bm{n}(t)}{\sum_jn_j(t)}\notag\\
    &=\lim_{t\to\infty}\frac{\sum_ic_ie^{\mu_i t}\uu^{(i)}}{\sum_j\sum_ic_ie^{\mu_i t}u^{(i)}_j}\notag\\
    &=\frac{\uu^{(1)}}{\sum_ju^{(1)}_j}=:\ppi,
\end{align}
where we used Eq.~\eqref{eq:cor_PF_1} in the third equality.
Equation~\eqref{eq:p_converge} means $\bm{p}(t)$ converges to $\ppi=\uu^{(1)}/\sum_ju^{(1)}_j$ regardless of the initial condition.
Furthermore, the second statement of Corollary~\ref{cor:PF} implies that all components of $\ppi=\uu^{(1)}/\sum_ju^{(1)}_j$ are strictly positive, which results in the positivity of $\ppi$. If the matrix $\mathsf{R}+\mathsf{M}$ is not diagonalizable, we also obtain a similar expression for $\boldsymbol{n} (t)$ whose dominant term in the long-time limit is $\boldsymbol{n}(t) \propto e^{\mu_1 t} \boldsymbol{u}^{(1)}$~\cite{antsaklis1997linear}. Thus, we again obtain Eq.~(\ref{eq:p_converge}) in general.

We show that the eigenvalue $\mu_1$ is equal to the reproductive rate $\langle R \rangle_{\bm \pi}$.
Noting $\ppi\propto\uu^{(1)}$, we find that $\ppi$ is the right eigenvector of $\RR+\MM$ with eigenvalue $\mu
_1$.
Combining this fact and Eq.~\eqref{eq:right_eigvec}, we obtain
\begin{align}\label{eq:first_eigval}
    \mu_1=\meanfit.
\end{align}

We prove that there exists the vector $\zzeta$ that satisfies Eqs.~\eqref{eq:left_eigvec}, \eqref{eq:normal}, and whose every element is positive.
Applying the Perron-Frobenius theorem to $\mathsf{A}^\top=(\RR+\MM-a\mathsf{I})^\top$ with $a =\max_i[-(R_i+M_{ii})]$, we can state that there exists the left eigenvector of $\RR+\MM$, $({\vv^{(1)}})^{\top}$, whose all components are strictly positive.
Therefore, if we define $\zzeta\in\mathbb{R}^N$ as $\zzeta:=\vv^{(1)}/\{(\vv^{(1)})^\top\ppi\}$, we find that all components of $\zzeta$ are strictly positive, and $\zzeta$ satisfies Eq.~\eqref{eq:normal}.
Since $\zzeta$ is the left eigenvector of $\RR+\MM$ and Eq.~\eqref{eq:first_eigval} holds, Eq.~\eqref{eq:left_eigvec} is also proved.

\section{Derivation of equations~\eqref{eq:resp_R} and~\eqref{eq:resp_M}}
\label{app:resp}
We here derive Eq.~\eqref{eq:resp_R}.
Multiplying Eq.~\eqref{eq:right_eigvec} by $\zzeta^\top$ from the left and using Eq.~\eqref{eq:normal}, we obtain
\begin{align}\label{eq:meanfit}
    \meanfit=\zzeta^\top(\RR+\MM)\ppi.
\end{align}
By differentiating this equation with respect to $R_i$, we obtain
\begin{align}\label{eq:resp_R_deriv}
    \frac{\partial\meanfit}{\partial R_i}&=\frac{\partial\zzeta^\top}{\partial R_i}(\RR+\MM)\ppi+\zeta_i\pi_i+\zzeta^\top(\RR+\MM)\frac{\partial\ppi}{\partial R_i}.
\end{align} 
The first and third terms in Eq.~\eqref{eq:resp_R_deriv} cancel out as follows,
\begin{align}\label{eq:vanish}
    &\frac{\partial\zzeta^\top}{\partial R_i}(\RR+\MM)\ppi+\zzeta^\top(\RR+\MM)\frac{\partial\ppi}{\partial R_i}\notag\\
    &=\meanfit\left(\frac{\partial\zzeta^\top}{\partial R_i}\ppi+\zzeta^\top\frac{\partial\ppi}{\partial R_i}\right)\notag\\
    &=\meanfit\frac{\partial}{\partial R_i}(\zzeta^\top\ppi)=0.
\end{align}
In the second line, we used Eqs.~\eqref{eq:right_eigvec} and \eqref{eq:left_eigvec}. 
In the third line, we used Eq.~\eqref{eq:normal}.
Substituting Eq.~\eqref{eq:vanish} into Eq.~\eqref{eq:resp_R_deriv}, we obtain Eq.~\eqref{eq:resp_R}.

Equation~\eqref{eq:resp_M} can be obtained in a similar way.
Differentiating Eq.~\eqref{eq:meanfit} with respect to $M_{ij}$, we obtain Eq.~\eqref{eq:resp_M} as follows,
\begin{align}\label{eq:resp_M_deriv}
    \frac{\partial\meanfit}{\partial M_{ij}}\!&=\!\frac{\partial\zzeta^\top}{\partial M_{ij}}\!(\RR+\MM)\ppi\!+\!\zeta_i\pi_j\!-\!\zeta_j\pi_j+\!\zzeta^\top(\RR+\MM)\frac{\partial\ppi}{\partial M_{ij}} \notag\\
    &=\meanfit\frac{\partial}{\partial M_{ij}}(\zzeta^\top\ppi) + \zeta_i\pi_j - \zeta_j\pi_j \notag \\
    &= (\zeta_i - \zeta_j )\pi_j,
\end{align}
where we used $\partial M_{jj}/\partial M_{ij} =-1$ and a result similar to Eq.~\eqref{eq:vanish}. 

\section{Construction of rooted $0$/$1$ loop forests}
\label{app:enumeration}
\begin{figure}[htbp]
    \centering
    \includegraphics[width=\linewidth]{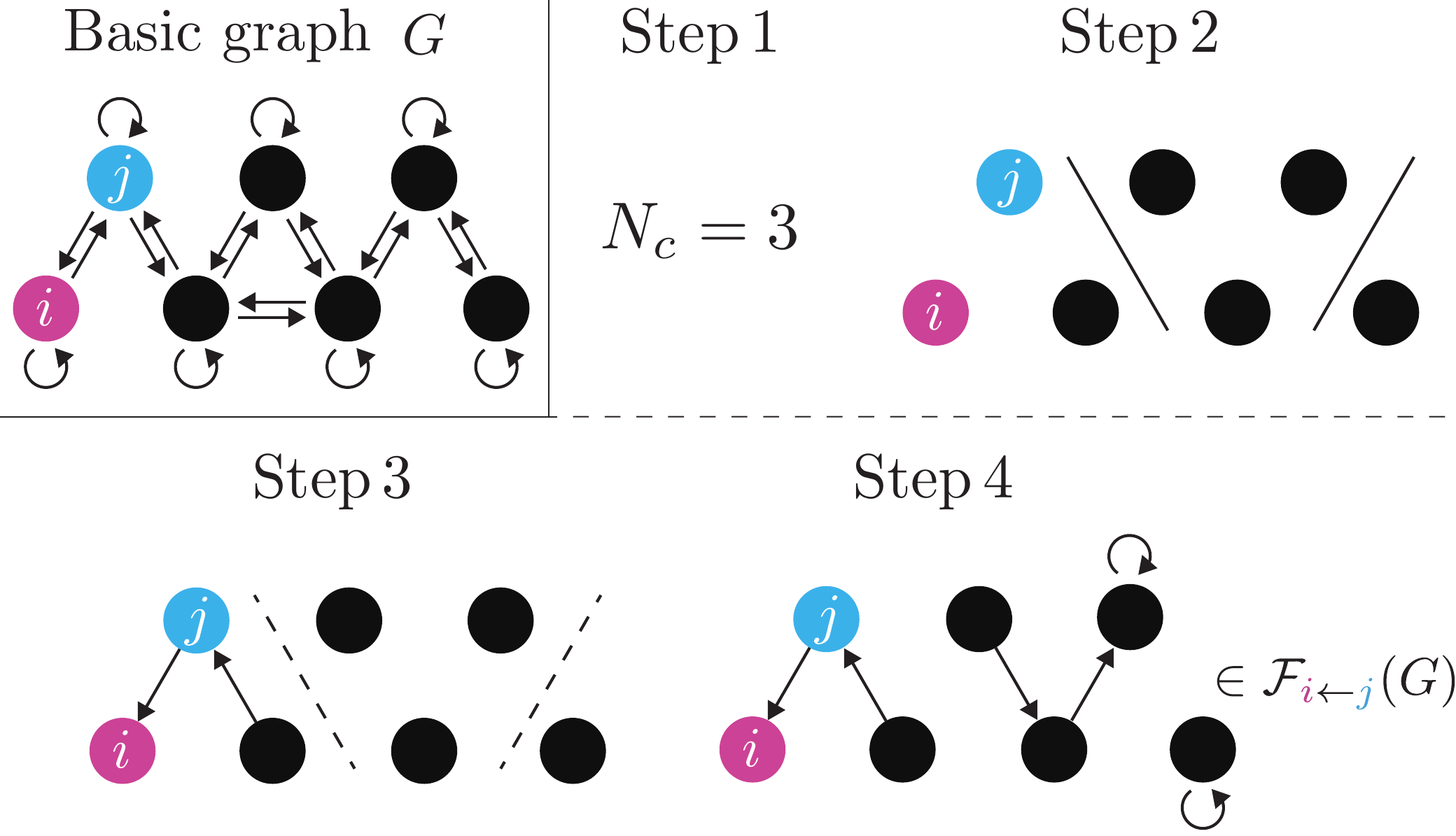}
    \caption{Illustration of the construction of a rooted forest in $\mathcal{F}_{i\leftarrow j}(G)$. A basic graph $G$ is shown in the top left.
    In Step 1, $N_{\mathrm{c}}$ is set to $3$.
    In Step 2, the vertices of $G$ are partitioned into $N_{\mathrm{c}}=3$ subsets {such that} vertex $i$ and vertex $j$ belong to the same subset.
    In Step 3, a spanning tree rooted at vertex $i$ is constructed using the edges of $G$ in the subset containing vertex $i$.
    In Step 4, spanning trees rooted at a selected vertex are constructed in the remaining subsets, with a loop attached to each root.
    }
    \label{fig:construction}
\end{figure}

We present a method for constructing all rooted $0$/$1$ loop forests of a given basic graph.
Given a basic graph $G$, a graph in $\mathcal{F}_{i\leftarrow j}(G)$ can be constructed using the following procedure (see Fig.~\ref{fig:construction} for an example).
\begin{enumerate}[label=Step\,\arabic*., leftmargin=1.1cm]
    \item Determine the number of components $N_{\mathrm{c}}$.
    \item Partition the vertices of $G$ into $N_{\mathrm{c}}$ subsets {such} that vertex $i$ and vertex $j$ belong to the same subset.
    \item In the subset containing vertex $i$, construct a spanning tree rooted at vertex $i$ using the edges of $G$.
    \item For each subset that does not contain vertex $i$, select one vertex.
    Then construct a spanning tree rooted at that vertex using the edges of $G$ and attach a loop to the root.
\end{enumerate}
By performing all possibilities for these four procedures, we can construct all rooted forests. Note that Step 3 and Step 4 are not always achievable for any partition of vertices.

Similarly, we can construct all rooted forests in $\mathcal{F}_{i\not\leftarrow j}(G)$ and $\mathcal{F}_i(G)$.
To construct all rooted forests in $\mathcal{F}_{i\not\leftarrow j}(G)$, replace Step 2 with Step 2', which is described as follows:
\begin{enumerate}[leftmargin=1.3cm]
    \item[Step\, 2'.] Partition the vertices of $G$ into $N_{\mathrm{c}}$ classes {such} that vertex $i$ and vertex $j$ belong to different classes.
\end{enumerate}
To construct all rooted forests in $\mathcal{F}_i(G)$, replace Step 2 with Step 2'', which is described as follows:
\begin{enumerate}[leftmargin=1.3cm]
    \item[Step\, 2''.] Partition the vertices of $G$ into $N_{\mathrm{c}}$ classes.
\end{enumerate}

\section{Derivation of Lemma~\ref{lem:Lem1}}
\label{app:proof_main}
In this section, we present the proof of Lemma~\ref{lem:Lem1}.
For convenience, instead of proving Eq.~\eqref{eq:Lem1} itself, we prove the following equation obtained by swapping the indices $i$ and $j$.
\begin{align}\label{eq:Lem1_ij}
    \zeta_i\pi_j =\frac{1}{Z}\sum_{H\in\mathcal{F}_{j\leftarrow i}(G)}w(H).
\end{align}
The proof in this section is a generalization of the proof of the Markov chain tree theorem in Ref.~\cite{avanzini2024methods}.

The outline of the proof is as follows.
In Sec.~\ref{app:zeta_pi_rewrite}, we express $\zeta_i\pi_j$ in terms of the matrix $-\left(\RR-\meanfit\mathsf{I}+\MM\right)$. 
In Sec.~\ref{app:incidence_matrix}, we rewrite the expression for $\zeta_i\pi_j$ derived in Sec.~\ref{app:zeta_pi_rewrite} as a sum of several contributions.
We show that each contribution corresponds to a graph.
In Sec.~\ref{app:equivalent}, we give the necessary and sufficient conditions for rooted $0$/$1$ loop forests that are suitable for our proof.
In Sec.~\ref{app:conditions}, we derive the necessary conditions for the contribution to be nonzero.
These conditions allow us to restrict the range of the summation in Sec.~\ref{app:incidence_matrix}.
We also show that these necessary conditions are related to some conditions for rooted $0$/$1$ loop forests given in Sec.~\ref{app:equivalent}.
In Sec.~\ref{app:block_diagonal}, we transform each matrix corresponding to a contribution into block diagonal matrix.
This makes it easy to calculate each contribution.
Each block of the block diagonal matrix represents a component of the graph corresponding to each contribution.
In Sec.~\ref{app:detB_detGamma}, we express each contribution in terms of the weight of a graph by using the block diagonalization given in Sec.~\ref{app:block_diagonal}.
In Sec.~\ref{app:zeta_pi_explicit}, we express $\zeta_i\pi_j$ in terms of the weights of rooted $0$/$1$ loop forests by combining the above results.

\subsection{An expression for $\zeta_i\pi_j$}
\label{app:zeta_pi_rewrite}
We derive an expression for $\zeta_i\pi_j$ in terms of the matrix $\LL:=-\left(\RR-\meanfit\mathsf{I}+\MM\right)$.
The adjugate matrix $\adj (\LL)$ is defined as $(\adj(\LL))_{ji}:=(-1)^{i+j}\det(\LL_{\backslash(i, j)})$, where $\LL_{\backslash(i, j)}$ is the matrix obtained by removing the $i$-th row and $j$-th column from $\LL$.
The adjugate matrix $\adj (\LL)$ satisfies $\LL \adj(\LL)=\det(\LL) \II$. Note that $\det(\LL)=0$ because Eq.~\eqref{eq:right_eigvec} implies that $\LL$ has zero eigenvalue.
Combining these equations, we obtain $\LL \adj(\LL)=\mathsf{O}$ where $\mathsf{O}$ is the zero matrix.
This means that every column of $\adj(\LL)$ is in the kernel $\ker(\LL)$.
Equation~\eqref{eq:right_eigvec} implies $\ppi\in\ker(\LL)$.
Moreover, the eigenvalue $\meanfit$ of the matrix $\RR+\MM$ is simple, and the corresponding eigenvector is $\ppi$ as discussed in Eqs.~\eqref{eq:cor_PF_1} and \eqref{eq:first_eigval}. This fact implies that $\ker(\LL)$ is spanned by $\ppi$, i.e., $\ker(\LL)=\{c\ppi\mid c\in\mathbb{C}\}$. Thus, each column of $\adj(\LL)$ is proportional to $\ppi$.  
Similarly, the adjugate matrix $\adj(\LL^{\top})$ also satisfies $\LL^{\top} \adj(\LL^{\top})=\det(\LL) \II$, and we also obtain $\LL^\top\adj(\LL^\top)=\mathsf{O}$ and $\ker(\LL^\top)=\{c\zzeta \mid c\in\mathbb{C}\}$. Thus, each column of $\adj(\LL^{\top}) =(\adj(\LL))^{\top}$ is proportional to $\zzeta$.
Combining these facts, we find $\adj(\LL)\propto\ppi\zzeta^\top$.
Since Eq.~\eqref{eq:normal} gives $\mathrm {tr}(\ppi\zzeta^\top) := \sum_i (\ppi\zzeta^\top)_{ii} = \zzeta^{\top} \ppi =1$, we obtain $\ppi\zzeta^\top={\adj(\LL)}/{{\rm tr} (\adj(\LL))}$. 
Therefore, we can rewrite $\zeta_i\pi_j=(\ppi\zzeta^\top)_{ji}$ as 
\begin{align}\label{eq:rewrite_zeta_pi}
    \zeta_i\pi_j=\frac{(\adj(\LL))_{ji}}{\rm tr(\adj(\LL))}=(-1)^{i+j}\frac{\det(\LL_{\backslash(i, j)})}{\sum_k\det(\LL_{\backslash(k, k)})}.
\end{align}

\subsection{An expression for $\det(\LL_{\backslash(i, j)})$ by generalizations of incidence matrix and weighted incidence matrix}
\label{app:incidence_matrix}
\begin{figure*}[htbp]
    \centering
    \includegraphics[width=\linewidth]{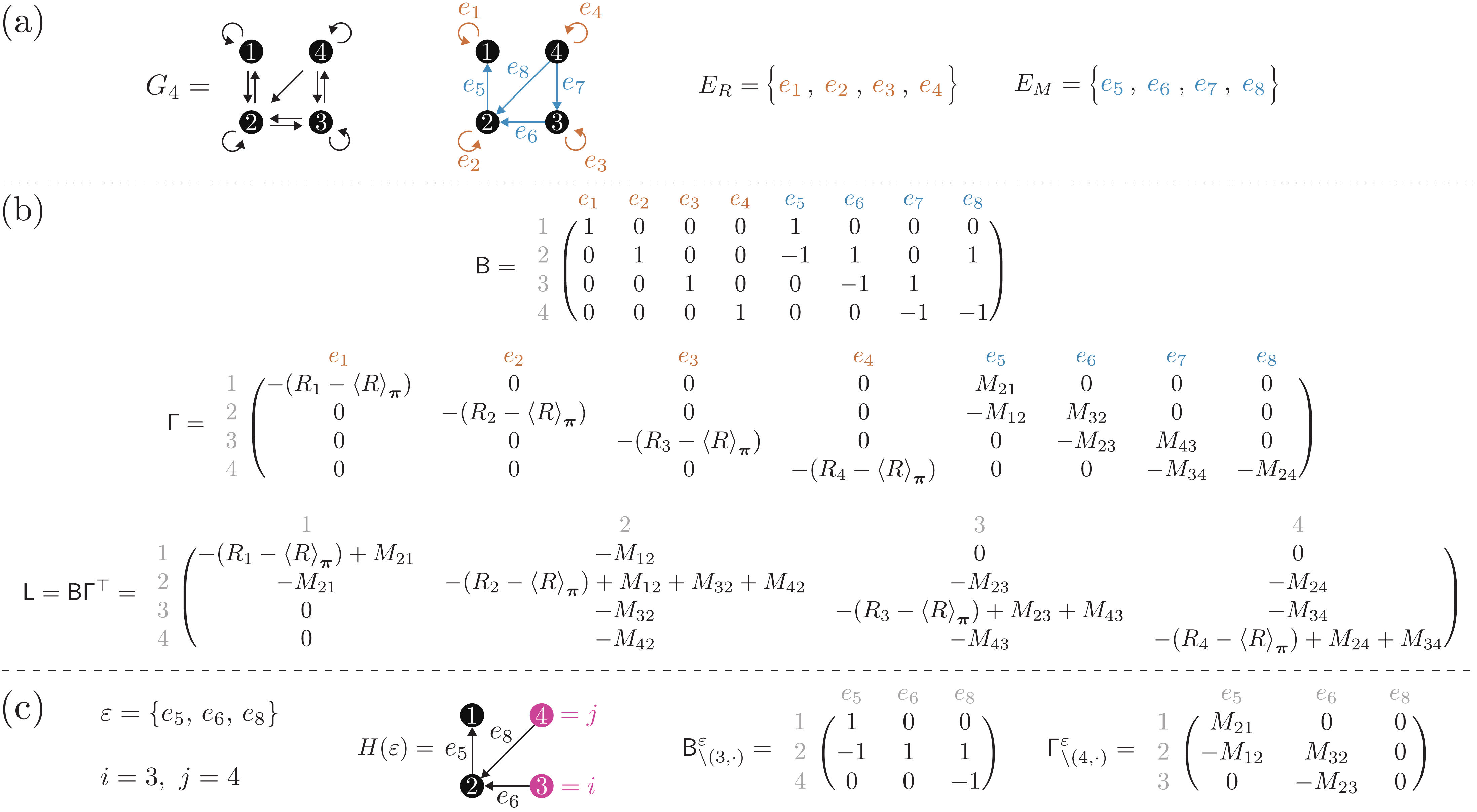}
    \caption{(a) An example of 
    $E_R$ and $E_M$ for the basic graph $G_4$.
    (b) An example of $\BB$, $\GGamma$, and $\LL$ for the basic graph $G_4$.
    Equation~\eqref{eq:L_decomp} can be confirmed.
    (c) An example of {$H(\varepsilon)$}, $\BB^\varepsilon_{\backslash(i, \cdot)}$, and $\GGamma^\varepsilon_{\backslash(j, \cdot)}$ for the basic graph $G_4$.}
    \label{fig:L_decomp}
\end{figure*}

We introduce generalizations of the incidence matrix and the weighted incidence matrix to express $\LL$ as the product of these matrices.
Since these matrices represent the structure of the basic graph, this fact allows us to apply methods from graph theory to compute $\LL_{\backslash(i, j)}$ in Eq.~\eqref{eq:rewrite_zeta_pi}.
To introduce these two matrices, we define two sets of edges $E_R$ and $E_M$ as follows (see also Fig.~\ref{fig:L_decomp}(a) for an example).
\begin{align}
    E_R&:=\{(i\leftarrow i)\mid i\in V(G) \},\label{eq:E_R}\\
    E_M&:=\{(i\leftarrow j)\mid i<j,\, M_{ij}\ne 0 \:\text{or}\: M_{ji}\ne 0,\notag\\
    &\hspace{3.4cm} i \in V(G),\, j\in V(G)\}.\label{eq:E_M}
\end{align}
Note that the element of $E_M$ is not necessarily in $E(G)$ while the element of $E_R$ is in $E(G)$ because $(i\leftarrow j)$ is in $E_M$ when {and $i<j$}, $M_{ij}= 0$, and $M_{ji} \neq 0$.
By arbitrarily labeling the edges in $E_R\cup E_M$ as $E_R\cup E_M=\{e_1,\cdots, e_{|E_R|+|E_M|}\}$, we define the \textit{generalized incidence matrix} $\BB\in\mathbb{R}^{N\times(|E_R|+|E_M|)}$ and the \textit{generalized weighted incidence matrix} $\GGamma\in\mathbb{R}^{N\times(|E_R|+|E_M|)}$ as 
\begin{equation}\label{eq:def_B}
(\BB)_{i\alpha}:= 
\left\{ 
\begin{aligned}
  &\delta_{i\rmt(e_\alpha)} && \text{if} \: e_\alpha\in E_R, \\
  &\delta_{i\rmt(e_\alpha)}-\delta_{i\rms(e_\alpha)} && \text{if} \: e_\alpha\in E_M,
\end{aligned}
\right.
\end{equation}

\begin{equation}\label{eq:def_Gamma}
(\GGamma)_{i\alpha}:= 
\left\{ 
\begin{aligned}
  &-(R_i-\meanfit)\delta_{i\rmt(e_\alpha)}\\
  &\hspace{4cm} \text{if} \: e_\alpha\in E_R, \\
  &M_{\rms(e_\alpha)\rmt(e_\alpha)}\delta_{i\rmt(e_\alpha)}-M_{\rmt(e_\alpha)\rms(e_\alpha)}\delta_{i\rms(e_\alpha)}\\
  &\hspace{4cm} \text{if} \: e_\alpha\in E_M,
\end{aligned}
\right.
\end{equation}
respectively, where {$N=|V(G)|$, and} $\rms(e)$ and $\rmt(e)$ are the source and target vertices of an edge $e=\left(\rmt(e)\leftarrow\rms(e)\right)$, respectively (see also Fig.~\ref{fig:L_decomp}(b) for an example). Note that while the element of the conventional incidence matrix corresponding to a loop is set to zero, the element of the generalized incidence matrix $(\BB)_{i\alpha}$ can be nonzero even if the edge $e_{\alpha}$ is a loop.

Using $\BB$ and $\GGamma$, we can express $\LL$ as follows (see also Fig.~\ref{fig:L_decomp}(b) for an example):
\begin{align}\label{eq:L_decomp}
    \LL&=\BB\GGamma^\top.
\end{align}
This equation can be verified by direct calculation as follows.
\begin{align*}
    &(\BB\GGamma^\top)_{ij}\\
    =&\sum_{\alpha\mid e_\alpha\in E_R\cup E_M}(\BB)_{i\alpha}(\GGamma)_{j\alpha}\\
    =&\!-\!\sum_{ e_\alpha\in E_R}(R_j-\meanfit)\delta_{i\rmt(e_\alpha)}\delta_{j\rmt(e_\alpha)}  \\
    &\!+\!\sum_{ e_\alpha\in E_M} (M_{\rms(e_\alpha)\rmt(e_\alpha)}\delta_{j\rmt(e_\alpha)} \delta_{i\rmt(e_\alpha)} \!\notag\\
    &\hspace{1.5cm}+\!M_{\rmt(e_\alpha)\rms(e_\alpha)}\delta_{j\rms(e_\alpha)} \delta_{i\rms(e_\alpha)})\\
    &\!-\!\sum_{ e_\alpha\in E_M}(M_{\rms(e_\alpha)\rmt(e_\alpha)}\delta_{j\rmt(e_\alpha)}\delta_{i\rms(e_\alpha)}\notag\\
    &\hspace{1.5cm}\!+ \!M_{\rmt(e_\alpha)\rms(e_\alpha)}\delta_{j\rms(e_\alpha)} \delta_{i\rmt(e_\alpha)})\\
    &=\delta_{ij}\left\{-(R_j-\meanfit)-M_{ii}\right\}+(1-\delta_{ij})(-M_{ij})\\
    &=\LL_{ij}.
\end{align*}
We refer to $\LL$ as the \textit{generalized Laplacian matrix} because the conventional Laplacian matrix for the weighted graph can be written as the product of the incidence matrix and the weighted incidence matrix~\cite{godsil2013algebraic}. The matrix $\LL_{\backslash(i, j)}$ is also given by 
\begin{align}\label{eq:L_decomp2}
    \LL_{\backslash(i, j)}&=\BB_{\backslash(i, \cdot)} (\GGamma_{\backslash( j, \cdot )})^\top,
\end{align}
where $\BB_{\backslash(i, \cdot)}$ and $\GGamma_{\backslash(i, \cdot )}$ are the matrices obtained by removing the $i$-th row from $\BB$ and $\GGamma$, respectively.

We obtain an expression for $\det(\LL_{\backslash(i, j)})$.
{Let} $\BB^\varepsilon_{\backslash(i, \cdot)}\in\mathbb{R}^{(N-1)\times(N-1)}$ and $\GGamma^\varepsilon_{\backslash(i, \cdot)}\in\mathbb{R}^{(N-1)\times(N-1)}$ be the matrices obtained by removing all columns except the selected $N-1$ columns corresponding to the set of the edges $\varepsilon$ from $\BB_{\backslash(i, \cdot)}$ and $\GGamma_{\backslash(i, \cdot)}$, respectively (see also Fig.~\ref{fig:L_decomp}(c) for an example).
Applying the Cauchy--Binet formula to Eq.~\eqref{eq:L_decomp2}, we obtain
\begin{align}\label{eq:CB_formula}
    \det(\LL_{\backslash(i, j)})=\sum_{\varepsilon}\det(\BB^\varepsilon_{\backslash(i, \cdot)})\det(\GGamma^\varepsilon_{\backslash(j, \cdot)}),
\end{align}
where the sum over $\varepsilon$ is taken over all possible subsets of $E_R\cup E_M$ consisting of $N-1$ edges.

As shown below, each term in the summation in Eq.~\eqref{eq:CB_formula} corresponds to a graph.
For each edge set $\varepsilon$ in Eq.~\eqref{eq:CB_formula}, we define the corresponding graph $H(\varepsilon)$ as the graph whose edge set is $\varepsilon$ and whose vertices are the endpoints of the edges in $\varepsilon$ {(see also Fig.~\ref{fig:L_decomp}(c) for an example)}.
We then interpret the right-hand side of Eq.~\eqref{eq:CB_formula} as the summation of the contributions from all such graphs $H(\varepsilon)$.
We note that $H(\varepsilon)$ is not necessarily a subgraph of $G$.

\subsection{Conditions for a rooted $0$/$1$ loop forest}
\label{app:equivalent}
We introduce some conditions imposed on a graph. Some of these conditions are used to characterize a rooted $0$/$1$ loop forest as shown in Eqs.\eqref{eq:Fji_def} and \eqref{eq:Fji_equivalence}. Moreover, some of these conditions imposed on $H(\varepsilon)$ are related to the conditions for $\det(\BB^\varepsilon_{\backslash(i, \cdot)})\det(\GGamma^\varepsilon_{\backslash(j, \cdot)})\ne 0$ in Eq.~\eqref{eq:CB_formula}.
Therefore, these conditions connect $H(\varepsilon)$ to a rooted $0$/$1$ loop forest.

We now introduce the following notations, $\mathrm{C}_1$, $\mathrm{C}_2$, $\mathrm{C}_3(i)$, $\mathrm{C}_4(i)$,  $\mathrm{C}_5(i)$, $\mathrm{C}_6(i, j)$, and $\mathrm{C}_7$ for various conditions imposed on a graph $H$.
\begin{itemize}
    \item $H$ satisfies $\mathrm{C}_1$.
    $\overset{\mathrm{def}}{\iff} H$ contains no cycles.
    \item $H$ satisfies $\mathrm{C}_2$.
    $\overset{\mathrm{def}}{\iff} V(H)=V(G)$. 
    \item $H$ satisfies $\mathrm{C}_3(i)$.
    $\overset{\mathrm{def}}{\iff}$ In each component of $H$ that does not contain vertex $i$, there is exactly one loop.
    \item $H$ satisfies $\mathrm{C}_4(i)$.
    $\overset{\mathrm{def}}{\iff}$ 
    In the component of $H$ that contains vertex $i$, there are no loops. 
    \item $H$ satisfies $\mathrm{C}_5(i)$.
    $\overset{\mathrm{def}}{\iff}$ In the component of $H$ that contains vertex $i$, all edges are directed to vertex $i$.
    In each component of $H$ that does not contain vertex $i$, all edges are directed to the vertex with the loop.
    \item $H$ satisfies $\mathrm{C}_6(i, j)$.
    $\overset{\mathrm{def}}{\iff}$ Vertex $i$ and vertex $j$ are connected {in} $H$.
    \item $H$ satisfies $\mathrm{C}_7$.
    $\overset{\mathrm{def}}{\iff} H$ has $|V(G)|-1$ edges. 
\end{itemize}

Using these notations, the definition of $\mathcal{F}_{j\leftarrow i}(G)$ given in Sec.~\ref{subsec:rooted_loop_forest} can be rewritten as follows.
\begin{align}
    \phantom{\overset{\mathrm{def}}{\iff}} &H\in\mathcal{F}_{j\leftarrow i}(G)\notag\\
    \overset{\mathrm{def}}{\iff} &H\subseteq G\:\text{and}\notag\\
    &H \text{satisfies}\: \mathrm{C}_1, \mathrm{C}_2, \mathrm{C}_3(j), \mathrm{C}_4(j), \mathrm{C}_5(j),\: \text{and}\: \mathrm{C}_6(i, j).
    \label{eq:Fji_def}
\end{align}
The condition $\mathrm{C}_2$ can be replaced with $\mathrm{C}_7$ as follows. 
\begin{align}
\phantom{\iff} &H\in\mathcal{F}_{j\leftarrow i}(G)\notag\\
\iff &H\subseteq G\:\text{and}\notag\\
&H \text{satisfies}\: \mathrm{C}_1, \mathrm{C}_3(j), \mathrm{C}_4(j), \mathrm{C}_5(j), \mathrm{C}_6(i, j),\: \text{and}\: \mathrm{C}_7.
\label{eq:Fji_equivalence}
\end{align}
The equivalent expression for $\mathcal{F}_{j\leftarrow i}(G)$ in Eq.~\eqref{eq:Fji_equivalence} is used later to prove Lemma.~\ref{lem:Lem1}, rather than its definition in Eq.~\eqref{eq:Fji_def}.

The statement in Eq.~\eqref{eq:Fji_equivalence} can be proved as follows.
As mentioned below Eq.~\eqref{eq:H_union}, any $H$ in $\mathcal{F}_{j\leftarrow i}(G)$ defined by Eq.~\eqref{eq:Fji_def} has $|V(G)|-1$ edges. 
In other words, if $H$ satisfies all the conditions in Eq.~\eqref{eq:Fji_def}, then it also satisfies $\mathrm{C}_7$, and thus it satisfies all the conditions in Eq.~\eqref{eq:Fji_equivalence}. Next, we show that if $H$ satisfies all the conditions in Eq.~\eqref{eq:Fji_equivalence}, then it also satisfies all the conditions in Eq.~\eqref{eq:Fji_def}.
Let $N_{\mathrm{c}}$ be the number of components of $H$.
Since $\mathrm{C}_3(j)$ is satisfied, $H$ can be expressed as $H=H^{(1)}\cup\bigcup_{\alpha=2}^{N_{\mathrm{c}}}\left(H^{(\alpha)}\cup (r^{(\alpha)}\leftarrow r^{(\alpha)})\right)$, where $H^{(1)}$ is the component containing vertex $j$, $r^{(\alpha)}$ is the vertex that has the unique loop in the $\alpha$-th component, and $H^{(\alpha)}$ is the graph obtained by removing the loop $(r^{(\alpha)}\leftarrow r^{(\alpha)})$ from the $\alpha$-th component {for $\alpha\ge 2$}.
Since $\mathrm{C}_1$, $\mathrm{C}_3(j)$, and $\mathrm{C}_4(j)$ are satisfied, each $H^{(\alpha)}$ is connected and contains neither cycles nor loops.
Therefore, $H^{(\alpha)}$ is a rooted spanning tree for any $\alpha$, and  $|E(H^{(\alpha)})|=|V(H^{(\alpha)})|-1$ is satisfied.
Using this equation and $\mathrm{C}_7$, we obtain $|V(G)|-1=|E(H)|=|E(H^{(1)})|+\sum_{\alpha=2}^{N_{\mathrm{c}}}(|E(H^{(\alpha)})|+1)=|V(H^{(1)})|-1+\sum_{\alpha=2}^{N_{\mathrm{c}}}|V(H^{(\alpha)})|=|V(H)|-1$. Because $H$ is a subgraph of $G$, $|V(G)|=|V(H)|$ leads to $V(H)=V(G)$, that is the condition $\mathrm{C}_2$.
Therefore, $H$ satisfying all the conditions in Eq.~\eqref{eq:Fji_equivalence} satisfies all the conditions in Eq.~\eqref{eq:Fji_def}. From the above, the necessary and sufficient conditions in Eq.~\eqref{eq:Fji_def} have been proved.

\subsection{Conditions for $\det(\BB^\varepsilon_{\backslash(i, \cdot)})\det(\GGamma^\varepsilon_{\backslash(j, \cdot)})\ne0$}
\label{app:conditions}
We consider the conditions for $\det(\BB^\varepsilon_{\backslash(i, \cdot)})\det(\GGamma^\varepsilon_{\backslash(j, \cdot)})\ne 0$ because $\varepsilon$ satisfying $\det(\BB^\varepsilon_{\backslash(i, \cdot)})\det(\GGamma^\varepsilon_{\backslash(j, \cdot)})\ne 0$ only contributes to $\sum_{\varepsilon}\det(\BB^\varepsilon_{\backslash(i, \cdot)})\det(\GGamma^\varepsilon_{\backslash(j, \cdot)})$ in Eq.~\eqref{eq:CB_formula}.
Here, we show that $\det(\BB^\varepsilon_{\backslash(i, \cdot)})\det(\GGamma^\varepsilon_{\backslash(j, \cdot)})= 0$ if the corresponding graph $H(\varepsilon)$ does not satisfy any of $\mathrm{C}_1$, $\mathrm{C}_3(j)$, $\mathrm{C}_4(j)$, and $\mathrm{C}_6(i, j)$. Note that $H(\varepsilon)$ always satisfies $\mathrm{C}_7$ since $\varepsilon$ consists of $N-1$ edges with $N=|V(G)|$.
This fact allows us to restrict the sum to those $\varepsilon$ such that the corresponding graph $H(\varepsilon)$ satisfies these conditions, which are related to  $\mathcal{F}_{j\leftarrow i}(G)$.

\subsubsection{$\mathrm{C}_1$}
\label{app:C1}
\begin{figure*}[htbp]
    \centering
    \includegraphics[width=\linewidth]{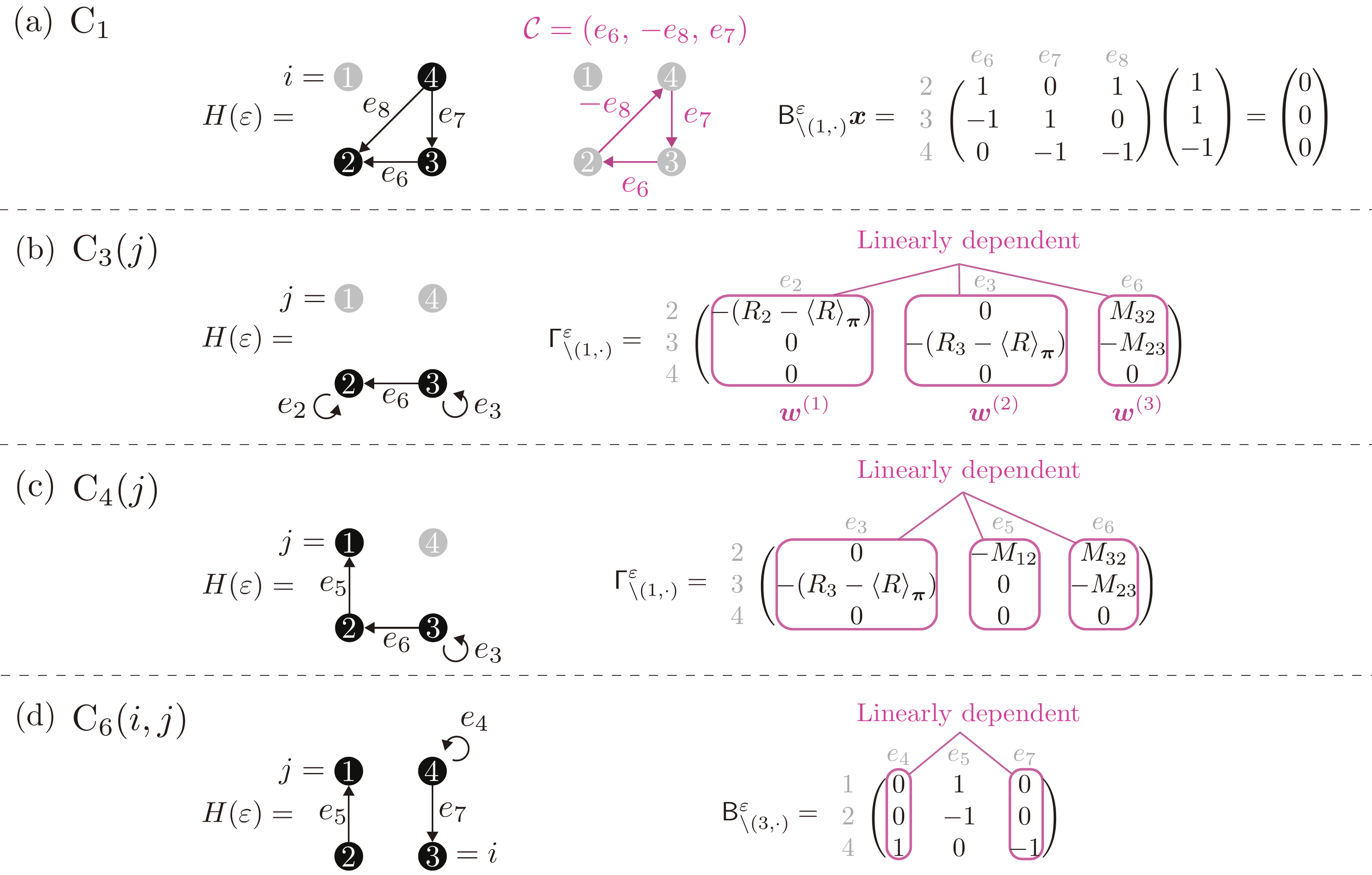}
    \caption{Examples of graphs that do not satisfy $\mathrm{C}_1$, $\mathrm{C}_3(j)$, $\mathrm{C}_4(j)$, or $\mathrm{C}_6(i, j)$.
    (a) This $H(\varepsilon)$ does not satisfy $\mathrm{C}_1$ because it has a cycle $(e_6,\, e_8,\, e_7)$.
    A directed cycle $\mathcal{C}=(e_6,\, -e_8,\, e_7)$ is obtained by reversing the direction of $e_8$. The vector
    $\bm{x}$ defined in Eq.~\eqref{eq:x} is $\bm{x}=(1,\, 1,\, -1)^\top$ because $e_6,\ e_7\in\mathcal{C}$ and $-e_8\in\mathcal{C}$.
    Then, $\BB^\varepsilon_{\backslash(1, \cdot)}\bm{x}=\bm{0}$ holds.
    (b) This $H(\varepsilon)$ does not satisfy $\mathrm{C}_3(1)$ because it contains two loops, $e_2$ and $e_3$, in the component that does not contain $j=1$.
    Then, the three columns in $\GGamma^\varepsilon_{\backslash(1, \cdot)}$, $\ww^{(1)}:=(-\delR{2}, 0, 0)^\top$, $\ww^{(2)}:=(0, -\delR{3}, 0)^\top$, and $\ww^{(3)}:=(M_{32}, -M_{23}, 0)^\top$, are linearly dependent.
    (c) This $H(\varepsilon)$ does not satisfy $\mathrm{C}_4(1)$ because it has a loop, $e_3$, in the component containing $j=1$.
    Then the three columns in $\GGamma^\varepsilon_{\backslash(1, \cdot)}$ are linearly dependent.
    (d) This $H(\varepsilon)$ does not satisfy $\mathrm{C}_6(1, 2)$ because $i=3$ and $j=1$ belong to different components.
    Then, two columns in $\BB^\varepsilon_{\backslash(3, \cdot)}$ are linearly dependent.
    }
    \label{fig:Conditions}
\end{figure*}

We show that $\det(\BB^\varepsilon_{\backslash(i, \cdot)})\det(\GGamma^\varepsilon_{\backslash(j, \cdot)})=0$ if $H(\varepsilon)$ does not satisfy $\mathrm{C}_1$. If $H(\varepsilon)$ does not satisfy $\mathrm{C}_1$, {then} $H(\varepsilon)$ contains a cycle. Here, we introduce the notation $E(H(\varepsilon))=\varepsilon=\{e_{\alpha_1},\cdots, e_{\alpha_{N-1}} \}$ where $e_{\alpha_k}$ corresponds to the $k$-th column of $\BB^\varepsilon_{\backslash(i, \cdot)}$ (or $\GGamma^\varepsilon_{\backslash(j, \cdot)}$).
Suppose that $H(\varepsilon)$ contains a cycle $(e'_{1},\cdots, e'_{L})$ where $e'_k\in E(H(\varepsilon))$ for $k=1, \dots, L$ and $L\ge 3$. Note that a cycle does not contain a loop, and the target vertex of the edge $e'_k$ is different from the source vertex of the edge $e'_k$ ($\rms(e'_k) \neq \rmt(e'_k)$). We introduce $-e:=\left(\rms(e)\leftarrow\rmt(e)\right)$ as the reverse of the edge $e = \left(\rmt(e)\leftarrow\rms(e)\right)$.
By appropriately considering the reverse of some edges in the cycle $(e'_{1},\cdots, e'_{L})$, we obtain a \textit{directed cycle} $\mathcal{C}=(\bar{e}'_1,\cdots, \bar{e}'_L)$ satisfying $\rms(\bar{e}'_{k+1})=\rmt(\bar{e}'_{k})$ for any $k=1, \dots, L$ where $\rms(\bar{e}'_{L+1}):=\rms(\bar{e}'_1)$.
Using the directed cycle $\mathcal{C}$, we define the vector $\bm{x}=(x_1, \dots, x_{N-1})^{\top}\in\mathbb{R}^{N-1}$ as
\begin{equation}\label{eq:x}
x_k:= 
\left\{ 
\begin{aligned}
  &1 && \text{if} \: e_{\alpha_k}\in\mathcal{C}, \\
  &-1 && \text{if} \: -e_{\alpha_k}\in\mathcal{C},\\
  &0 && \text{otherwise}.
\end{aligned}
\right.
\end{equation}
Therefore, the following relation for the directed cycle and the incidence matrix~\cite{west2001introduction},
\begin{align}\label{cycleincidence}
    \BB^\varepsilon_{\backslash(i, \cdot)}\bm{x}=\bm{0},
\end{align}
is satisfied (see also Fig.~\ref{fig:Conditions}(a) for an example) because the contributions of loops do not exist in $\BB^\varepsilon_{\backslash(i, \cdot)}\bm{x}$, i.e., $x_k=0$ if $e_{\alpha_k} \in E_R$, and we only consider the contribution by the conventional incidence matrix for $e_{\alpha_k} \in E_M$. We can also verify Eq.~\eqref{cycleincidence} directly using Eqs.~\eqref{eq:def_B} and \eqref{eq:x}, and $\rmt(\bar{e}'_{k})=\rms(\bar{e}'_{k+1})$ for $k=1, \dots, L$.
Equation~\eqref{cycleincidence} implies that the columns of $\BB^\varepsilon_{\backslash(i, \cdot)}$ are linearly dependent and thus $\det(\BB^\varepsilon_{\backslash(i, \cdot)})=0$. Therefore, $\det(\BB^\varepsilon_{\backslash(i, \cdot)})\det(\GGamma^\varepsilon_{\backslash(j, \cdot)})=0$ if $H(\varepsilon)$ does not satisfy $\mathrm{C}_1$.
\subsubsection{$\mathrm{C}_3(j)$}
\label{app:C3}
We show that $\det(\BB^\varepsilon_{\backslash(i, \cdot)})\det(\GGamma^\varepsilon_{\backslash(j, \cdot)})=0$ if $H(\varepsilon)$ does not satisfy $\mathrm{C}_3(j)$ and satisfies $\mathrm{C}_1$. We assume that $H(\varepsilon)$ satisfies $\mathrm{C}_1$, since otherwise $\det(\BB^\varepsilon_{\backslash(i, \cdot)})\det(\GGamma^\varepsilon_{\backslash(j, \cdot)})=0$ already holds, as discussed in \ref{app:C1}. Therefore, we assume that the component does not contain cycles.

There are two cases where $\mathrm{C}_3(j)$ is not satisfied.
The first case is that $H(\varepsilon)$ contains a component with at least two loops, where this component does not contain vertex $j$.
The second case is that $H(\varepsilon)$ contains a component without loops, where this component does not contain vertex $j$.
We show $\det(\BB^\varepsilon_{\backslash(i, \cdot)})\det(\GGamma^\varepsilon_{\backslash(j, \cdot)})=0$ in both cases.

First, we consider the first case.
We arbitrarily choose such a component and let $V'$ and $E'$ denote the set of vertices and the set of edges for this component, respectively.
We can show $|E'|-1\ge |V'|$ because the number of edges in a component without cycles and with $n_{\rm l} (\geq 2)$ loops is $|E'| = |V'| -1 +n_{\rm l}$. Here, the number of edges that are not loops for a component without cycles is $|V'| -1$ because the component without cycles becomes a spanning tree for $V'$ with $|V'| -1$ edges when we remove the loops from the component. Using $|E'|-1\ge |V'|$, we can show that the columns of $\GGamma^\varepsilon_{\backslash(j, \cdot)}$ are linearly dependent as follows.
Note that all rows corresponding to $V'$ are present in $\GGamma^\varepsilon_{\backslash(j, \cdot)}$ because the component we are now considering does not contain vertex $j$.
We arbitrarily label the edges in $E'$ as $E'=\{e^{(1)},\cdots, e^{(|E'|)}\}$. Let $\ww^{(k)}\in\mathbb{R}^{|V(G)|-1}$ be the column of $\GGamma^\varepsilon_{\backslash(j, \cdot)}$ corresponding to $e^{(k)}$  (see also Fig.~\ref{fig:Conditions}(b) for examples of $\ww^{(k)}$).
By the definition of $\GGamma$ in Eq.~\eqref{eq:def_Gamma}, $(\GGamma)_{i\alpha}\ne 0$ if vertex $i$ is the target vertex or the source vertex of $e_{\alpha}$.
Therefore, the element of $\ww^{(k)}$ can be nonzero only if the corresponding node is in $V'$.
In other words, each column of $\GGamma^\varepsilon_{\backslash(j, \cdot)}$ corresponding to an edge in $E'$ can have nonzero elements only in the rows corresponding to $V'$.
Combining this fact with $|E'|-1\ge |V'|$, we find that $\ww^{(k)}$ has at most $|E'|-1$ nonzero elements in the rows corresponding to $V'$.
Thus, the $|E'|$ vectors, $\ww^{(1)},\cdots, \ww^{(|E'|)}$, are linearly dependent.
As a result, the columns of $\GGamma^\varepsilon_{\backslash(j, \cdot)}$ are linearly dependent, and hence $\det(\GGamma^\varepsilon_{\backslash(j, \cdot)})=0$ (see also Fig.~\ref{fig:Conditions}(b) for an example).
Thus, we conclude $\det(\BB^\varepsilon_{\backslash(i, \cdot)})\det(\GGamma^\varepsilon_{\backslash(j, \cdot)})=0$ in the first case.

Next, we consider the second case.
Let $V'$ and $E'$ denote the set of vertices and the set of edges for this component, respectively.
We have $|E'|=|V'|-1$ because the component does not contain loops and cycles, and the component is a spanning tree for $V'$ with $|V'|-1$ edges.
Using $|E'|=|V'|-1$, we can show that the rows of $\GGamma^\varepsilon_{\backslash(j, \cdot)}$ are linearly dependent as follows.
Note that all rows corresponding to $V'$ are present in $\GGamma^\varepsilon_{\backslash(j, \cdot)}$.
These rows can have nonzero elements only in the columns corresponding to $E'$.
Combining this fact with $|E'|=|V'|-1$, we find that each row of $\GGamma^\varepsilon_{\backslash(j, \cdot)}$ corresponding to an element of $V'$ can have at most $|V'|-1$ nonzero elements in the columns corresponding to $E'$.
Thus, the $|V'|$ rows of $\GGamma^\varepsilon_{\backslash(j, \cdot)}$ corresponding to the set $V'$ are linearly dependent. 
Therefore, we conclude $\det(\GGamma^\varepsilon_{\backslash(j, \cdot)})=0$, and thus $\det(\BB^\varepsilon_{\backslash(i, \cdot)})\det(\GGamma^\varepsilon_{\backslash(j, \cdot)})=0$ in the second case.

Combining the results of the two cases, we conclude that $\det(\BB^\varepsilon_{\backslash(i, \cdot)})\det(\GGamma^\varepsilon_{\backslash(j, \cdot)})=0$ holds when $H(\varepsilon)$ does not satisfy $\mathrm{C}_3(j)$ and satisfies $\mathrm{C}_1$.

\subsubsection{$\mathrm{C}_4(j)$}
\label{app:C4}
We show that $\det(\BB^\varepsilon_{\backslash(i, \cdot)})\det(\GGamma^\varepsilon_{\backslash(j, \cdot)})=0$ if $H(\varepsilon)$ does not satisfy $\mathrm{C}_4(j)$ and satisfies $\mathrm{C}_1$. If $H(\varepsilon)$ does not satisfy $\mathrm{C}_4(j)$, then $H(\varepsilon)$ contains a component with at least one loop and vertex $j$.
Let $V'$ and $E'$ be the set of vertices and the set of edges for this component, respectively.
We can show $|E'|\ge |V'|$ because the component with $n_{\rm l} (\geq 1)$ loops and without cycles, the number of the edges is $|E'| = |V'|-1+  n_{\rm l}$.
Using $|E'|\ge |V'|$, we show that the columns of $\GGamma^\varepsilon_{\backslash(j, \cdot)}$ are linearly dependent as follows.
Note that the component contains vertex $j$, while $\GGamma^\varepsilon_{\backslash(j, \cdot)}$ does not have the row corresponding to vertex $j$.
Therefore, each column of $\GGamma^\varepsilon_{\backslash(j, \cdot)}$ corresponding to an edge in $E'$ can have nonzero elements only in the rows corresponding to $V'\setminus\{j\}$.
Since $|E'|\ge |V'|$ implies $|E'|-1\ge|V'\setminus\{j\}|$, the $|E'|$ columns of $\GGamma^\varepsilon_{\backslash(j, \cdot)}$ corresponding to $E'$ can have at most $|E'|-1$ nonzero elements in the rows corresponding to $V'\setminus\{j\}$.
Thus, these $|E'|$ columns of $\GGamma^\varepsilon_{\backslash(j, \cdot)}$ corresponding to the set $E'$  are linearly dependent (see also Fig.~\ref{fig:Conditions}(c) for an example). Therefore, we conclude $\det(\GGamma^\varepsilon_{\backslash(j, \cdot)})=0$, and thus $\det(\BB^\varepsilon_{\backslash(i, \cdot)})\det(\GGamma^\varepsilon_{\backslash(j, \cdot)})=0$ if $H(\varepsilon)$ does not satisfy $\mathrm{C}_4(j)$ and satisfies $\mathrm{C}_1$.

\begin{figure*}[htbp]
    \centering
    \includegraphics[width=\linewidth]{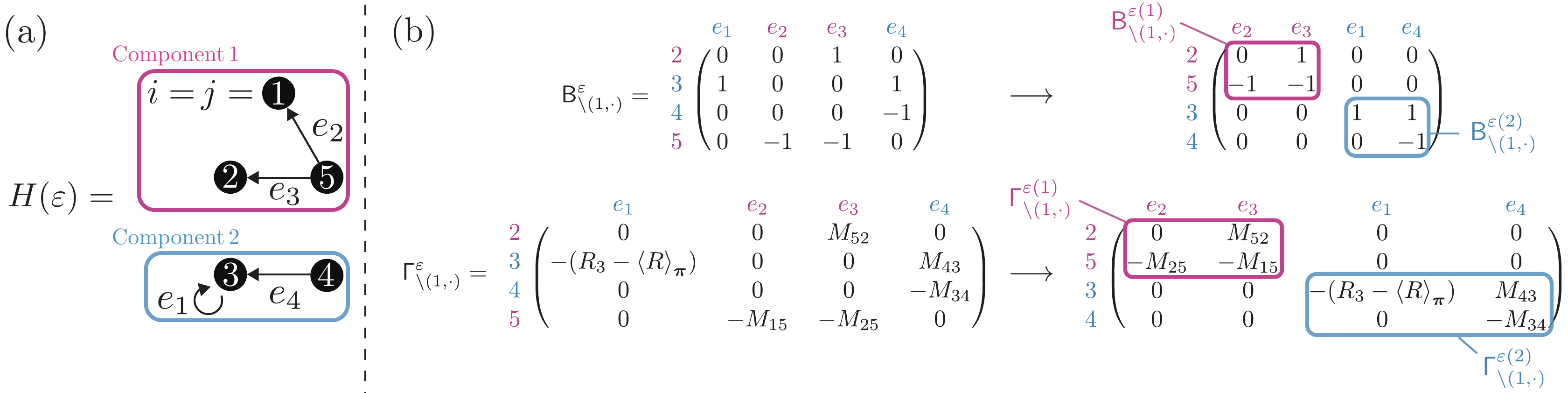}
    \caption{An example of block diagonalization of  $\BB^\varepsilon_{\backslash(1, \cdot)}$ and $\GGamma^\varepsilon_{\backslash(1, \cdot)}$.
    (a) This $H(\varepsilon)$ consists of two components.
    (b) By appropriately swapping the rows and columns, $\BB^\varepsilon_{\backslash(1, \cdot)}$ is transformed into a block diagonal matrix, and $\GGamma^\varepsilon_{\backslash(1, \cdot)}$ is also transformed into a block diagonal matrix. Each block corresponds to each component of the graph $H(\varepsilon)$.
    }
    \label{fig:BGamma_diag}
\end{figure*}
\subsubsection{$\mathrm{C}_6(i, j)$}
We show that $\det(\BB^\varepsilon_{\backslash(i, \cdot)})\det(\GGamma^\varepsilon_{\backslash(j, \cdot)})=0$ if $H(\varepsilon)$ does not satisfy $\mathrm{C}_6(i, j)$ and satisfies $\mathrm{C}_1$, $\mathrm{C}_3(j)$, and $\mathrm{C}_4(j)$. We assume that $H(\varepsilon)$ satisfies $\mathrm{C}_1$, $\mathrm{C}_3(j)$, and $\mathrm{C}_4(j)$, since otherwise $\det(\BB^\varepsilon_{\backslash(i, \cdot)})\det(\GGamma^\varepsilon_{\backslash(j, \cdot)})=0$ already holds.
Since $\mathrm{C}_6(i, j)$ is trivially satisfied when $j=i$, we consider only the case $j\ne i$.
If $H(\varepsilon)$ does not satisfy $\mathrm{C}_6(i, j)$ with $j\ne i$, then vertex $i$ and vertex $j$ belong to different components.
We focus on the component that contains vertex $i$ and does not contain vertex $j$. 
Let $V'$ and $E'$ be the set of vertices and the set of edges for this component, respectively. Since $\mathrm{C}_1$, $\mathrm{C}_3(j)$, and $\mathrm{C}_4(j)$ are satisfied, this component has a loop and no cycles.
We can show $|E'|=|V'|$ because the component is connected and contains a loop and no cycles.
Using $|E'|=|V'|$, we can show that the columns of $\BB^\varepsilon_{\backslash(i, \cdot)}$ are linearly dependent as follows.
The $|E'|$ columns of $\BB^\varepsilon_{\backslash(i, \cdot)}$ corresponding to $E'$ can have nonzero elements {only} in the rows corresponding to $V'\setminus\{i\}$, since $\BB^\varepsilon_{\backslash(i, \cdot)}$ does not have the row corresponding to vertex $i$.
Since $|E'|=|V'|$ implies $|E'|-1 = |V'\setminus\{i\}|$, each column corresponding to an edge in $E'$ can have at most $|E'|-1$ nonzero elements in the rows corresponding to $V'\setminus\{i\}$.
Therefore, the $|E'|$ columns of $\BB^\varepsilon_{\backslash(i, \cdot)}$ corresponding to the set $E'$ are linearly dependent (see also Fig.~\ref{fig:Conditions}(d) for an example).  Therefore, we conclude $\det(\BB^\varepsilon_{\backslash(i, \cdot)})=0$, and thus $\det(\BB^\varepsilon_{\backslash(i, \cdot)})\det(\GGamma^\varepsilon_{\backslash(j, \cdot)})=0$ if $H(\varepsilon)$ does not satisfy $\mathrm{C}_6(i, j)$ and satisfies $\mathrm{C}_1$, $\mathrm{C}_3(j)$, and $\mathrm{C}_4(j)$.

\subsubsection{Another expression for Eq.~\eqref{eq:CB_formula}}

\vspace{1\baselineskip}
Combining the discussion above, we can restrict the summation in Eq.~\eqref{eq:CB_formula} as
\begin{align}\label{eq:CB_formula_restricted}
    \det(\LL_{\backslash(i, j)})=\sum_{\substack{\varepsilon|H(\varepsilon)\:\text{satisfies}\:\mathrm{C}_1, \mathrm{C}_3(j),\\ \mathrm{C}_4(j), \mathrm{C}_6(i, j),\:\text{and}\:\mathrm{C}_7}}\det(\BB^\varepsilon_{\backslash(i, \cdot)})\det(\GGamma^\varepsilon_{\backslash(j, \cdot)}),
\end{align}
where the summation over $\varepsilon$ is taken over all subsets of $E_R\cup E_M$ such that the corresponding graph $H(\varepsilon)$ satisfies $\mathrm{C}_1$, $\mathrm{C}_3(j)$, $\mathrm{C}_4(j)$, $\mathrm{C}_6(i, j)$, and $\mathrm{C}_7$. 

\subsection{Block diagonalization of $\BB^\varepsilon_{\backslash(i, \cdot)}$ and $\GGamma^\varepsilon_{\backslash(j, \cdot)}$}
\label{app:block_diagonal}

We show that $\BB^\varepsilon_{\backslash(i, \cdot)}$ and $\GGamma^\varepsilon_{\backslash(j, \cdot)}$ can be transformed into block diagonal matrices if $H(\varepsilon)$ satisfies all the conditions $\mathrm{C}_1$, $\mathrm{C}_3(j)$, $\mathrm{C}_4(j)$, $\mathrm{C}_6(i, j)$, and $\mathrm{C}_7$, which are necessary conditions for $\det(\BB^\varepsilon_{\backslash(i, \cdot)})\det(\GGamma^\varepsilon_{\backslash(j, \cdot)})\ne 0$.
The block diagonalization makes it easy to compute nonzero values of $\det(\BB^\varepsilon_{\backslash(i, \cdot)})\det(\GGamma^\varepsilon_{\backslash(j, \cdot)})$.

If $H(\varepsilon)$ satisfies $\mathrm{C}_1$, $\mathrm{C}_3(j)$, $\mathrm{C}_4(j)$, $\mathrm{C}_6(i, j)$, and $\mathrm{C}_7$, we can introduce a procedure that transforms $\BB^\varepsilon_{\backslash(i, \cdot)}$ and $\GGamma^\varepsilon_{\backslash(j, \cdot)}$ into block diagonal matrices by swapping rows and columns as follows,
\begin{align}
    \BB^\varepsilon_{\backslash(i, \cdot)}\rightarrow
    \begin{pmatrix}
        \BB^{\varepsilon(1)}_{\backslash(i, \cdot)} & &\\
         & \ddots &\\
         & & \BB^{\varepsilon(N_{\rm c})}_{\backslash(i, \cdot)}
    \end{pmatrix},\label{eq:B_block}\\
        \GGamma^\varepsilon_{\backslash(j, \cdot)}\rightarrow
    \begin{pmatrix}
        \GGamma^{\varepsilon(1)}_{\backslash(j, \cdot)} & &\\
         & \ddots &\\
         & & \GGamma^{\varepsilon(N_{\rm c})}_{\backslash(j, \cdot)} \end{pmatrix},\label{eq:Gamma_block}
\end{align}
 (see also Fig.~\ref{fig:BGamma_diag} for an example). Note that $\BB^{\varepsilon(\alpha)}_{\backslash(i, \cdot)}$ or $\GGamma^{\varepsilon(\alpha)}_{\backslash(j, \cdot)}$ corresponds to the $\alpha$-th component $H(\varepsilon)^{(\alpha)}$ where $H(\varepsilon)$ satisfies $\mathrm{C}_1$, $\mathrm{C}_3(j)$, $\mathrm{C}_4(j)$, $\mathrm{C}_6(i, j)$, and $\mathrm{C}_7$. Thus, $N_{\mathrm{c}}$ is the number of components in $H(\varepsilon)$.
 If $i = j$, the rows and columns are swapped in the same way for both matrices $\BB^{\varepsilon}_{\backslash(i, \cdot)}$ and $\GGamma^\varepsilon_{\backslash(j, \cdot)}$. If $i \neq j$, the columns are swapped in the same way for both matrices $\BB^{\varepsilon}_{\backslash(i, \cdot)}$ and $\GGamma^\varepsilon_{\backslash(j, \cdot)}$, and the rows except the rows corresponding to $i$ and $j$ are swapped in the same way for both matrices $\BB^{\varepsilon}_{\backslash(i, \cdot)}$ and $\GGamma^\varepsilon_{\backslash(j, \cdot)}$. The row corresponding to $j$ for the matrix $\BB^{\varepsilon}_{\backslash(i, \cdot)}$ is placed in a row corresponding to the first block $\BB^{\varepsilon(1)}_{\backslash(i, \cdot)}$, and the row corresponding to $i$ for the matrix $\GGamma^{\varepsilon}_{\backslash(j, \cdot)}$ is placed in the same row as the row corresponding to $j$ for the matrix $\BB^{\varepsilon}_{\backslash(i, \cdot)}$. Thus, the row corresponding to $i$ for the matrix $\GGamma^{\varepsilon}_{\backslash(j, \cdot)}$ is placed in a row corresponds to the first block $\GGamma^{\varepsilon(1)}_{\backslash(j, \cdot)}$. The first component corresponding to $\BB^{\varepsilon(1)}_{\backslash(i, \cdot)}$ contains the vertcies $i$ and $j$ because of the condition $\mathrm{C}_6(i, j)$. Similarly,  the first component corresponding $\GGamma^{\varepsilon(1)}_{\backslash(j, \cdot)}$ contains the vertcies $j$ and $i$.

First, we discuss a procedure for the block diagonalization of $\BB^\varepsilon_{\backslash(i, \cdot)}$ in Eq.~\eqref{eq:B_block} more precisely. Let $V(H(\varepsilon)^{(\alpha)})$ be the set of vertices for the $\alpha$-th component corresponding to $\BB^{\varepsilon(\alpha)}_{\backslash(i, \cdot)}$ or $\GGamma^{\varepsilon(\alpha)}_{\backslash(j, \cdot)}$. Let $E(H(\varepsilon)^{(\alpha)})$ be the set of edges for the $\alpha$-th component corresponding to $\BB^{\varepsilon(\alpha)}_{\backslash(i, \cdot)}$ or $\GGamma^{\varepsilon(\alpha)}_{\backslash(j, \cdot)}$. For the first component, we have $|E(H(\varepsilon)^{(\alpha)})|=|V(H(\varepsilon)^{(\alpha)})|-1$ because $\mathrm{C}_1$ and $\mathrm{C}_4(i)$ imply that the first component is a spanning tree. We swap the $|V(H(\varepsilon)^{(1)})|-1$ rows of $\BB^\varepsilon_{\backslash(i, \cdot)}$ corresponding to  $V(H(\varepsilon)^{(1)})\setminus\{i\}$ with the top $|V(H(\varepsilon)^{(1)})|-1$ rows, preserving their relative order in $\BB^\varepsilon_{\backslash(i, \cdot)}$. We also swap the columns of $\BB^\varepsilon_{\backslash(i, \cdot)}$ corresponding to $E(H(\varepsilon)^{(1)})$ with the first $|E(H(\varepsilon)^{(1)})|$ columns, preserving their relative order in $\BB^\varepsilon_{\backslash(i, \cdot)}$. Note that the rows of $\BB^\varepsilon_{\backslash(i, \cdot)}$ corresponding to $V(H(\varepsilon)^{(1)})\setminus\{i\}$ can have nonzero elements only in the columns corresponding to $E(H(\varepsilon)^{(1)})$, and the columns of $\BB^\varepsilon_{\backslash(i, \cdot)}$ corresponding to $E(H(\varepsilon)^{(1)})$ can have nonzero elements only in the rows corresponding to $V(H(\varepsilon)^{(1)})\setminus\{i\}$. Because $|V(H(\varepsilon)^{(1)})\setminus\{i\}| = |V(H(\varepsilon)^{(1)})|-1=|E(H(\varepsilon)^{(1)})|$, we obtain a $|E(H(\varepsilon)^{(1)})|\times|E(H(\varepsilon)^{(1)})|$ matrix, that is $\BB^{\varepsilon(1)}_{\backslash(i, \cdot)}$. For the $\alpha$-th component with $\alpha \neq 1$, we have  $|V(H(\varepsilon)^{(\alpha)})|=|E(H(\varepsilon)^{(\alpha)})|$ because $\mathrm{C}_1$ and $\mathrm{C}_3(i)$ imply that the $\alpha$-th component contains no cycles and contains a loop with $\alpha \neq 1$. Note that $\BB^\varepsilon_{\backslash(i, \cdot)}$ 
contains all the rows corresponding to $V(H(\varepsilon)^{(\alpha)})$ since the $\alpha$-th component does not contain vertex $i$ with $\alpha \neq 1$. We swap the rows corresponding to $E(H(\varepsilon)^{(\alpha)})$ with the rows in the range from $(\sum_{\alpha'=1}^{\alpha-1}|E(H(\varepsilon)^{(\alpha')})| +1)$-th row to $\sum_{\alpha'=1}^{\alpha}|E(H(\varepsilon)^{(\alpha')})|$-th row, preserving their relative order in $\BB^\varepsilon_{\backslash(i, \cdot)}$. We also swap the columns corresponding to $V(H(\varepsilon)^{(\alpha)})$ with the columns in the same range, preserving their relative order in $\BB^\varepsilon_{\backslash(i, \cdot)}$. This swapping results in a $|E(H(\varepsilon)^{(\alpha)})|\times |E(H(\varepsilon)^{(\alpha)})|$ square matrix, which is $\BB^{\varepsilon(\alpha)}_{\backslash(i, \cdot)}$ with $\alpha \neq 1$. Note that the rows of $\BB^\varepsilon_{\backslash(i, \cdot)}$ corresponding to $V(H(\varepsilon)^{(\alpha)})$ can have nonzero elements only in the columns corresponding to $E(H(\varepsilon)^{(\alpha)})$, and the columns of $\BB^\varepsilon_{\backslash(i, \cdot)}$ corresponding to $E(H(\varepsilon)^{(\alpha)})$ can have nonzero elements only in the rows corresponding to $V(H(\varepsilon)^{(\alpha)})$ with $\alpha \neq 1$. Thus, the block diagonalization of $\BB^\varepsilon_{\backslash(i, \cdot)}$ in Eq.~\eqref{eq:B_block} is obtained in the above procedure. Note that this swapping can be done sequentially from $\alpha=1$ to $\alpha =N_{\rm c}$ because $V(H(\varepsilon)^{(\alpha)})\cap V(H(\varepsilon)^{(\beta)})=\emptyset$ and $E(H(\varepsilon)^{(\alpha)})\cap E(H(\varepsilon)^{(\beta)})=\emptyset$ hold for any $\alpha$ and $\beta(\ne \alpha)$.

Second, we discuss a procedure for the block diagonalization of $\GGamma^\varepsilon_{\backslash(j, \cdot)}$ in Eq.~\eqref{eq:Gamma_block}.  By swapping the rows of $\GGamma^\varepsilon_{\backslash(j, \cdot)}$, the row corresponding to the vertex $i$ is placed in the same relative position as the row corresponding to the vertex $j$ in $\BB^\varepsilon_{\backslash(i, \cdot)}$, and the other rows are also placed in the same relative positions as the other rows in $\BB^\varepsilon_{\backslash(i, \cdot)}$.
After swapping the rows, we additionally do the same swapping for the rows and columns of $\GGamma^\varepsilon_{\backslash(j, \cdot)}$ as we did for the rows and columns of $\BB^\varepsilon_{\backslash(i, \cdot)}$ in the block denationalization. Comparing the definition of $\BB$ in Eq.~\eqref{eq:def_B} with {that of} $\GGamma$ in Eq.~\eqref{eq:def_Gamma}, we find that $\GGamma^\varepsilon_{\backslash(i, \cdot)}$ can have nonzero elements only at the same positions where  $\BB^\varepsilon_{\backslash(j, \cdot)}$ can have nonzero elements, except for the rows corresponding to vertices $i$ and $j$.  Note that the row of $\GGamma^\varepsilon_{\backslash(j, \cdot)}$ corresponding to $i$ can have nonzero elements only in the columns corresponding to $E(H(\varepsilon)^{(1)})$.
Therefore, this swapping provides the block diagonalization of $\GGamma^\varepsilon_{\backslash(j, \cdot)}$ in Eq.~\eqref{eq:Gamma_block}.

The block diagonalizations in Eqs.~\eqref{eq:B_block} and~\eqref{eq:Gamma_block} provides
\begin{align}
\label{eq:det_BGamma_decomp_abs}
&\left|\det(\BB^\varepsilon_{\backslash(i, \cdot)})\det(\GGamma^\varepsilon_{\backslash(j, \cdot)}) \right| \notag \\
&= \prod_{\alpha=1}^{N_{\mathrm{c}}}\left|\det(\BB^{\varepsilon(\alpha)}_{\backslash(i, \cdot)})\det(\GGamma^{\varepsilon(\alpha)}_{\backslash(j, \cdot)})\right|, 
\end{align}
because swapping rows or columns of a matrix does not affect the absolute value of its determinant, and block diagonalization factorizes the determinant. 

We consider the sign when the absolute value symbols are removed from both sides of Eq.~\eqref{eq:det_BGamma_decomp_abs}. Note that swapping two rows multiplies the determinant by $-1$. By swapping two rows many times, we replace the row corresponding to $i$ in the matrix $\GGamma^\varepsilon_{\backslash(j, \cdot)}$ with the same position of the row corresponding to $j$ in the matrix $\BB^\varepsilon_{\backslash(i, \cdot)}$ while preserving the relative order of the other rows in the matrices $\GGamma^\varepsilon_{\backslash(j, \cdot)}$ and $\BB^\varepsilon_{\backslash(i, \cdot)}$. If the swapping is only between adjacent rows, this replacement can be achieved by swapping $|i-j|-1$ times for $i\neq j$ and by swapping $0$ times for $i= j$. After the replacement, we obtain the block diagonalizations in Eqs.~\eqref{eq:B_block} and~\eqref{eq:Gamma_block} using the same swapping procedure. When the same swapping is performed on two matrices,  the product of the determinants of the two matrices does not change sign even if the sign of the determinant of each matrix changes due to the swapping. Thus, we conclude that the sign change occurs only by swapping $|i-j|-1$ times for $i\neq j$ and does not occur for $i= j$.
Because the factor of the determinant for swapping $|i-j|-1$ times is $(-1)^{|i-j|-1}$, we obtain
\begin{align}
\label{eq:det_BGamma_decomp}
&\det(\BB^\varepsilon_{\backslash(i, \cdot)})\det(\GGamma^\varepsilon_{\backslash(j, \cdot)}) \notag \\
&=((-1)^{|i-j|} (\delta_{ij}-1)+ \delta_{ij})  \prod_{\alpha=1}^{N_{\mathrm{c}}}\det(\BB^{\varepsilon(\alpha)}_{\backslash(i, \cdot)})\det(\GGamma^{\varepsilon(\alpha)}_{\backslash(j, \cdot)}).
\end{align}

\begin{figure*}[htbp]
    \centering
    \includegraphics[width=\linewidth]{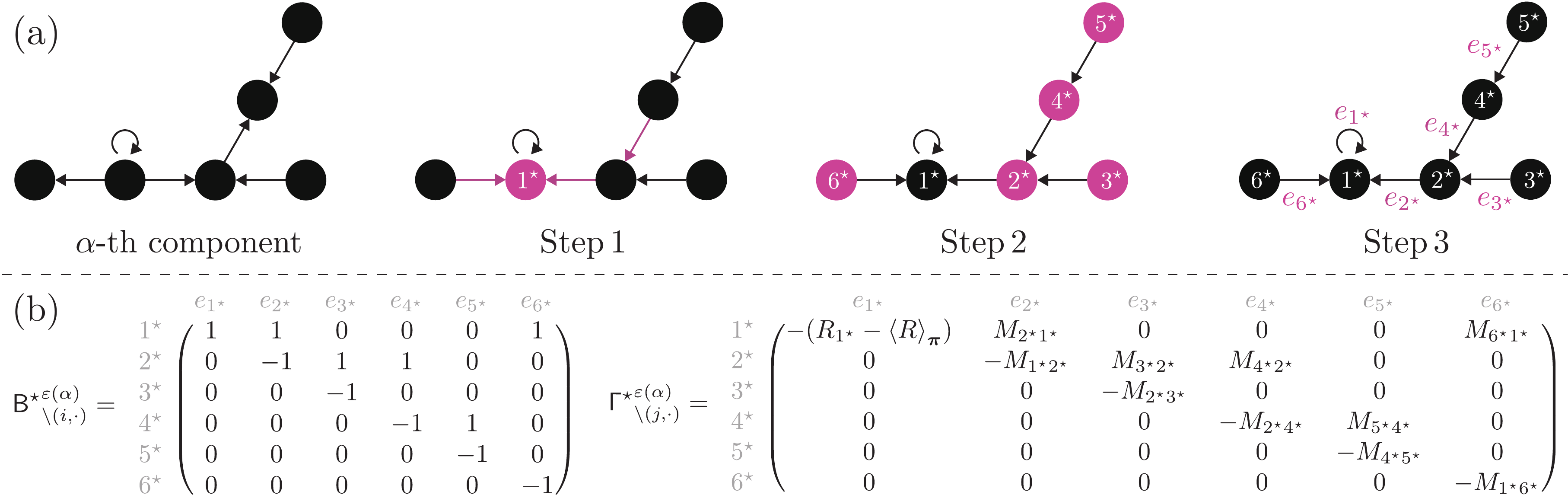}
    \caption{(a) An example of labeling the vertices and edges with $\alpha \neq 1$.
    Step 1: Label the vertex with the loop as $1^{\star}$ and change the direction of the edges so that they point toward $1^{\star}$.
        Step 2:  Starting from $l^{\star}=1^{\star}$, label the other vertices.
    Step 3: Label the loop as $e_{1^{\star}}$, and the edge from $k^{\star}$ to $l^{\star} (\prec k^{\star})$ as $e_{k^{\star}}$ for $ k^{\star} \in\{2^{\star}, \cdots, |E(H(\varepsilon)^{(\alpha)})|^{\star}\}$.
    (b) Upper triangular matrices ${\BB^\star}^{\varepsilon(\alpha)}_{\backslash(i, \cdot)}$ and ${\GGamma^\star}^{\varepsilon(\alpha)}_{\backslash(j, \cdot)}$ obtained by the procedure in (a).}
    \label{fig:labeling_loop}
\end{figure*}

\subsection{Expression for $\det(\BB^\varepsilon_{\backslash(i, \cdot)})\det(\GGamma^\varepsilon_{\backslash(j, \cdot)})$ in terms of the weight of a graph}
\label{app:detB_detGamma}
We express $\det(\BB^\varepsilon_{\backslash(i, \cdot)})\det(\GGamma^\varepsilon_{\backslash(j, \cdot)})$ in Eq.~\eqref{eq:det_BGamma_decomp} in terms of the weight of a graph. Here we introduce the graph $\widetilde{H}(\varepsilon)$, which is defined as the graph obtained from $H(\varepsilon)$ by changing the edge directions according to the condition $\mathrm{C}_5(j)$.
Thus, $\widetilde{H}(\varepsilon)$ satisfies all of the conditions for a rooted $0$/$1$ loop forest in Eq.~\eqref{eq:Fji_equivalence}, except for the condition $\widetilde{H}(\varepsilon)\subseteq G$. Note that $\widetilde{H}(\varepsilon)$ is not necessarily a subgraph of $G$ because $\widetilde{H}(\varepsilon)$ can contain an edge $(k\leftarrow l)$ such that $k<l$, $M_{kl}=0$, and $M_{lk}\ne 0$, and then $E(\widetilde{H}(\varepsilon))\nsubseteq E(G)$. Therefore, $\widetilde{H}(\varepsilon)$ is not necessarily a rooted $0$/$1$ loop forest. However, the weight of $\widetilde{H}(\varepsilon)$ is related to the weight of a rooted $0$/$1$ loop forest.

To compute $\det(\BB^\varepsilon_{\backslash(i, \cdot)})\det(\GGamma^\varepsilon_{\backslash(j, \cdot)})$, we consider each $\det(\BB^{\varepsilon(\alpha)}_{\backslash(i, \cdot)})\det(\GGamma^{\varepsilon(\alpha)}_{\backslash(i, \cdot)})$ in Eq.~\eqref{eq:det_BGamma_decomp} individually.
We show that $\det(\BB^{\varepsilon(\alpha)}_{\backslash(i, \cdot)})\det(\GGamma^{\varepsilon(\alpha)}_{\backslash(j, \cdot)})$ can be expressed by the weight of $\widetilde{H}(\varepsilon)^{(\alpha)}$, which is the $\alpha$-th component of $\widetilde{H}(\varepsilon)$.
Thus, we find that $\det(\BB^{\varepsilon}_{\backslash(i, \cdot)})\det(\GGamma^{\varepsilon}_{\backslash(j, \cdot)})$ can be expressed by the weight of $\widetilde{H}(\varepsilon)$.
We will discuss the case $\alpha\ne 1$, the case $\alpha=1$ with $i=j$, and the case $\alpha=1$ with $i\neq j$ one by one, and summarize the result.  

\subsubsection{Expression for $\det(\BB^{\varepsilon(\alpha)}_{\backslash(i, \cdot)})\det(\GGamma^{\varepsilon(\alpha)}_{\backslash(j, \cdot)})$ for $\alpha\ne 1$}
\label{subsubsec:not_contain_j}
 First, we label the vertices and edges of the $\alpha$-th component $H(\varepsilon)^{(\alpha)}$ with $\alpha \neq 1$. Here, we use the notation $^\star$ for the labeled vertices $V(H(\varepsilon)^{(\alpha)}) = \{ 1^{\star}, \dots, |E(H(\varepsilon)^{(\alpha)})|^{\star} \} $. We also introduce the total order $\prec$ (or $\succ$), such that $1^{\star}\prec 2^{\star} \prec\dots \prec |E(H(\varepsilon)^{(\alpha)}|^{\star}$ (or $|E(H(\varepsilon)^{(\alpha)}|^{\star} \succ \dots \succ 2^{\star} \succ 1^{\star} $).
 We label the vertices and edges, and 
 change the edge directions according to the following steps (see also Fig.~\ref{fig:labeling_loop}(a) for an example).
\begin{enumerate}[label=Step\,\arabic*., leftmargin=1.1cm]
    \item Label the vertex containing the loop as vertex $1^{\star}$. Change the direction of all edges to point toward vertex $1^{\star}$. Because of the conditions $\mathrm{C}_1$ and $\mathrm{C}_3(j)$, these processes are uniquely determined.
    \item Repeat the following process from $k^{\star}=1^{\star}$ until all the vertices are labeled. If we can find a vertex that is the source vertex of an edge whose target vertex is $k^{\star}$, label this vertex as vertex $(k+1)^{\star}$. If we cannot find a vertex that is the source vertex of an edge whose target vertex is $k^{\star}$, find the source vertex of an edge whose target vertex is $l^{\star}$ with $l^{\star}\prec k^{\star}$ and label this vertex as vertex $(k+1)^{\star}$. 
    \item Label the loop as $e_{1^{\star}}$. Label the edge from vertex $k^{\star}$ to vertex $l^{\star} (\prec k^{\star})$ as $e_{k^{\star} }$ for each $k^{\star} \in\{2^{\star}, \cdots, |E(H(\varepsilon)^{(\alpha)})|^{\star}\}$.
\end{enumerate}
Note that this procedure is not unique, but its non-unique nature does not violate the validity of the later discussions.
Note also that $E(\widetilde{H}(\varepsilon)^{(\alpha)})=\{ e_{k^{\star}}| k^{\star}=1^{\star}, \dots, |E({H}(\varepsilon)^{(\alpha)}) |^{\star} \}$ because the edges in $E(H(\varepsilon)^{(\alpha)})$ are inverted in this procedure according to the direction corresponding to the condition $\mathrm{C}_5(j)$, and $\widetilde{H}(\varepsilon)$ is defined as the graph obtained from $H(\varepsilon)$ by changing the edge directions according to the condition $\mathrm{C}_5(j)$. 

\begin{figure*}[htbp]
    \centering
    \includegraphics[width=\linewidth]{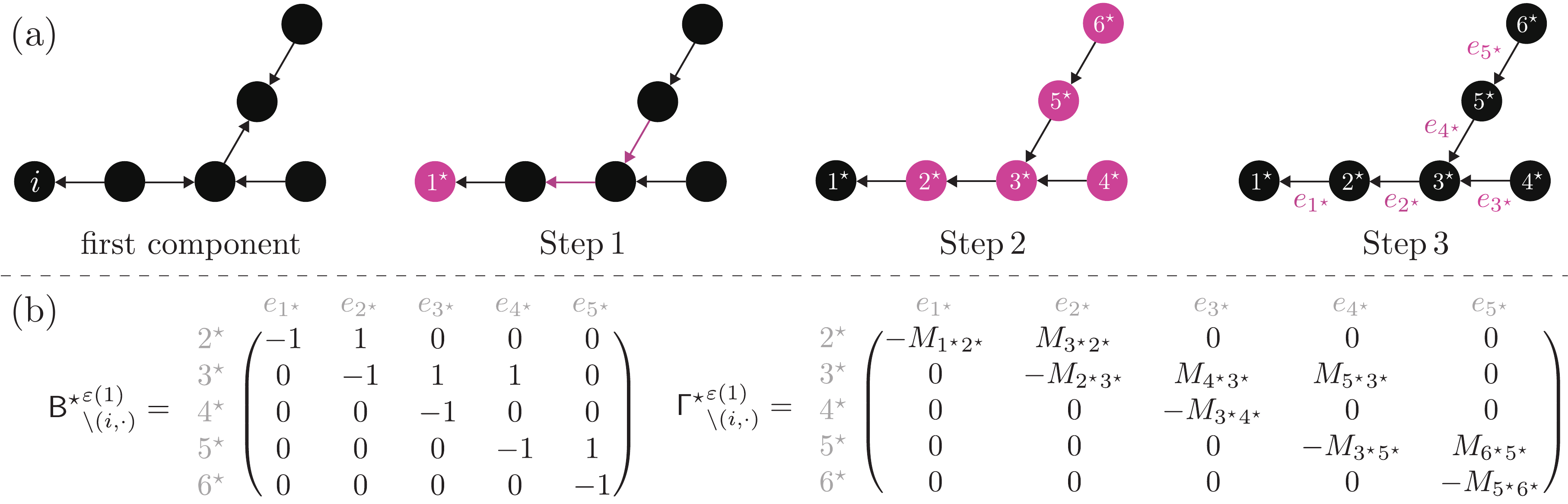}
    \caption{(a) An example of labeling the vertices and edges with $\alpha = 1$. 
    Step 1: Label vertex $i$ as $1^\star$ and change the direction of the edges so that they point toward $1^\star$.
    Step 2: Starting from vertex $1^{\star}$, label the other vertices.
    Step 3: Label the edge from $(k+1)^{\star}$ to $l^{\star} (\prec (k+1)^\star)$ as $e_{k^\star}$.
    (b) Upper triangular matrices ${\BB^\star}^{\varepsilon(1)}_{\backslash(i, \cdot)}$ and ${\GGamma^\star}^{\varepsilon(1)}_{\backslash(i, \cdot)}$ obtained by the procedure in (a).}
    \label{fig:labeling_wo_loop} 
\end{figure*}
This procedure is used to define the upper triangle matrices
${\BB^{\star}}^{\varepsilon(\alpha)}_{\backslash(i, \cdot)}$ and ${\GGamma^{\star}}^{\varepsilon(\alpha)}_{\backslash(j, \cdot)}$, which are the generalized incidence matrix and the generalized weighted incidence matrix corresponding to $\widetilde{H}(\varepsilon)^{(\alpha)}$, respectively. 
The matrices ${\BB^{\star}}^{\varepsilon(\alpha)}_{\backslash(i, \cdot)}$ and ${\GGamma^{\star}}^{\varepsilon(\alpha)}_{\backslash(j, \cdot)}$ for $\alpha \neq 1$ are obtained by applying the transformation into $\BB^{\varepsilon(\alpha)}_{\backslash(i, \cdot)}$ and $\GGamma^{\varepsilon(\alpha)}_{\backslash(j, \cdot)}$ according to the procedure, respectively. Let $\sigma(k)$ be the vertex $k^{\star}$ for indices $k=1, \dots, |E(H(\varepsilon)^{(\alpha)})|$.
We rearrange the rows and columns of $\BB^{\varepsilon(\alpha)}_{\backslash(i, \cdot)}$ and $\GGamma^{\varepsilon(\alpha)}_{\backslash(j, \cdot)}$ so that the $k$-th row corresponds to vertex $\sigma(k)$ and the $l$-th column corresponds to $e_{\sigma(l)}$.
Additionally, we also multiply the columns of $\BB^{\varepsilon(\alpha)}_{\backslash(i, \cdot)}$ and $\GGamma^{\varepsilon(\alpha)}_{\backslash(j, \cdot)}$ corresponding to the edge  $e_{k^{\star}}$ by $-1$ if the edge $e_{k^{\star}}$ is inverted from the corresponding edge in $H(\varepsilon)^{(\alpha)}$ in Step 1. This is because inverting the edge $e \to -e$ gives the factor $-1$ for the elements of $\BB^{\varepsilon(\alpha)}_{\backslash(i, \cdot)}$ and $\GGamma^{\varepsilon(\alpha)}_{\backslash(j, \cdot)}$ in Eqs.~\eqref{eq:def_B} and \eqref{eq:def_Gamma} as $\delta_{i\rmt(-e)}-\delta_{i\rms(-e)} = (-1) (\delta_{i\rmt(e)}-\delta_{i\rms(e)})$ and $M_{\rms(-e)\rmt(-e)}\delta_{i\rmt(-e)}-M_{\rmt(-e)\rms(-e)}\delta_{i\rms(-e)} = (-1)(M_{\rms(e)\rmt(e)}\delta_{i\rmt(e)}-M_{\rmt(e)\rms(e)}\delta_{i\rms(e)})$ for $e\in E_M$.

We show that ${\BB^{\star}}^{\varepsilon(\alpha)}_{\backslash(i, \cdot)}$ and ${\GGamma^{\star}}^{\varepsilon(\alpha)}_{\backslash(j, \cdot)}$ are upper triangular matrices. Here, we define $R_{k^{\star}}$ and $M_{k^{\star} l^{\star}}$ as $R_{k^{\star}}=R_{i}$ and  $M_{k^{\star} l^{\star}}=M_{ij}$ where  $i \in V(G)$ and $j \in V(G)$ are the vertices before relabeling, corresponding to $k^{\star}$ and $l^{\star}$, respectively.
The diagonal elements of ${\BB^{\star}}^{\varepsilon(\alpha)}_{\backslash(i, \cdot)}$ and ${\GGamma^{\star}}^{\varepsilon(\alpha)}_{\backslash(j, \cdot)}$ are given by $({\BB^{\star}}^{\varepsilon(\alpha)}_{\backslash(i, \cdot)})_{11}=1$, $({\BB^{\star}}^{\varepsilon(\alpha)}_{\backslash(i, \cdot)})_{kk}=-1$ for $k\ge2$, $({\GGamma^{\star}}^{\varepsilon(\alpha)}_{\backslash(j, \cdot)})_{11}=-\delR{\sigma(1)}$, and $({\GGamma^{\star}}^{\varepsilon(\alpha)}_{\backslash(j, \cdot)})_{kk}=-M_{\rmt(e_{\sigma(k)})\rms(e_{\sigma(k)})}$ for $k\ge 2$ (see also Fig.~\ref{fig:labeling_loop}(b) for an example). Since the vertex with the loop is labeled $1^{\star}$ and the loop is labeled $e_{1^{\star}}$, the first column of ${\BB^{\star}}^{\varepsilon(\alpha)}_{\backslash(i, \cdot)}$ and ${\GGamma^{\star}}^{\varepsilon(\alpha)}_{\backslash(j, \cdot)}$ have nonzero elements only in the first row, and these nonzero elements are given by $({\BB^{\star}}^{\varepsilon(\alpha)}_{\backslash(i, \cdot)})_{11}=1$ and $({\GGamma^{\star}}^{\varepsilon(\alpha)}_{\backslash(j, \cdot)})_{11}=-\delR{\sigma(1)}$. Since the labeling in Step 3 implies $\rms(e_{k^{\star}})=k^{\star}$ for $k^{\star} \in\{2^{\star}, \cdots, |E(H(\varepsilon)^{(\alpha)})|^{\star}\}$, we obtain $({\BB^{\star}}^{\varepsilon(\alpha)}_{\backslash(i, \cdot)})_{kk}=-1$ and {$({\GGamma^{\star}}^{\varepsilon(\alpha)}_{\backslash(j, \cdot)})_{kk}=-M_{\rmt(e_{\sigma(k)})\rms(e_{\sigma(k)})}$} for $k\ge 2$. Since the vertices are labeled in ascending order, we find that $\rms(e_{k^{\star}})=k^{\star}$ and $\rmt(e_{k^{\star}}) \prec
 k^{\star}$ for $k^{\star} \in\{2^{\star}, \cdots, |E(H(\varepsilon)^{(\alpha)})|^{\star}\}$. Therefore, the $k$-th column of ${\BB^{\star}}^{\varepsilon(\alpha)}_{\backslash(i, \cdot)}$ and ${\GGamma^{\star}}^{\varepsilon(\alpha)}_{\backslash(j, \cdot)}$ can have nonzero elements for the $l$-th row and the $k$-th row, where the $l$-th row corresponds to the row for vertex $\rmt(e_{k^{\star}})$ and thus $l < k$ is satisfied. Thus, ${\BB^{\star}}^{\varepsilon(\alpha)}_{\backslash(i, \cdot)}$ and ${\GGamma^{\star}}^{\varepsilon(\alpha)}_{\backslash(j, \cdot)}$ are upper triangular matrices.

Based on the transformations of $\BB^{\varepsilon(\alpha)}_{\backslash(i, \cdot)}$ and $\GGamma^{\varepsilon(\alpha)}_{\backslash(j, \cdot)}$ into ${\BB^\star}^{\varepsilon(\alpha)}_{\backslash(i, \cdot)}$ and ${\GGamma^\star}^{\varepsilon(\alpha)}_{\backslash(j, \cdot)}$, we provide the expression for $\det(\BB^{\varepsilon(\alpha)}_{\backslash(i, \cdot)})\det(\GGamma^{\varepsilon(\alpha)}_{\backslash(j, \cdot)})$ in terms of the weight of $\widetilde{H}(\varepsilon)^{(\alpha)}$.
Note that this transformation can be achieved by applying the same procedure to both $\BB^{\varepsilon(\alpha)}_{\backslash(i, \cdot)}$ and $\GGamma^{\varepsilon(\alpha)}_{\backslash(j, \cdot)}$, since their rows and columns are arranged in the same way. Because the transformation only changes the signs of the determinants of $\BB^{\varepsilon(\alpha)}_{\backslash(i, \cdot)}$ and $\GGamma^{\varepsilon(\alpha)}_{\backslash(j, \cdot)}$ in the same way, 
the transformation does not change the value of the product of the determinants, i.e., 
$\det(\BB^{\varepsilon(\alpha)}_{\backslash(i, \cdot)})\det(\GGamma^{\varepsilon(\alpha)}_{\backslash(j, \cdot)})=\det({\BB^\star}^{\varepsilon(\alpha)}_{\backslash(i, \cdot)})\det({\GGamma^\star}^{\varepsilon(\alpha)}_{\backslash(j, \cdot)})$.
Moreover, the determinants of ${\BB^\star}^{\varepsilon(\alpha)}_{\backslash(i, \cdot)}$ and ${\GGamma^\star}^{\varepsilon(\alpha)}_{\backslash(j, \cdot)}$ are given by the products of their diagonal elements, since they are upper triangular matrices.
Combining these facts, we obtain 
\begin{align}
&\det(\BB^{\varepsilon(\alpha)}_{\backslash(i, \cdot)})\det(\GGamma^{\varepsilon(\alpha)}_{\backslash(j, \cdot)}) \notag\\
&=-(R_{\sigma(1)}-\meanfit)\prod_{k=2}^{\left|E(H(\varepsilon)^{(\alpha)})\right|} M_{\rmt(e_{\sigma(k)})\rms(e_{\sigma(k)})}.
\end{align}
Note that $-(R_{\sigma(1)}-\meanfit) =-(R_{1^{\star}}-\meanfit)$ is the weight of the loop $e_{1^\star}$, and $M_{\rmt(e_{\sigma(k)})\rms(e_{\sigma(k)})}= M_{\rmt(e_{k^\star})\rms(e_{k^\star})}$ is the weight of edge $e_{k^\star}$ for $k^{\star} \in\{2^{\star}, \cdots, |E(H(\varepsilon)^{(\alpha)})|^{\star}\}$.
Combining these facts with the expression $E(\widetilde{H}(\varepsilon)^{(\alpha)})=\{ e_{k^{\star}}| k^{\star}=1^{\star}, \dots, |E({H}(\varepsilon)^{(\alpha)}) |^{\star} \}$, we find 
\begin{align}\label{eq:weight_with_loop}   
\det(\BB^{\varepsilon(\alpha)}_{\backslash(i, \cdot)})\det(\GGamma^{\varepsilon(\alpha)}_{\backslash(i, \cdot)})=w(\widetilde{H}(\varepsilon)^{(\alpha)}).
\end{align}
\subsubsection{Expression for $\det(\BB^{\varepsilon(1)}_{\backslash(i, \cdot)})\det(\GGamma^{\varepsilon(1)}_{\backslash(i, \cdot)})$}
\label{app:det_B1_det_Gamma1}
For $i=j$, we introduce the procedure to obtain  ${\BB^{\star}}^{\varepsilon(1)}_{\backslash(i, \cdot)}$ and ${\GGamma^{\star}}^{\varepsilon(1)}_{\backslash(j, \cdot)} = {\GGamma^{\star}}^{\varepsilon(1)}_{\backslash(i, \cdot)}$, which are upper triangular matrices corresponding to $\widetilde{H}(\varepsilon)^{(1)}$.
First, we label the vertices and edges of ${H}(\varepsilon)^{(1)}$, and change the edge directions according to the following steps (see also Fig.~\ref{fig:labeling_wo_loop}(a) for an example).
\begin{enumerate}[label=Step\,\arabic*., leftmargin=1.1cm]
    \item Label vertex $i$ as vertex $1^\star$.
    Change the direction of all edges so that they point toward vertex $1^\star$.
    \item Repeat the process introduced in Step 2 in Sec.~\ref{subsubsec:not_contain_j} from $k^{\star}=1^{\star}$ until all the vertices are labeled.
    \item Label the edge from vertex $(k+1)^{\star}$ to $l^{\star} (\prec (k+1)^\star)$ as $e_{k^\star}$ for each $k^\star\in\{1^\star,\cdots, |E(H(\varepsilon)^{(1)})|^\star\}$.
\end{enumerate}
Note that $E(\widetilde{H}(\varepsilon)^{(1)})=\{ e_{k^{\star}}| k^{\star}=1^\star, \dots, |E({H}(\varepsilon)^{(1)}) |^\star \}$ because the edges in $E(H(\varepsilon)^{(1)})$ are inverted in this procedure according to the direction corresponding to the condition $\mathrm{C}_5(j)=\mathrm{C}_5(i)$.
Then, we rearrange the rows and columns of $\BB^{\varepsilon(1)}_{\backslash(i, \cdot)}$ and $\GGamma^{\varepsilon(1)}_{\backslash(i, \cdot)}$ so that the $k$-th row corresponds to vertex $(k+1)^\star$, and $l$-th column corresponds to edge $e_{l^\star}$. Additionally, we also multiply the columns of $\BB^{\varepsilon(1)}_{\backslash(i, \cdot)}$ and $\GGamma^{\varepsilon(1)}_{\backslash(i, \cdot)}$ corresponding to the edge $e_{k^{\star}}$ by $-1$ if the edge $e_{k^{\star}}$ is inverted from the corresponding edge in $H(\varepsilon)^{(1)}$ in Step 1.
Through this process, we obtain ${\BB^\star}^{\varepsilon(1)}_{\backslash(i, \cdot)}$ and ${\GGamma^\star}^{\varepsilon(1)}_{\backslash(i, \cdot)}$ from $\BB^{\varepsilon(1)}_{\backslash(i, \cdot)}$ and $\GGamma^{\varepsilon(1)}_{\backslash(i, \cdot)}$, respectively. For the same reason as in the case of $\alpha\ne 1$, ${\BB^\star}^{\varepsilon(1)}_{\backslash(i, \cdot)}$ and ${\GGamma^\star}^{\varepsilon(1)}_{\backslash(i, \cdot)}$ are upper triangular matrices, and their diagonal elements are given by $({\BB^\star}^{\varepsilon(1)}_{\backslash(i, \cdot)})_{kk}=-1$ and $({\GGamma^\star}^{\varepsilon(1)}_{\backslash(i, \cdot)})_{kk}=-M_{\rmt(e_{\sigma(k)})\rms(e_{\sigma(k)})}$ (see also Fig.~\ref{fig:labeling_wo_loop}(b) for an example).

We then express $\det(\BB^{\varepsilon(1)}_{\backslash(i, \cdot)})\det(\GGamma^{\varepsilon(1)}_{\backslash(i, \cdot)})$ in terms of the weight of $\widetilde{H}(\varepsilon)^{(1)}$.
Note that this transformation can be achieved by applying the same procedure to both $\BB^{\varepsilon(1)}_{\backslash(i, \cdot)}$ and $\GGamma^{\varepsilon(1)}_{\backslash(i, \cdot)}$. For the same reason as in the case of $\alpha\ne 1$,
we obtain $\det(\BB^{\varepsilon(1)}_{\backslash(i, \cdot)})\det(\GGamma^{\varepsilon(1)}_{\backslash(i, \cdot)})=\det({\BB^\star}^{\varepsilon(1)}_{\backslash(i, \cdot)})\det({\GGamma^\star}^{\varepsilon(1)}_{\backslash(i, \cdot)})$.
Furthermore, since the matrices ${\BB^\star}^{\varepsilon(1)}_{\backslash(i, \cdot)}$ and ${\GGamma^\star}^{\varepsilon(1)}_{\backslash(i, \cdot)}$ are upper triangular, their determinants are equal to the product of their diagonal elements.
Combining these facts with the expression $E(\widetilde{H}(\varepsilon)^{(1)})=\{ e_{k^{\star}}| k^{\star}=1^{\star}, \dots, |E({H}(\varepsilon)^{(1)}) |^{\star} \}$, we obtain
\begin{align}\label{eq:weight_without_loop_i=j}   
\det(\BB^{\varepsilon(1)}_{\backslash(i, \cdot)})\det(\GGamma^{\varepsilon(1)}_{\backslash(i, \cdot)})&=\prod_{k=1}^{\left|E(H(\varepsilon)^{(1)})\right|} M_{\rmt(e_{\sigma(k)})\rms(e_{\sigma(k)})}\notag\\
&=w(\widetilde{H}(\varepsilon)^{(1)}).
\end{align}

\begin{figure*}[htbp]
    \centering
    \includegraphics[width=\linewidth]{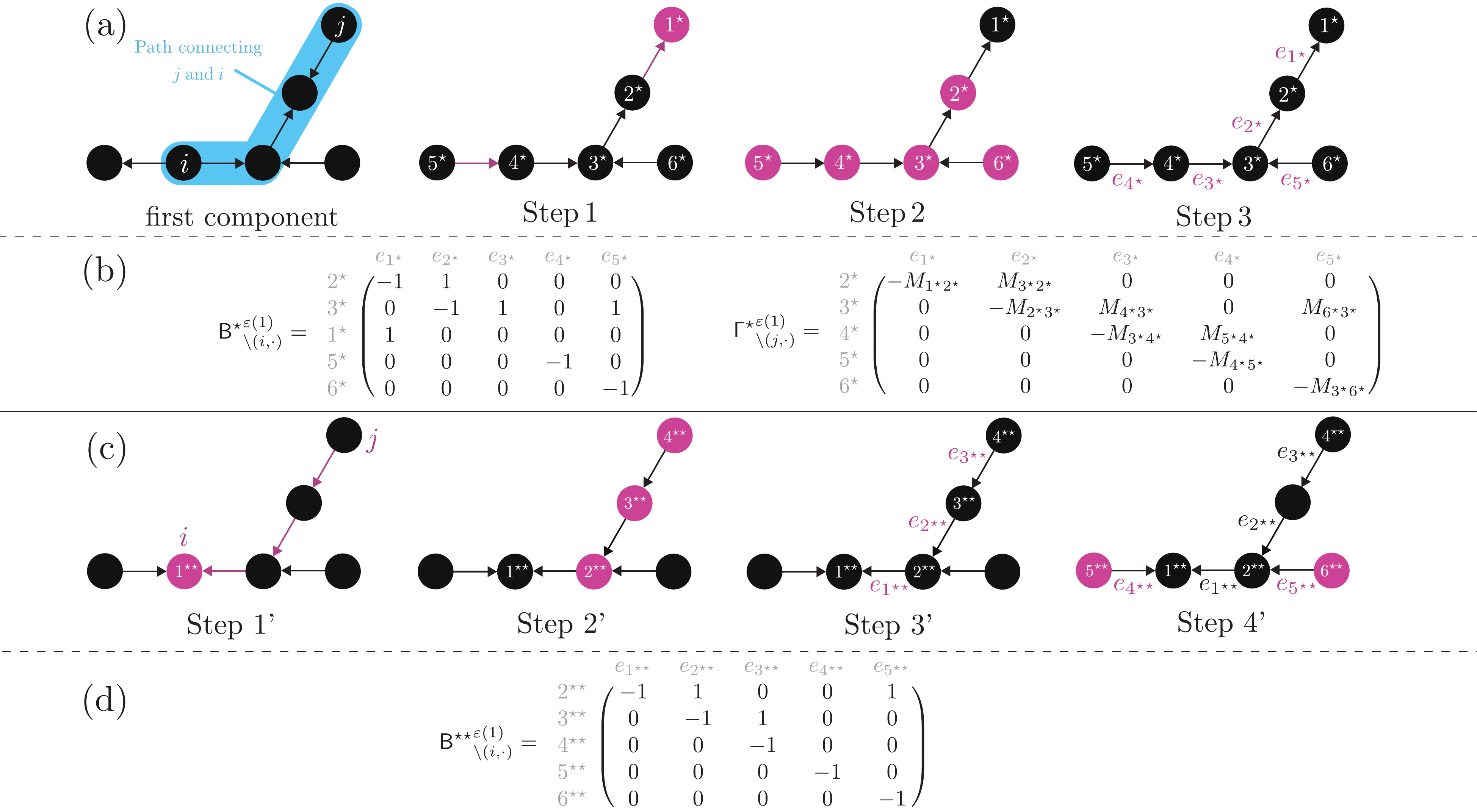}
    \caption{(a) An example of labeling the vertices and edges with $\alpha = 1$ and $j \neq i$.
    Step 1: Label vertex $j$ as $1^\star$ and change the direction of the edges so that they point toward $1^\star$.
    Step 2: Starting from vertex $1^{\star}$, label the other vertices.
    Step 3: Label the edge from $(k+1)^\star$ to $m^{\star} (\prec (k+1)^\star)$ as $e_{k^\star}$.
    (b) The procedure in (a) yields an upper triangular matrix ${\GGamma^\star}^{\varepsilon(1)}_{\backslash(i, \cdot)}$, while the yealding matrix ${\BB^\star}^{\varepsilon(1)}_{\backslash(i, \cdot)}$ is not an upper triangular matrix. Here, the number of edges in the path from vertex $i$ to vertex $j$ is $L=3$. (c) An example of labeling the vertices and edges.
    Step 1': Label vertex $i$ as $1^{\star \star}$ and change the direction of the edges so that they point toward $1^{\star \star}$.
    Step 2': Starting from vertex $1^{\star\star}$, label the vertices in the path from $j$ to $i$.
    Step 3': Label the edge from $(k+1)^{\star\star}$ to $m^{\star\star} ( \prec (k+1)^{\star\star})$ as $e_{k^{\star \star}}$. Step 4': Label the remaining vertices and edges according to the procedure in (a). (d) The procedure in (c) yields an upper triangular matrix ${\BB^{\star\star} }^{\varepsilon(1)}_{\backslash(i, \cdot)}$
    }
    \label{fig:relabeling_jnei} 
\end{figure*}

\subsubsection{Expression for $\det(\BB^{\varepsilon(1)}_{\backslash(i, \cdot)})\det(\GGamma^{\varepsilon(1)}_{\backslash(j, \cdot)})$ with $j\ne i$}
For $j\ne i$, we first introduce the procedure for transforming $\GGamma^{\varepsilon(1)}_{\backslash(j, \cdot)}$ into an upper triangular matrix ${\GGamma^{\star}}^{\varepsilon(1)}_{\backslash(j, \cdot)}$. We label the vertices and edges of $H(\varepsilon)^{(1)}$ and change the edge directions according to the following steps (see also Fig.~\ref{fig:relabeling_jnei}(a) for an example)
\begin{enumerate}[label=Step\,\arabic*., leftmargin=1.1cm]
    \item Label vertex $j$ as vertex $1^\star$.
    Change the direction of all the edges so that they point toward vertex $1^\star$.
    \item Find a path from vertex $i$ to vertex $j$. Following the path, label the vertices in ascending order $1^{\star}, 2^{\star}, \dots, (L+1)^{\star}$, where vertex $i$ is labeled as vertex $(L+1)^{\star}$.
    Repeat the process introduced in Step 2 in Sec.~\ref{subsubsec:not_contain_j} from $k^{\star}=(L+1)^{\star}$ until all the vertices are labeled. 
    Because of $\mathrm{C}_6(i, j)$, these processes are possible.
    \item Label the edge from vertex $(k+1)^{\star}$ to $m^{\star} (\prec (k+1)^\star)$ as $e_{k^\star}$ for each $k^{\star}\in\{1^{\star},\cdots, |E(H(\varepsilon)^{(1)})|^{\star}\}$.
\end{enumerate}
Note that $E(\widetilde{H}(\varepsilon)^{(1)})=\{ e_{k^{\star}}| k^{\star}=1^{\star}, \dots, |E({H}(\varepsilon)^{(1)}) |^{\star} \}$ because the edges in $E(H(\varepsilon)^{(1)})$ are inverted in this procedure according to the direction corresponding to the condition $\mathrm{C}_5(j)$.

The matrices ${\BB^{\star}}^{\varepsilon(1)}_{\backslash(i, \cdot)}$ and ${\GGamma^{\star}}^{\varepsilon(1)}_{\backslash(j, \cdot)}$ for $j \neq i$ are obtained by applying the transformations of $\BB^{\varepsilon(1)}_{\backslash(i, \cdot)}$ and $\GGamma^{\varepsilon(1)}_{\backslash(j, \cdot)}$ according to the procedure, respectively. We rearrange the rows and columns of $\BB^{\varepsilon(1)}_{\backslash(i, \cdot)}$ and $\GGamma^{\varepsilon(1)}_{\backslash(j, \cdot)}$ so that the $k$-th row corresponds to vertex $(k+1)^{\star}$ and the $l$-th column corresponds to $e_{l^{\star}}$.
Additionally, we also multiply the columns of $\BB^{\varepsilon(1)}_{\backslash(i, \cdot)}$ and $\GGamma^{\varepsilon(1)}_{\backslash(j, \cdot)}$ corresponding to the edge  $e_{k^{\star}}$ by $-1$ if the edge $e_{k^{\star}}$ is inverted from the corresponding edge in $H(\varepsilon)^{(1)}$ in Step 1. As a result, ${\GGamma^\star}^{\varepsilon(1)}_{\backslash(j, \cdot)}$ becomes an upper triangular matrix whose diagonal elements are given by $({\GGamma^\star}^{\varepsilon(1)}_{\backslash(j, \cdot)})_{kk}=-M_{\rmt(e_{\sigma(k)})\rms(e_{\sigma(k)})}$ (see also Fig.~\ref{fig:relabeling_jnei}(b) for an example). 
This fact can be shown in the same way as in Sec.~\ref{subsubsec:not_contain_j}.

Here, ${\BB^\star}^{\varepsilon(1)}_{\backslash(i, \cdot)}$ is not an upper triangular matrix in general (see also Fig.~\ref{fig:relabeling_jnei}(b) for an example). For $k\ge L+1$ and $l<k$, $({\BB^\star}^{\varepsilon(1)}_{\backslash(i, \cdot)})_{kk}=-1$ and $({\BB^\star}^{\varepsilon(1)}_{\backslash(i, \cdot)})_{kl}=0$ hold where $L$ is the number of edges in the path from vertex $i$ to vertex $j$ (see also Fig.~\ref{fig:relabeling_jnei}(b) for an example).
These facts can be shown as follows.
The rows and columns in $\BB^{\varepsilon(1)}_{\backslash(i, \cdot)}$ and $\GGamma^{\varepsilon(1)}_{\backslash(j, \cdot)}$ are arranged in the same way, except that the row for vertex $j$ in $\BB^{\varepsilon(1)}_{\backslash(i, \cdot)}$ is in the same position as the row for vertex $i$ in $\GGamma^{\varepsilon(1)}_{\backslash(j, \cdot)}$. 
The row corresponding to vertex $i$ is in the $L$-th row in ${\GGamma^\star}^{\varepsilon(1)}_{\backslash(j, \cdot)}$ because in Step 2, we label the vertices on the path from vertex $j$ to vertex $i$ as $1^\star, \cdots, (L+1)^\star$ and the row corresponding to vertex $j$ is missing in ${\GGamma^\star}^{\varepsilon(1)}_{\backslash(j, \cdot)}$.
Therefore, the $L$-th row of ${\BB^\star}^{\varepsilon(1)}_{\backslash(i, \cdot)}$ corresponds to vertex $j$. Since the target vertex of $e_{1^\star}$ is vertex $j$, we find $({\BB^\star}^{\varepsilon(1)}_{\backslash(i, \cdot)})_{L 1}=1$, which means that ${\BB^\star}^{\varepsilon(1)}_{\backslash(i, \cdot)}$ is not an upper triangular matrix if $L\geq 2$.
Note that the rows and columns from the $(L+1)$-th to the $|E(H(\varepsilon)^{(1)})|$-th in ${\BB^\star}^{\varepsilon(1)}_{\backslash(i, \cdot)}$ are arranged in the same order as in ${\GGamma^\star}^{\varepsilon(1)}_{\backslash(j, \cdot)}$.
Thus, as in the case of $j=i$ in Sec.~\ref{app:det_B1_det_Gamma1}, we have $({\BB^\star}^{\varepsilon(1)}_{\backslash(i, \cdot)})_{kk}=-1$ and $({\BB^\star}^{\varepsilon(1)}_{\backslash(i, \cdot)})_{kl}=0$ for $k\ge L+1$ and $l<k$.

We consider the transformation of ${\BB^\star}^{\varepsilon(1)}_{\backslash(i, \cdot)}$ into an upper triangular matrix. We relabel the vertices and edges of $\widetilde{H}(\varepsilon)^{(1)}$. First, we relabel the vertices and edges of ${H}(\varepsilon)^{(1)}$. Here, we use the notation $^{\star \star}$ for the relabeled vertices $V(H(\varepsilon)^{(\alpha)}) = \{ 1^{\star \star}, \dots, |E(H(\varepsilon)^{(\alpha)})|^{\star \star} \} $. We also introduce the total order $\prec$ (or $\succ$), such that $1^{\star \star}\prec 2^{\star \star} \prec\dots \prec |E(H(\varepsilon)^{(\alpha)}|^{\star \star}$ (or $|E(H(\varepsilon)^{(\alpha)}|^{\star \star} \succ \dots \succ 2^{\star \star} \succ 1^{\star \star} $). We relabel the vertices and edges, and change the edge directions according to the following steps (see also Fig.~\ref{fig:relabeling_jnei}(c) for an example),
\begin{enumerate}[leftmargin=1.1cm]
    \item[Step\, 1'.] Label vertex $i$ as vertex $1^{\star\star}$.
    Change the direction of all edges so that they point toward vertex $1^{\star\star}$.
    \item[Step\, 2'.] Find a path from vertex $j$ to vertex $i$. Following the path, label the vertices in ascending order $1^{\star\star}, 2^{\star\star}, \dots, (L+1)^{\star\star}$, where vertex $j$ is labeled as vertex $(L+1)^{\star\star}$.
    \item[Step\, 3'.] Label the edge from vertex $(k+1)^{\star \star}$ to $m^{\star \star} (\prec (k+1)^{\star \star})$ as $e_{k^{\star\star}}$ for each $k^{\star \star}\in\{1^{\star \star},\cdots, L^{\star \star}\}$.
    \item[Step\, 4'.] Label the remaining vertex $k^\star$ as $k^{\star\star}$ and the remaining edge $e_{k^\star}$ as $e_{k^{\star\star}}$, respectively, for each $k^{\star}\in\{(L+1)^{\star},\cdots, |E(H(\varepsilon)^{(1)})|^{\star}\}$.
\end{enumerate}
We rearrange the rows and columns of ${\BB^\star}^{\varepsilon(1)}_{\backslash(i, \cdot)}$ so that the $k$-th row corresponds to vertex $k^{\star\star}$ and the $l$-th row corresponds to $e_{l^{\star\star}}$.
Additionally, we also multiply the columns of ${\BB^\star}^{\varepsilon(1)}_{\backslash(i, \cdot)}$ corresponding to the edge  $e_{k^{\star\star}}$ by $-1$ if the edge $e_{k^{\star\star}}$ is inverted from the corresponding edge in $\widetilde{H}(\varepsilon)^{(1)}$ in Step 1'. Let ${\BB^{\star\star}}^{\varepsilon(1)}_{\backslash(i, \cdot)}$ denote the resulting matrix.

We show that ${\BB^{\star\star}}^{\varepsilon(1)}_{\backslash(i, \cdot)}$ is an upper triangular matrix whose diagonal elements are given by $({\BB^{\star\star}}^{\varepsilon(1)}_{\backslash(i, \cdot)})_{kk}=-1$  (see also Fig.~\ref{fig:relabeling_jnei}(d) for an example).
As implied in step 4', this transformation does not affect the $(L+1)$-th through the $|E(H(\varepsilon)^{(1)})|$-th rows and columns.
Therefore, ${\BB^{\star\star}}^{\varepsilon(1)}_{\backslash(i, \cdot)}$ satisfies $({\BB^{\star\star}}^{\varepsilon(1)}_{\backslash(i, \cdot)})_{kk}=-1$ and $({\BB^{\star\star}}^{\varepsilon(1)}_{\backslash(i, \cdot)})_{kl}=0$ for $k\ge L+1$ and $l<k$, as ${\BB^\star}^{\varepsilon(1)}_{\backslash(i, \cdot)}$ {does}.
Furthermore, we show that $({\BB^{\star\star}}^{\varepsilon(1)}_{\backslash(i, \cdot)})_{kk}=-1$ and $({\BB^{\star\star}}^{\varepsilon(1)}_{\backslash(i, \cdot)})_{kl}=0$ for $k\in\{1,\cdots, L\}$ and $l<k$ as follows.
Since the source vertex of $e_{k^{\star\star}}$ is vertex $(k+1)^{\star\star}$ corresponding to the $k$-th row for $k\in\{1,\cdots, L\}$, we obtain $({\BB^{\star\star}}^{\varepsilon(1)}_{\backslash(i, \cdot)})_{kk}=-1$.
Since $e_{1^{\star\star}},\cdots, e_{L^{\star\star}}$ forms the path from vertex $(L+1)^{\star\star}$ to vertex $1^{\star\star}$, $e_{l^{\star\star}}=(l^{\star\star}\leftarrow (l+1)^{\star\star})$ holds for all $l^{\star \star}\in\{1^{\star \star},\cdots, L^{\star \star}\}$.
Therefore, $({\BB^{\star\star}}^{\varepsilon(1)}_{\backslash(i, \cdot)})_{kl}=0$ also holds for $k\in\{1,\cdots, L\}$ and $l<k$.

We show that the transformation of ${\BB^\star}^{\varepsilon(1)}_{\backslash(i, \cdot)}$ into ${\BB^{\star\star}}^{\varepsilon(1)}_{\backslash(i, \cdot)}$  reverses the sign of $\det({\BB^\star}^{\varepsilon(1)}_{\backslash(i, \cdot)})$, i.e., 
\begin{align}\label{eq:det*_det**}
    \det({\BB^\star}^{\varepsilon(1)}_{\backslash(i, \cdot)})=-\det({\BB^{\star\star}}^{\varepsilon(1)}_{\backslash(i, \cdot)}).
\end{align}
In Step 1', we invert $L$ edges, $e_{1^\star}, \cdots, e_{L^\star}$.
This procedure corresponds to multiplying the first $L$ columns in ${\BB^\star}^{\varepsilon(1)}_{\backslash(i, \cdot)}$ by $-1$.
In Step 2', we label the vertices $2^\star,\cdots,L^\star$ as $L^{\star\star},\cdots,2^{\star\star}$, respectively.
This procedure rearranges the first to the $(L-1)$-th rows in ${\BB^\star}^{\varepsilon(1)}_{\backslash(i, \cdot)}$ in reverse order, which can be achieved by swapping two rows $(L-2)/2$ times (or $(L-1)/2$ times) if $L$ is even (or odd).
In Step 3', we relabel the edges $e_{1^\star},\cdots,e_{L^\star}$ as $e_{L^{\star\star}},\cdots,e_{1^{\star\star}}$.
This procedure rearranges the first to the $L$-th columns in ${\BB^\star}^{\varepsilon(1)}_{\backslash(i, \cdot)}$ in reverse order, which can be achieved by swapping two columns $L/2$ times (or $(L-1)/2$ times) if $L$ is even (or odd).
Overall, the transformation of ${\BB^\star}^{\varepsilon(1)}_{\backslash(i, \cdot)}$ into ${\BB^{\star\star}}^{\varepsilon(1)}_{\backslash(i, \cdot)}$ reverses the sign of $\det({\BB^\star}^{\varepsilon(1)}_{\backslash(i, \cdot)})$ exactly $2L-1$ times, regardless of whether $L$ is odd or even.
Therefore, this procedure multiplies $\det({\BB^\star}^{\varepsilon(1)}_{\backslash(i, \cdot)})$ by $(-1)^{2L-1}=-1$, as shown in Eq.~\eqref{eq:det*_det**}.

We now express $\det(\BB^{\varepsilon(1)}_{\backslash(i, \cdot)})\det(\GGamma^{\varepsilon(1)}_{\backslash(j, \cdot)})$ in terms of the weight of $\widetilde{H}(\varepsilon)^{(1)}$.
Note that $\det(\BB^{\varepsilon(1)}_{\backslash(i, \cdot)})\det(\GGamma^{\varepsilon(1)}_{\backslash(j, \cdot)})=\det({\BB^\star}^{\varepsilon(1)}_{\backslash(i, \cdot)})\det({\GGamma^\star}^{\varepsilon(1)}_{\backslash(j, \cdot)})$ holds because the same procedure is applied to $\BB^{\varepsilon(1)}_{\backslash(i, \cdot)}$ and $\GGamma^{\varepsilon(1)}_{\backslash(j, \cdot)}$.
Combining this equality with Eq.~\eqref{eq:det*_det**}, we obtain
\begin{align}\label{eq:weight_without_loop_inej} 
\det(\BB^{\varepsilon(1)}_{\backslash(i, \cdot)})\det(\GGamma^{\varepsilon(1)}_{\backslash(j, \cdot)})&= - \det({\BB^{\star\star}}^{\varepsilon(1)}_{\backslash(i, \cdot)})\det({\GGamma^{\star}}^{\varepsilon(1)}_{\backslash(j, \cdot)}) \notag\\
&=-\prod_{k=1}^{\left|E(H(\varepsilon)^{(1)})\right|} M_{\rmt(e_{\sigma(k)})\rms(e_{\sigma(k)})}\notag\\
&=-w(\widetilde{H}(\varepsilon)^{(1)}),
\end{align}
where we used the fact that ${\BB^{\star\star}}^{\varepsilon(1)}_{\backslash(i, \cdot)}$ and ${\GGamma^\star}^{\varepsilon(1)}_{\backslash(j, \cdot)}$ are upper triangular matrices.
The third equality follows from $E(\widetilde{H}(\varepsilon)^{(1)})=\{ e_{k^{\star}}| k^{\star}=1^\star, \dots, |E({H}(\varepsilon)^{(1)})|^\star \}$.

\subsubsection{Expression for $\det(\BB^\varepsilon_{\backslash(i, \cdot)})\det(\GGamma^\varepsilon_{\backslash(j, \cdot)})$ in terms of the weight of $\widetilde{H}(\varepsilon)$}
We now express $\det(\BB^\varepsilon_{\backslash(i, \cdot)})\det(\GGamma^\varepsilon_{\backslash(j, \cdot)})$ for $H(\varepsilon)$ which satisfies $\mathrm{C}_1$, $\mathrm{C}_3(j)$, $\mathrm{C}_4(j)$, $\mathrm{C}_6(i, j)$, and $\mathrm{C}_7$ in terms of the weight of $\widetilde{H}(\varepsilon)$. Here, $\widetilde{H}(\varepsilon)$ is expressed as $\widetilde{H}(\varepsilon)=\bigcup_{\alpha=1}^{N_\mathrm{c}}\widetilde{H}(\varepsilon)^{(\alpha)}$, where $N_\mathrm{c}$ is the number of components in $\widetilde{H}(\varepsilon)$. Combining the result in Eq.~\eqref{eq:weight_without_loop_i=j} for $i=j$ with the result in Eq.~\eqref{eq:weight_without_loop_inej} for $i \neq j$, we obtain an expression 
\begin{align}\label{eq:weight_without_loop} 
\det(\BB^{\varepsilon(1)}_{\backslash(i, \cdot)})\det(\GGamma^{\varepsilon(1)}_{\backslash(j, \cdot)})=(2\delta_{ij}-1)w(\widetilde{H}(\varepsilon)^{(1)}).
\end{align}
By combining Eqs.~\eqref{eq:det_BGamma_decomp} and~\eqref{eq:weight_with_loop} with \eqref{eq:weight_without_loop}, we obtain
\begin{align}\label{eq:detB_detGamma}
    &\det(\BB^\varepsilon_{\backslash(i, \cdot)})\det(\GGamma^\varepsilon_{\backslash(j, \cdot)})\notag\\
    &\quad=(\delta_{ij}+(1-\delta_{ij})(-1)^{|i-j|})w(\widetilde{H}(\varepsilon)),
\end{align}
where we used $w(\widetilde{H}(\varepsilon))=\prod_{\alpha=1}^{N_\mathrm{c}}w(\widetilde{H}(\varepsilon)^{(\alpha)})$, which follows from $\widetilde{H}(\varepsilon)=\bigcup_{\alpha=1}^{N_\mathrm{c}}\widetilde{H}(\varepsilon)^{(\alpha)}$.
\subsection{Expression for $\zeta_i\pi_j$ in terms of the weights of rooted $0$/$1$ loop forests}
\label{app:zeta_pi_explicit}

We now derive Eq.~\eqref{eq:Lem1_ij}, which is an expression for $\zeta_i\pi_j$ in terms of the weights of rooted $0$/$1$ loop forests. To derive Eq.~\eqref{eq:Lem1_ij}, we express $\det(\LL_{\backslash(i, j)})$ in Eq.~\eqref{eq:rewrite_zeta_pi} in terms of the weights of rooted $0$/$1$ loop forests.

We express $\det(\LL_{\backslash(i, j)})$ in terms of the weights of graphs. 
By substituting Eq.~\eqref{eq:detB_detGamma} into Eq.~\eqref{eq:CB_formula_restricted}, we obtain
\begin{align}\label{eq:det_L_ii_F_tilde}
    \det(\LL_{\backslash(i, j)})=(\delta_{ij}+(1-\delta_{ij})(-1)^{|i-j|})\sum_{H\in\widetilde{\mathcal{F}}_{j\leftarrow i}(G)}w(H),
\end{align}
where $\widetilde{\mathcal{F}}_{j\leftarrow i}(G)$ is the set of all $\widetilde{H}(\varepsilon)$ satisfying $\mathrm{C}_1$, $\mathrm{C}_3(j)$, $\mathrm{C}_4(j)$, $\mathrm{C}_5(j)$, $\mathrm{C}_6(i, j)$, and $\mathrm{C}_7$.
We can restrict the summation in Eq.~\eqref{eq:det_L_ii_F_tilde} to graphs $H (\in \widetilde{\mathcal{F}}_{j\leftarrow i}(G))$ such that $\tilde{w}(e)\ne 0$ holds for every edge $e\in E(H)$ that is not a loop.
This is because the weight $w(H)$ also vanishes, and thus such $H$ does not contribute to $\det(\LL_{\backslash(i, j)})$ if $H\in\widetilde{\mathcal{F}}_{j\leftarrow i}(G)$ contains an edge $e$ that is not a loop and whose weight vanishes.
Therefore, we rewrite Eq.~\eqref{eq:det_L_ii_F_tilde} as 
\begin{align}\label{eq:det_L_ji_F^+}
    \det(\LL_{\backslash(i, j)})=(\delta_{ij}+(1-\delta_{ij})(-1)^{|i-j|})\sum_{H\in\widetilde{\mathcal{F}}_{j\leftarrow i}^{\,+}(G)}w(H),
\end{align}
where $\widetilde{\mathcal{F}}_{j\leftarrow i}^{\,+}(G)$ is the set of all graphs $H\in\widetilde{\mathcal{F}}_{j\leftarrow i}(G)$ such that $\tilde{w}(e)\ne 0$ holds for every edge $e\in E(H)$ that is not a loop.

We prove the identity $\widetilde{\mathcal{F}}_{j\leftarrow i}^{\,+}(G)=\mathcal{F}_{j\leftarrow i}(G)$, which allows us to replace $\widetilde{\mathcal{F}}_{j\leftarrow i}^{\,+}(G)$ by $\mathcal{F}_{j\leftarrow i}(G)$ in Eq.~\eqref{eq:det_L_ji_F^+}.
Suppose that $H$ belongs to $\widetilde{\mathcal{F}}_{j\leftarrow i}^{\,+}(G)$.
Here, $H (\in \widetilde{\mathcal{F}}_{j\leftarrow i}^{\,+}(G))$ satisfies $\mathrm{C}_1$, $\mathrm{C}_3(j)$, $\mathrm{C}_4(j)$, $\mathrm{C}_5(j)$, $\mathrm{C}_6(i, j)$, and $\mathrm{C}_7$.
Since every edge in $H$ that is not a loop has non-zero weight, it belongs to $E(G)$ by definition Eq.~\eqref{eq:def_E}.
Furthermore, every loop in $H$ belongs to $E(G)$, and every vertex of $H$ belongs to $V(G)$.
Therefore, we have $H\subseteq G$.
Combining these facts, we find that $H$ satisfies all the conditions in Eq.~\eqref{eq:Fji_equivalence}, which means $H\in\mathcal{F}_{j\leftarrow i}(G)$.
Next, suppose that $H$ belongs to $\mathcal{F}_{j\leftarrow i}(G)$.
We consider that the edge directions of $H$ are changed such that every non-loop edge $e=(k\leftarrow l)$ satisfies {$k<l$}.
The resulting graph is $H(\varepsilon)$ for a set of edges $\varepsilon\in E_R\cup E_M$ that satisfies $\mathrm{C}_1$, $\mathrm{C}_3(j)$, $\mathrm{C}_4(j)$, $\mathrm{C}_6(i, j)$, and $\mathrm{C}_7$.
We consider that the edge directions of $H(\varepsilon)$  are changed according to $\mathrm{C}_5(j)$.
The resulting graph is $\widetilde{H}(\varepsilon)$ corresponding to $H(\varepsilon)$.
Note that this graph $\widetilde{H}(\varepsilon)$ is nothing but the original graph $H \in \mathcal{F}_{j\leftarrow i}(G)$ because the edge directions are recovered by changing the edge directions according to $\mathrm{C}_5(j)$.
Thus, we find that $H = \widetilde{H}(\varepsilon)\in\widetilde{\mathcal{F}}_{j\leftarrow i}(G)$.
Moreover, since every edge in $H$ that is not a loop has nonzero weight, we obtain $H\in\widetilde{\mathcal{F}}_{j\leftarrow i}^{\,+}(G)$. Therefore, $\mathcal{F}_{j\leftarrow i}(G) \subseteq \widetilde{\mathcal{F}}_{j\leftarrow i}^{\,+}(G)$ and $ \widetilde{\mathcal{F}}_{j\leftarrow i}^{\,+}(G) \subseteq  \mathcal{F}_{j\leftarrow i}(G)$ are proved, and thus $\widetilde{\mathcal{F}}_{j\leftarrow i}^{\,+}(G)=\mathcal{F}_{j\leftarrow i}(G)$.

Finally, we derive Eq.~\eqref{eq:Lem1_ij}.
Since $\widetilde{\mathcal{F}}_{j\leftarrow i}^{\,+}(G)=\mathcal{F}_{j\leftarrow i}(G)$ holds as shown above, we can rewrite Eq.~\eqref{eq:det_L_ji_F^+} as
\begin{align}\label{eq:det_L_ii_F_i}
    \det(\LL_{\backslash(i, j)})=(\delta_{ij}+(1-\delta_{ij})(-1)^{|i-j|})\sum_{H\in\mathcal{F}_{j \leftarrow i}(G)}w(H).
\end{align}
Substituting this equality into Eq.~\eqref{eq:rewrite_zeta_pi}, we obtain
\begin{align}
\zeta_i\pi_j=\frac{1}{Z}\sum_{H\in\mathcal{F}_{j\leftarrow i}(G)}w(H),
\end{align}
with $Z:=\sum_{i\in V(G)}\sum_{H\in\mathcal{F}_i(G)}w(H)$, where we used $(\delta_{ij}+(1-\delta_{ij})(-1)^{|i-j|}) (-1)^{i+j} =1$ and $Z=\sum_k \sum_{H\in\mathcal{F}_{k \leftarrow k}(G)}w(H)$.
This equation is Eq.~\eqref{eq:Lem1_ij}, that is Lemma~\ref{lem:Lem1}.

\section{Maximum number of rooted $0$/$1$ loop forests}
\label{app:max_number}
We determine the maximum number of rooted $0$/$1$ loop forests for a given basic graph.
The number of graphs in $\mathcal{F}_i(G)$ is bounded by
\begin{align}\label{eq:upper_bound}
    |\mathcal{F}_i(G)|\le 2(N+1)^{N-2},
\end{align}
with $N=|V(G)|$.
Since $\mathcal{F}_{i\leftarrow j}(G)$ and $\mathcal{F}_{i\not\leftarrow j}(G)$ are subsets of $\mathcal{F}_i(G)$, the maximum number of them is also bounded by $2(N+1)^{N-2}$.
Note that the number of rooted spanning trees with $N$ vertices is bounded by $N^{N-2}$, i.e., $|\mathcal{T}_i(G)|\le N^{N-2}$ holds with $N=|V(G)|$, which is known as Cayley's formula~\cite{west2001introduction}.

The inequality~\eqref{eq:upper_bound} is proved as follows.
To derive the upper bound on $|\mathcal{F}_i(G)|$, we consider the \textit{complete graph} with $N$ vertices $G^{\mathrm{comp}}_N$, which is defined as a basic graph where each vertex has a loop and there are edges in both directions between each pair of vertices.
The set of $0$/$1$ loop forests  $\mathcal{F}_i(G^{\mathrm{comp}}_N)$ satisfies $\mathcal{F}_i(G)\subseteq\mathcal{F}_i(G^{\mathrm{comp}}_N)$ for any basic graph $G$ with $N$ vertices.
Therefore, we obtain
\begin{align}  |\mathcal{F}_i(G)|&\le|\mathcal{F}_i(G^{\mathrm{comp}}_N)|=\sum_{H\in\mathcal{F}_i(G^{\mathrm{comp}}_N)}1.
\end{align}
If we compare the right-hand side of this inequality with the right-hand side of Eq.~\eqref{eq:det_L_ii_F_i} when $i=j$, we notice that $\sum_{H\in\mathcal{F}_i(G^{\mathrm{comp}}_N)}1$ can be obtained by replacing every $R_k-\meanfit$ with $-1$ and every $M_{kl}$ with $1$ for $k\ne l$, because this replacement turns every $w(H)$ in eq.~\eqref{eq:det_L_ii_F_i} to $1$.
Because $M_{kk} = -\sum_{l (\neq k)}M_{lk}$ is replaced with $-(N-1)$, applying this replacement transforms $\LL=-\left(\RR-\meanfit\mathsf{I}+\MM\right)$ into $\tilde{\mathsf{L}}\in\mathbb{R}^{N\times N}$, which is defined as $\tilde{L}_{kl}:=N$ if $k=l$, and $\tilde{L}_{kl}:=-1$ if $k\ne l$.
Then, we obtain
\begin{align*}    |\mathcal{F}_i(G)|&\le\det(\tilde{\mathsf{L}}_{\backslash(i, i)})\\
&=
\begin{vmatrix}
    N & -1 & \cdots & \cdots & -1\\
    -1 & N & -1 & \cdots & -1\\
    \vdots &-1 &\ddots & \ddots &\vdots\\
    \vdots &\vdots &\ddots &\ddots &-1\\
    -1 & -1 & \cdots & -1 & N
\end{vmatrix}\\
&=
\begin{vmatrix}
    2 & -1 & \cdots & \cdots & -1\\
    2 & N & -1 & \cdots & -1\\
    \vdots &-1 &\ddots & \ddots &\vdots\\
    \vdots &\vdots &\ddots &\ddots &-1\\
    2 & -1 & \cdots & -1 & N
\end{vmatrix}\\
&=
\begin{vmatrix}
    2 & 0 & \cdots & \cdots & 0\\
    2 & N+1 & 0 & \cdots & 0\\
    \vdots &0 &\ddots & \ddots &\vdots\\
    \vdots &\vdots &\ddots &\ddots &0\\
    2 & 0 & \cdots & 0 & N+1
\end{vmatrix}\\
&=2(N+1)^{N-2}.
\end{align*}
In the third line, we added the columns from the second to the $(N-1)$-th column to the first column.
In the fourth line, we added half of the first column, $(1, \cdots, 1)^\top\in\mathbb{R}^{N-1}$, to each of the columns from the second to the $(N-1)$-th column.

\clearpage
\bibliography{biblio.bib}

@book{antsaklis1997linear,
  title={Linear systems},
  author={Antsaklis, Panos J and Michel, Anthony N},
  volume={8},
  year={1997},
  url={https://doi.org/10.1007/0-8176-4435-0},
  publisher={Springer}
}

@book{van1992stochastic,
  title={Stochastic processes in physics and chemistry},
  author={Van Kampen, Nicolaas Godfried},
  volume={1},
  year={1992},
  publisher={Elsevier}
}

@article{khodabandehlou2022trees,
  title={Trees and forests for nonequilibrium purposes: an introduction to graphical representations},
  author={Khodabandehlou, Faezeh and Maes, Christian and Neto{\v{c}}n{\`y}, Karel},
  journal={Journal of Statistical Physics},
  volume={189},
  number={3},
  pages={41},
  year={2022},
  publisher={Springer},
  url={https://link.springer.com/article/10.1007/s10955-022-03003-4}
}

@article{gardner1970connectance,
  title={Connectance of large dynamic (cybernetic) systems: critical values for stability},
  author={Gardner, Mark R and Ashby, W Ross},
  journal={Nature},
  volume={228},
  number={5273},
  pages={784--784},
  year={1970},
  publisher={Nature Publishing Group UK London},
 url={https://www.nature.com/articles/228784a0}
}

@article{may1972will,
  title={Will a large complex system be stable?},
  author={May, Robert M},
  journal={Nature},
  volume={238},
  number={5364},
  pages={413--414},
  year={1972},
  publisher={Nature Publishing Group UK London},
 url={https://www.nature.com/articles/238413a0}
}

@article{allesina2012stability,
  title={Stability criteria for complex ecosystems},
  author={Allesina, Stefano and Tang, Si},
  journal={Nature},
  volume={483},
  number={7388},
  pages={205--208},
  year={2012},
  publisher={Nature Publishing Group UK London},
 url={https://www.nature.com/articles/nature10832}
}

@article{furusawa2015global,
  title={Global relationships in fluctuation and response in adaptive evolution},
  author={Furusawa, Chikara and Kaneko, Kunihiko},
  journal={Journal of The Royal Society Interface},
  volume={12},
  number={109},
  pages={20150482},
  year={2015},
  publisher={The Royal Society},
  url={https://royalsocietypublishing.org/doi/full/10.1098/rsif.2015.0482}
}

@article{sala2000global,
  title={Global biodiversity scenarios for the year 2100},
  author={Sala, Osvaldo E and Stuart Chapin, FIII and Armesto, Juan J and Berlow, Eric and Bloomfield, Janine and Dirzo, Rodolfo and Huber-Sanwald, Elisabeth and Huenneke, Laura F and Jackson, Robert B and Kinzig, Ann and others},
  journal={science},
  volume={287},
  number={5459},
  pages={1770--1774},
  year={2000},
  publisher={American Association for the Advancement of Science},
 url={https://www.science.org/doi/10.1126/science.287.5459.1770}
}

@article{lenski2017experimental,
  title={Experimental evolution and the dynamics of adaptation and genome evolution in microbial populations},
  author={Lenski, Richard E},
  journal={The ISME journal},
  volume={11},
  number={10},
  pages={2181--2194},
  year={2017},
  publisher={Oxford University Press},
  url={https://www.nature.com/articles/ismej201769}
}

@article{balaban2004bacterial,
  title={Bacterial persistence as a phenotypic switch},
  author={Balaban, Nathalie Q and Merrin, Jack and Chait, Remy and Kowalik, Lukasz and Leibler, Stanislas},
  journal={Science},
  volume={305},
  number={5690},
  pages={1622--1625},
  year={2004},
  publisher={American Association for the Advancement of Science},
 url={https://www.science.org/doi/10.1126/science.1099390}
}

@article{aldridge2012asymmetry,
  title={Asymmetry and aging of mycobacterial cells lead to variable growth and antibiotic susceptibility},
  author={Aldridge, Bree B and Fernandez-Suarez, Marta and Heller, Danielle and Ambravaneswaran, Vijay and Irimia, Daniel and Toner, Mehmet and Fortune, Sarah M},
  journal={Science},
  volume={335},
  number={6064},
  pages={100--104},
  year={2012},
  publisher={American Association for the Advancement of Science},
 url={https://www.science.org/doi/10.1126/science.1216166}
}

@article{wakamoto2013dynamic,
  title={Dynamic persistence of antibiotic-stressed mycobacteria},
  author={Wakamoto, Yuichi and Dhar, Neeraj and Chait, Remy and Schneider, Katrin and Signorino-Gelo, Fran{\c{c}}ois and Leibler, Stanislas and McKinney, John D},
  journal={Science},
  volume={339},
  number={6115},
  pages={91--95},
  year={2013},
  publisher={American Association for the Advancement of Science},
 url={https://www.science.org/doi/10.1126/science.1229858}
}

@article{maltas2025dynamic,
  title={Dynamic collateral sensitivity profiles highlight opportunities and challenges for optimizing antibiotic treatments},
  author={Maltas, Jeff and Huynh, Anh and Wood, Kevin B},
  journal={Plos Biology},
  volume={23},
  number={1},
  pages={e3002970},
  year={2025},
  publisher={Public Library of Science San Francisco, CA USA},
 url={https://journals.plos.org/plosbiology/article?id=10.1371/journal.pbio.3002970}
}

@article{markov2023evolution,
  title={{The evolution of SARS-CoV-2}},
  author={Markov, Peter V and Ghafari, Mahan and Beer, Martin and Lythgoe, Katrina and Simmonds, Peter and Stilianakis, Nikolaos I and Katzourakis, Aris},
  journal={Nature Reviews Microbiology},
  volume={21},
  number={6},
  pages={361--379},
  year={2023},
  publisher={Nature Publishing Group UK London},
  url={https://www.nature.com/articles/s41579-023-00878-2}
}

@article{van1962integration,
  title={The integration of chemical and biological control of arthropod pests},
  author={van den Bosch, Robert and Stern, VM},
  journal={Annual Review of Entomology},
  volume={7},
  number={1},
  pages={367--386},
  year={1962},
  publisher={Annual Reviews 4139 El Camino Way, PO Box 10139, Palo Alto, CA 94303-0139, USA},
  url={https://www.annualreviews.org/content/journals/10.1146/annurev.en.07.010162.002055}
}

@article{georghiou1972evolution,
  title={The evolution of resistance to pesticides},
  author={Georghiou, George P},
  journal={Annual Review of Ecology and Systematics},
  pages={133--168},
  year={1972},
  publisher={JSTOR},
  url={https://www.jstor.org/stable/2096845?seq=1}
}

@article{Euler,
  title={Recherches g{\'{e}}n{\'{e}}rales sur la mortalit{\'{e}} et la multiplication du genre humain},
  author={Leonhard Euler},
  journal={M{\'{e}}moires de {l'}acad{\'{e}}mie des sciences de Berlin},
  volume={16},
  pages={144--164},
  year={1767},
  url={https://gallica.bnf.fr/ark:/12148/bpt6k814735.image}
}

@article{lotka1907relation,
  title={Relation between birth rates and death rates},
  author={Lotka, Alfred J},
  journal={Science},
  volume={26},
  number={653},
  pages={21--22},
  year={1907},
  publisher={American Association for the Advancement of Science},
  url={https://www.science.org/doi/10.1126/science.26.653.21.b}
}

@article{powell1956growth,
  title={Growth rate and generation time of bacteria, with special reference to continuous culture},
  author={Powell, E O},
  journal={Microbiology},
  volume={15},
  number={3},
  pages={492--511},
  year={1956},
  publisher={Microbiology Society},
  url={https://www.microbiologyresearch.org/content/journal/micro/10.1099/00221287-15-3-492}
}

@article{pigolotti2021generalized,
  title={{Generalized Euler-Lotka equation for correlated cell divisions}},
  author={Pigolotti, Simone},
  journal={Physical Review E},
  volume={103},
  number={6},
  pages={L060402},
  year={2021},
  publisher={APS},
  url={https://journals.aps.org/pre/abstract/10.1103/PhysRevE.103.L060402}
}

@book{fisher,
	title = {The genetical theory of natural selection},
	copyright = {Not provided. Contact Holding Institution to verify copyright status.},
	url = {https://www.biodiversitylibrary.org/item/69976},
	publisher = {Oxford, Clarendon Press},
	author = {Fisher, Ronald Aylmer},
	year = {1930}
}

@article{EWENS1989167,
title = {An interpretation and proof of the fundamental theorem of natural selection},
author={W. J. Ewens},
journal = {Theoretical Population Biology},
volume = {36},
number = {2},
pages = {167-180},
year = {1989},
issn = {0040-5809},
doi = {https://doi.org/10.1016/0040-5809(89)90028-2},
url = {https://www.sciencedirect.com/science/article/pii/0040580989900282}
}

@article{frank1997price,
  title={{The Price equation, Fisher's fundamental theorem, kin selection, and causal analysis}},
  author={Frank, Steven A},
  journal={Evolution},
  volume={51},
  number={6},
  pages={1712--1729},
  year={1997},
  publisher={Blackwell Publishing Inc Malden, USA},
  url={https://academic.oup.com/evolut}
}

@article{eigen1989molecular,
  title={The molecular quasi-species},
  author={Eigen, Manfred and McCaskill, John and Schuster, Peter},
  journal={Advances in chemical physics},
  volume={75},
  pages={149--263},
  year={1989},
  url={https://pubs.acs.org/doi/10.1021/j100335a010}
}

@article{sughiyama2015pathwise,
  title={Pathwise thermodynamic structure in population dynamics},
  author={Sughiyama, Yuki and Kobayashi, Tetsuya J and Tsumura, Koji and Aihara, Kazuyuki},
  journal={Physical Review E},
  volume={91},
  number={3},
  pages={032120},
  year={2015},
  publisher={APS},
  url={https://journals.aps.org/pre/abstract/10.1103/PhysRevE.91.032120}
}

@article{miyahara2022steady,
  title={Steady-state thermodynamics for population dynamics in fluctuating environments with side information},
  author={Miyahara, Hideyuki},
  journal={Journal of Statistical Mechanics: Theory and Experiment},
  volume={2022},
  number={1},
  pages={013501},
  year={2022},
  publisher={IOP Publishing},
  url={https://iopscience.iop.org/article/10.1088/1742-5468/ac42cc}
}

@article{leibler2010individual,
  title={Individual histories and selection in heterogeneous populations},
  author={Leibler, Stanislas and Kussell, Edo},
  journal={Proceedings of the National Academy of Sciences},
  volume={107},
  number={29},
  pages={13183--13188},
  year={2010},
  publisher={National Academy of Sciences},
  url={https://www.pnas.org/doi/full/10.1073/pnas.0912538107}
}

@article{kobayashi2015fluctuation,
  title={Fluctuation relations of fitness and information in population dynamics},
  author={Kobayashi, Tetsuya J and Sughiyama, Yuki},
  journal={Physical Review Letters},
  volume={115},
  number={23},
  pages={238102},
  year={2015},
  publisher={APS},
  url={https://journals.aps.org/prl/abstract/10.1103/PhysRevLett.115.238102}
}

@article{nozoe2017inferring,
  title={Inferring fitness landscapes and selection on phenotypic states from single-cell genealogical data},
  author={Nozoe, Takashi and Kussell, Edo and Wakamoto, Yuichi},
  journal={PLoS genetics},
  volume={13},
  number={3},
  pages={e1006653},
  year={2017},
  publisher={Public Library of Science San Francisco, CA USA},
  url={https://journals.plos.org/plosgenetics/article?id=10.1371/journal.pgen.1006653}
}

@article{genthon2021universal,
  title={Universal constraints on selection strength in lineage trees},
  author={Genthon, Arthur and Lacoste, David},
  journal={Physical Review Research},
  volume={3},
  number={2},
  pages={023187},
  year={2021},
  publisher={APS},
  url={https://journals.aps.org/prresearch/abstract/10.1103/PhysRevResearch.3.023187}
}

@article{garcia2019linking,
  title={Linking lineage and population observables in biological branching processes},
  author={Garc{\'\i}a-Garc{\'\i}a, Reinaldo and Genthon, Arthur and Lacoste, David},
  journal={Physical Review E},
  volume={99},
  number={4},
  pages={042413},
  year={2019},
  publisher={APS},
  url={https://journals.aps.org/pre/abstract/10.1103/PhysRevE.99.042413}
}

@article{genthon2020fluctuation,
  title={Fluctuation relations and fitness landscapes of growing cell populations},
  author={Genthon, Arthur and Lacoste, David},
  journal={Scientific Reports},
  volume={10},
  number={1},
  pages={11889},
  year={2020},
  publisher={Nature Publishing Group UK London},
  url={https://www.nature.com/articles/s41598-020-68444-x}
}

@article{scott2010interdependence,
  title={Interdependence of cell growth and gene expression: origins and consequences},
  author={Scott, Matthew and Gunderson, Carl W and Mateescu, Eduard M and Zhang, Zhongge and Hwa, Terence},
  journal={Science},
  volume={330},
  number={6007},
  pages={1099--1102},
  year={2010},
  publisher={American Association for the Advancement of Science},
  url={https://www.science.org/doi/10.1126/science.1192588}
}

@article{wang2010robust,
  title={Robust growth of Escherichia coli},
  author={Wang, Ping and Robert, Lydia and Pelletier, James and Dang, Wei Lien and Taddei, Francois and Wright, Andrew and Jun, Suckjoon},
  journal={Current biology},
  volume={20},
  number={12},
  pages={1099--1103},
  year={2010},
  publisher={Elsevier},
  url={https://www.sciencedirect.com/science/article/pii/S0960982210005245}
}

@article{lambert2015quantifying,
  title={Quantifying selective pressures driving bacterial evolution using lineage analysis},
  author={Lambert, Guillaume and Kussell, Edo},
  journal={Physical review X},
  volume={5},
  number={1},
  pages={011016},
  year={2015},
  publisher={APS},
  url={https://journals.aps.org/prx/abstract/10.1103/PhysRevX.5.011016}
}

@article{hashimoto2016noise,
  title={Noise-driven growth rate gain in clonal cellular populations},
  author={Hashimoto, Mikihiro and Nozoe, Takashi and Nakaoka, Hidenori and Okura, Reiko and Akiyoshi, Sayo and Kaneko, Kunihiko and Kussell, Edo and Wakamoto, Yuichi},
  journal={Proceedings of the National Academy of Sciences},
  volume={113},
  number={12},
  pages={3251--3256},
  year={2016},
  publisher={National Academy of Sciences},
  url={https://www.pnas.org/doi/10.1073/pnas.1519412113}
}

@book{crow2017introduction,
  title={An introduction to population genetics theory},
  author={Crow, James Franklin and Kimura, Motoo},
  year={2017},
  publisher={Scientific Publishers}
}

@article{hermisson2002mutation,
  title={Mutation--selection balance: ancestry, load, and maximum principle},
  author={Hermisson, Joachim and Redner, Oliver and Wagner, Holger and Baake, Ellen},
  journal={Theoretical population biology},
  volume={62},
  number={1},
  pages={9--46},
  year={2002},
  publisher={Elsevier},
  url={https://www.sciencedirect.com/science/article/pii/S0040580902915820?via%3Dihub}
}

@article{baake2007mutation,
  title={Mutation, selection, and ancestry in branching models: a variational approach},
  author={Baake, Ellen and Georgii, Hans-Otto},
  journal={Journal of mathematical biology},
  volume={54},
  pages={257--303},
  year={2007},
  publisher={Springer},
  url={https://link.springer.com/article/10.1007/s00285-006-0039-5}
}

@article{baake1997ising,
  title={Ising quantum chain is equivalent to a model of biological evolution},
  author={Baake, Ellen and Baake, Michael and Wagner, Holger},
  journal={Physical Review Letters},
  volume={78},
  number={3},
  pages={559},
  year={1997},
  publisher={APS},
  url={https://journals.aps.org/prl/pdf/10.1103/PhysRevLett.78.559?casa_token=1Qbi1WNq26IAAAAA%3AzLy8whKESDRuXzEy9S0nsO_Uat6LTH_T8oF7-IOUleMD0WApnjgy9hgnl9Aw48E_j0wEK9t9_WyG91JR}
}

@article{saakian2007new,
  title={A new method for the solution of models of biological evolution: Derivation of exact steady-state distributions},
  author={Saakian, David B},
  journal={Journal of statistical physics},
  volume={128},
  pages={781--798},
  year={2007},
  publisher={Springer},
  url={https://link.springer.com/article/10.1007/s10955-007-9334-9}
}

@article{saakian2008dynamics,
  title={Dynamics of the Eigen and the Crow-Kimura models for molecular evolution},
  author={Saakian, David B and Rozanova, Olga and Akmetzhanov, Andrei},
  journal={Physical Review E},
  volume={78},
  number={4},
  pages={041908},
  year={2008},
  publisher={APS},
  url={https://journals.aps.org/pre/abstract/10.1103/PhysRevE.78.041908}
}

@article{munoz2009solution,
  title={Solution of the Crow-Kimura and Eigen models for alphabets of arbitrary size by Schwinger spin coherent states},
  author={Mu{\~n}oz, Enrique and Park, Jeong-Man and Deem, Michael W},
  journal={Journal of Statistical Physics},
  volume={135},
  pages={429--465},
  year={2009},
  publisher={Springer},
  url={https://link.springer.com/article/10.1007/s10955-009-9732-2}
}

@article{bratus2014linear,
  title={Linear algebra of the permutation invariant Crow--Kimura model of prebiotic evolution},
  author={Bratus, Alexander S and Novozhilov, Artem S and Semenov, Yuri S},
  journal={Mathematical biosciences},
  volume={256},
  pages={42--57},
  year={2014},
  publisher={Elsevier},
  url={https://www.sciencedirect.com/science/article/pii/S0025556414001503?casa_token=fart9JLmKQoAAAAA:GY9u4uv9b-bC_8GORHDFijxZqam4so9MaGTDZxP5hNhJiDTJ_eyCpx41ZnlZvRmqb7iTyjo9w9Sw}
}

@article{SEMENOV20151,
title = {{Exact solutions for the selection–mutation equilibrium in the Crow–Kimura evolutionary model}},
journal = {Mathematical Biosciences},
volume = {266},
pages = {1-9},
year = {2015},
issn = {0025-5564},
author = {Yuri S. Semenov and Artem S. Novozhilov},
url={https://www.sciencedirect.com/science/article/pii/S0025556415001029}
}

@book{moon1970counting,
  title={Counting Labelled Trees, by J.W. Moon},
  author={Moon, J.W.},
  series={Canadian mathematical monographs},
  url={https://books.google.co.jp/books?id=B6dDcgAACAAJ},
  year={1970}
}

@article{chaiken1982combinatorial,
  title={A combinatorial proof of the all minors matrix tree theorem},
  author={Chaiken, Seth},
  journal={SIAM Journal on Algebraic Discrete Methods},
  volume={3},
  number={3},
  pages={319--329},
  year={1982},
  publisher={SIAM},
  url={https://epubs.siam.org/doi/10.1137/0603033}
}

@article{Cayley1856,
author = {Cayley, A.},
journal = {Journal f\"{u}r die reine und angewandte Mathematik},
pages = {276-284},
title = {Note sur une formule pour la reversion des s{\'{e}}ries.},
url = {http://eudml.org/doc/147652},
volume = {52},
year = {1856},
}

@article{borchardt1860ueber,
  title={{Ueber eine der Interpolation entsprechende Darstellung der Eliminations-Resultante.}},
  author={Borchardt, Carl Wilhelm},
  year={1860},
  journal={Journal für die reine und angewandte Mathematik},
  url={https://eudml.org/doc/147779}
}

@article{Kirchhoff,
author = {Kirchhoff, G.},
title = {Ueber die Auflösung der Gleichungen, auf welche man bei der Untersuchung der linearen Vertheilung galvanischer Ströme geführt wird},
journal = {Annalen der Physik},
volume = {148},
number = {12},
pages = {497-508},
doi = {https://doi.org/10.1002/andp.18471481202},
url = {https://onlinelibrary.wiley.com/doi/abs/10.1002/andp.18471481202},
year = {1847}
}

@book{maxwell1892treatise,
  author    = {J. C. Maxwell},
  title     = {A Treatise on Electricity and Magnetism},
  edition   = {3rd},
  publisher = {Clarendon Press},
  address   = {Oxford},
  year      = {1892}
}

@article{hill1966studies,
  title={Studies in irreversible thermodynamics IV. Diagrammatic representation of steady state fluxes for unimolecular systems},
  author={Hill, Terrell L},
  journal={Journal of theoretical biology},
  volume={10},
  number={3},
  pages={442--459},
  year={1966},
  publisher={Elsevier},
  url={https://www.sciencedirect.com/science/article/pii/0022519366901378}
}

@article{schnakenberg1976network,
  title={Network theory of microscopic and macroscopic behavior of master equation systems},
  author={Schnakenberg, J{\"u}rgen},
  journal={Reviews of Modern physics},
  volume={48},
  number={4},
  pages={571},
  year={1976},
  publisher={APS},
  url={https://journals.aps.org/rmp/abstract/10.1103/RevModPhys.48.571}
}

@article{avanzini2024methods,
  title={Methods and conversations in (post) modern thermodynamics},
  author={Avanzini, Francesco and Bilancioni, Massimo and Cavina, Vasco and Dal Cengio, Sara and Esposito, Massimiliano and Falasco, Gianmaria and Forastiere, Danilo and Freitas, Jose Nahuel and Garilli, Alberto and Harunari, Pedro E and others},
  journal={SciPost Physics Lecture Notes},
  pages={080},
  year={2024},
  url={https://scipost.org/10.21468/SciPostPhysLectNotes.80}
}

@article{polettini2017effective,
  title={Effective thermodynamics for a marginal observer},
  author={Polettini, Matteo and Esposito, Massimiliano},
  journal={Physical Review Letters},
  volume={119},
  number={24},
  pages={240601},
  year={2017},
  publisher={APS},
  url={https://journals.aps.org/prl/abstract/10.1103/PhysRevLett.119.240601}
}

@inproceedings{maes2013heat,
  title={Heat bounds and the blowtorch theorem},
  author={Maes, Christian and Neto{\v{c}}n{\`y}, Karel},
  booktitle={Annales Henri Poincar{\'e}},
  volume={14},
  pages={1193--1202},
  year={2013},
  organization={Springer},
  url={https://link.springer.com/article/10.1007/s00023-012-0214-8}
}

@article{owen2020universal,
  title={Universal thermodynamic bounds on nonequilibrium response with biochemical applications},
  author={Owen, Jeremy A and Gingrich, Todd R and Horowitz, Jordan M},
  journal={Physical Review X},
  volume={10},
  number={1},
  pages={011066},
  year={2020},
  publisher={APS},
  url={https://journals.aps.org/prx/abstract/10.1103/PhysRevX.10.011066}
}

@article{fernandes2023topologically,
  title={Topologically constrained fluctuations and thermodynamics regulate nonequilibrium response},
  author={Fernandes Martins, Gabriela and Horowitz, Jordan M},
  journal={Physical Review E},
  volume={108},
  number={4},
  pages={044113},
  year={2023},
  publisher={APS},
  url={https://journals.aps.org/pre/abstract/10.1103/PhysRevE.108.044113}
}

@article{dal2023geometry,
  title={Geometry of nonequilibrium reaction networks},
  author={Dal Cengio, Sara and Lecomte, Vivien and Polettini, Matteo},
  journal={Physical Review X},
  volume={13},
  number={2},
  pages={021040},
  year={2023},
  publisher={APS},
  url={https://journals.aps.org/prx/abstract/10.1103/PhysRevX.13.021040}
}

@article{chun2023trade,
  title={Trade-offs between number fluctuations and response in nonequilibrium chemical reaction networks},
  author={Chun, Hyun-Myung and Horowitz, Jordan M},
  journal={The Journal of Chemical Physics},
  volume={158},
  number={17},
  year={2023},
  publisher={AIP Publishing},
  url={https://pubs.aip.org/aip/jcp/article/158/17/174115/2888610/Trade-offs-between-number-fluctuations-and}
}

@article{liang2024thermodynamic,
  title={Thermodynamic bounds on symmetry breaking in linear and catalytic biochemical systems},
  author={Liang, Shiling and De Los Rios, Paolo and Busiello, Daniel Maria},
  journal={Physical Review Letters},
  volume={132},
  number={22},
  pages={228402},
  year={2024},
  publisher={APS},
  url={https://journals.aps.org/prl/abstract/10.1103/PhysRevLett.132.228402}
}

@article{harunari2024mutual,
  title={Mutual linearity of nonequilibrium network currents},
  author={Harunari, Pedro E and Dal Cengio, Sara and Lecomte, Vivien and Polettini, Matteo},
  journal={Physical Review Letters},
  volume={133},
  number={4},
  pages={047401},
  year={2024},
  publisher={APS},
  url={https://journals.aps.org/prl/abstract/10.1103/PhysRevLett.133.047401}
}

@article{floyd2024learning,
  title={Learning to control non-equilibrium dynamics using local imperfect gradients},
  author={Floyd, Carlos and Dinner, Aaron R and Vaikuntanathan, Suriyanarayanan},
  journal={arXiv preprint arXiv:2404.03798},
  year={2024},
  url={https://arxiv.org/abs/2404.03798}
}

@article{floyd2024limits,
  title={Limits on the computational expressivity of non-equilibrium biophysical processes},
  author={Floyd, Carlos and Dinner, Aaron R and Murugan, Arvind and Vaikuntanathan, Suriyanarayanan},
  journal={arXiv preprint arXiv:2409.05827},
  year={2024},
  url={https://arxiv.org/abs/2409.05827}
}

@article{pillai2005perron,
  title={{The Perron-Frobenius theorem: some of its applications}},
  author={Pillai, S Unnikrishna and Suel, Torsten and Cha, Seunghun},
  journal={IEEE Signal Processing Magazine},
  volume={22},
  number={2},
  pages={62--75},
  year={2005},
  publisher={IEEE},
  url={https://ieeexplore.ieee.org/stamp/stamp.jsp?arnumber=1406483}
}

@book{godsil2013algebraic,
  title={Algebraic graph theory},
  author={Godsil, Chris and Royle, Gordon F},
  volume={207},
  year={2013},
  publisher={Springer Science \& Business Media},
  url={https://link.springer.com/book/10.1007/978-1-4613-0163-9}
}

@book{west2001introduction,
  title={Introduction to graph theory},
  author={West, Douglas Brent},
  volume={2},
  year={2001},
  publisher={Prentice hall Upper Saddle River}
}

@article{baym2016multidrug,
  title={Multidrug evolutionary strategies to reverse antibiotic resistance},
  author={Baym, Michael and Stone, Laura K and Kishony, Roy},
  journal={Science},
  volume={351},
  number={6268},
  pages={aad3292},
  year={2016},
  publisher={American Association for the Advancement of Science},
  url={https://www.science.org/doi/10.1126/science.aad3292}
}

@article{tyers2019drug,
  title={Drug combinations: a strategy to extend the life of antibiotics in the 21st century},
  author={Tyers, Mike and Wright, Gerard D},
  journal={Nature Reviews Microbiology},
  volume={17},
  number={3},
  pages={141--155},
  year={2019},
  publisher={Nature Publishing Group UK London},
  url={https://www.google.com/search?client=safari&rls=en&q=Drug+combinations%3A+a+strategy+to+extend+the+life+of+antibiotics+in+the+21st+century&ie=UTF-8&oe=UTF-8}
}

@article{molina2025optimization,
  title={Optimization of sequential therapies to maximize extinction of resistant bacteria through collateral sensitivity},
  author={Molina-Hern{\'a}ndez, Javier and Cuesta, Jos{\'e} A and Pascual-Escudero, Beatriz and Ares, Sa{\'u}l and Catal{\'a}n, Pablo},
  journal={arXiv preprint arXiv:2510.01808},
  year={2025},
  url={https://arxiv.org/abs/2510.01808}
}

@article{mokhtari2017combination,
  title={Combination therapy in combating cancer},
  author={Mokhtari, Reza Bayat and Homayouni, Tina S and Baluch, Narges and Morgatskaya, Evgeniya and Kumar, Sushil and Das, Bikul and Yeger, Herman},
  journal={Oncotarget},
  volume={8},
  number={23},
  pages={38022},
  year={2017},
  url={https://pmc.ncbi.nlm.nih.gov/articles/PMC5514969/}
}

@article{tamma2012combination,
  title={Combination therapy for treatment of infections with gram-negative bacteria},
  author={Tamma, Pranita D and Cosgrove, Sara E and Maragakis, Lisa L},
  journal={Clinical microbiology reviews},
  volume={25},
  number={3},
  pages={450--470},
  year={2012},
  publisher={American Society for Microbiology 1752 N St., NW, Washington, DC},
  url={https://pmc.ncbi.nlm.nih.gov/articles/PMC3416487/}
}

@article{quinn2015pancreatic,
  title={Pancreatic cancer combination therapy using a BH3 mimetic and a synthetic tetracycline},
  author={Quinn, Bridget A and Dash, Rupesh and Sarkar, Siddik and Azab, Belal and Bhoopathi, Praveen and Das, Swadesh K and Emdad, Luni and Wei, Jun and Pellecchia, Maurizio and Sarkar, Devanand and others},
  journal={Cancer research},
  volume={75},
  number={11},
  pages={2305--2315},
  year={2015},
  publisher={American Association for Cancer Research},
  url={https://aacrjournals.org/cancerres/article/75/11/2305/599825/Pancreatic-Cancer-Combination-Therapy-Using-a-BH3}
}

@article{chou1981generalized,
  title={Generalized equations for the analysis of inhibitions of Michaelis-Menten and higher-order kinetic systems with two or more mutually exclusive and nonexclusive inhibitors},
  author={CHOU, Ting-Chao and TaLaLay, Paul},
  journal={European journal of biochemistry},
  volume={115},
  number={1},
  pages={207--216},
  year={1981},
  publisher={Wiley Online Library},
  url={https://febs.onlinelibrary.wiley.com/doi/pdf/10.1111/j.1432-1033.1981.tb06218.x}
}

@article{pena2014testing,
  title={Testing the optimality properties of a dual antibiotic treatment in a two-locus, two-allele model},
  author={Pena-Miller, Rafael and Fuentes-Hernandez, Ayari and Reding, Carlos and Gudelj, Ivana and Beardmore, Robert},
  journal={Journal of The Royal Society Interface},
  volume={11},
  number={96},
  pages={20131035},
  year={2014},
  publisher={The Royal Society},
  url={https://pmc.ncbi.nlm.nih.gov/articles/PMC4032525/pdf/rsif20131035.pdf}
}

@article{david1970probability,
  title={Probability distribution of drug-resistant mutants in unselected populations of Mycobacterium tuberculosis},
  author={David, Hugo L},
  journal={Applied microbiology},
  volume={20},
  number={5},
  pages={810--814},
  year={1970},
  url={https://journals.asm.org/doi/pdf/10.1128/am.20.5.810-814.1970}
}

@article{bergval2009resistant,
  title={Resistant mutants of Mycobacterium tuberculosis selected in vitro do not reflect the in vivo mechanism of isoniazid resistance},
  author={Bergval, Indra L and Schuitema, Anja RJ and Klatser, Paul R and Anthony, Richard M},
  journal={Journal of Antimicrobial Chemotherapy},
  volume={64},
  number={3},
  pages={515--523},
  year={2009},
  publisher={Oxford University Press},
  url={https://academic.oup.com/jac/article-pdf/64/3/515/13759219/dkp237.pdf}
}

\end{document}